%% file: AML.tex
\documentclass{article}

\usepackage{macros}
\usepackage{macros-gen-en}
\usepackage{macros-math}
\usepackage{macros-logic}
\usepackage{macros-ML-2025-05-15}

\usepackage[toc]{appendix}

\begin{document}

\title{Notes on applicative matching logic}
\author{Lauren{\c t}iu Leu{\c s}tean${}^{a,b,c}$\\[2mm]	
	\footnotesize ${}^{a}$ Institute for Logic and Data Science, Bucharest\\[1mm]
	\footnotesize ${}^{b}$ LOS, Faculty of Mathematics and Computer Science, University of Bucharest,\\[1mm]
	\footnotesize ${}^{c}$ Simion Stoilow Institute of Mathematics of the Romanian Academy\\
}

\date{}

\maketitle

\setlength{\parindent}{0em}

\tableofcontents

\newpage

\input{intro}

\newpage


\input{AML-language}

\input{AML-language-unique-read}

\input{AML-language-recursion}
\input{AML-language-subpatterns}

\input{AML-language-free-bound}

\input{AML-language-positive}

\input{AML-substitution}

\input{AML-substitution-svar}

\input{AML-substitution-free}

\input{AML-language-contexts}

\input{AML-language-derived-conventions}

\input{AML-semantics}

\input{AML-semantics-models-valid-log-equiv}

\input{AML-semantics-finite-conj-disj}

\input{AML-semantics-consequence}
\input{AML-semantics-sets-patterns}

\input{AML-semantics-sets-consequence}

\input{AML-semantics-closed-patterns}
\input{AML-semantics-predicates}

\input{AML-semantics-tautologies}

\input{AML-semantics-tautologies-examples}

\input{AML-semantics-change-val}
\input{AML-semantics-substitution}

\input{AML-semantics-substitution-svar}

\input{AML-semantics-nu}
\input{AML-semantics-lfn-lgn}

\input{AML-semantics-examples}

\input{AML-semantics-examples-rules}

\input{AML-semantics-examples-application}
\input{AML-semantics-examples-contexts}

\input{AML-semantics-examples-set-variables}
\input{AML-semantics-examples-mu-nu}

\input{AML-syntax}

\input{AML-syntax-derived-rules}

\input{AML-soundness}

\input{AML-semantics-def-equal-tot}

\newpage

\appendix

\input{appendix-proof-system-AML}

\input{appendix-set-theory}
\input{appendix-set-theory-app}
\input{appendix-KnasterTarski}

\end{document}

%% file: intro.tex
\section{Introduction}

Matching logic was introduced by Grigore Ro\c{s}u \cite{RosEllSch10,Ros17}. 

and collaborators   
as a logic  for defining the formal semantics of programming languages and for specifying  
and reasoning about the behavior of programs. Applicative Matching Logic (AML), a functional variant  of ML,
was introduced  recently \cite{CheRos19a} and developed further in \cite{CheRos20,CheLucRos21}

These lecture notes present basic definitions and results on AML, with very rigorous and  extremely 
detailed proofs. They can be used as an introductory text in the theory of AML.  We point out that the definition of 
patterns in these notes is more general than the one from  \cite{CheRos19a}, as we do not require the positivity of 
the pattern in the definition of the $\mu$ binder. 

Monk's textbook on mathematical logic \cite{Mon76} has an enormous influence on the notes. A number of notations and results for first-order logic from Monk's book are adapted to AML.
The set-theoretic notions and properties used in the notes are given in Appendix~\ref{app-set-theory}.

\section*{Acknowledgements}

This research was supported by sponsorships from Runtime Verification Inc. and Pi Squared Inc.

The author is grateful to Dafina Trufa\c{s}, Traian \c{S}erb\u{a}nu\c{t}\u{a}, Grigore Ro\c{s}u, 
Andrei Sipo\c{s}, and C\u{a}t\u{a}lina Isache for reading different
versions of these notes and providing corrections and suggestions for improvements.

%% file: AML-language.tex
\section{Language}

\bdfn	An \defnterm{applicative matching logic (AML) signature} (or simply \defnterm{signature}) is a triple 
$\AMLsig=(EVar, SVar, \Sigma)$, 
where 
\be
\item $EVar=\{v_n\mid n\in\N\}$ is a  countable set of \defnterm{element variables}.
\item $SVar=\{V_n\mid n\in\N\}$ is a  countable set of \defnterm{set variables}.
\item $\Sigma$ is a set of \defnterm{constants}.
\ee
The sets $EVar$, $SVar$ and $\Sigma$ are pairwise disjoint.
\edfn

We denote element variables  by $x,y,z,x_1, x_2, \ldots$, set variables by $X, Y, Z, X_1, X_2, \ldots$
and symbols by $\sigma, f,g, \ldots$. \\

In the sequel, $\AMLsig=(EVar, SVar, \Sigma)$ is a signature.

\bdfn
The set $Sym_{\AMLsig}$ of \defnterm{$\AMLsig$-symbols}  is defined as 
$$Sym_{\AMLsig}=EVar\cup SVar\cup\{\to,\exists,\mu,\Appl\} \cup \Sigma.$$
\edfn

\bdfn
The set $Expr_{\AMLsig}$ of \defnterm{$\AMLsig$-expressions} is the set of all expressions
over $Sym_{\AMLsig}$.
\edfn

\bdfn	
An \defnterm{atomic $\AMLsig$-pattern} is an element variable, a set variable or a constant. 
We shall use the notation $\setatomicpat_\AMLsig$ for the set of atomic $\AMLsig$-patterns.
\edfn

\bdfn\label{def-pats}
The \defnterm{$\AMLsig$-patterns}  are the $\AMLsig$-expressions inductively defined as follows:
\be
\item Every atomic $\AMLsig$-pattern is a $\AMLsig$-pattern.
\item If $\vp$ and $\psi$ are $\AMLsig$-patterns, then $\Appl\vp\psi$ is a $\AMLsig$-pattern.
\item If $\vp$ and $\psi$ are $\AMLsig$-patterns, then $\to\vp\psi$ is a $\AMLsig$-pattern.
\item If  $\vp$ is a $\AMLsig$-pattern and $x$ is an element variable, then  $\exists x\vp$ is a $\AMLsig$-pattern.
\item If  $\vp$ is a $\AMLsig$-pattern and $X$ is a set variable, then  $\mu X\vp$ is a $\AMLsig$-pattern.
\item Only the expressions obtained by applying the above rules  are $\AMLsig$-patterns.
\ee
\edfn

We use the Polish notation in the definition of $\AMLsig$-patterns as this notation allows us 
to obtain the unique readability of $\AMLsig$-patterns (see Proposition~\ref{unique-read-patterns}), 
a fundamental property.

The set of $\AMLsig$-patterns is denoted by $\setpat_\AMLsig$ and  $\AMLsig$-patterns are denoted 
by  $\vp,\psi,\chi, \ldots$. 

For any $\AMLsig$-pattern $\vp$, we denote by \defnterm{$EVar(\vp)$} 
the set of element variables occuring in $\vp$ and by \defnterm{$SVar(\vp)$} the set 
of set variables occuring in $\vp$.

\bfact
The definition of $\AMLsig$-patterns can be written using the BNF notation:
\begin{align*}
\vp  ::= &  x\in EVar \mid X\in SVar \mid \sigma \in \Sigma \mid  Appl\vp \vp \mid \to\vp \vp \\[1mm]
&  \mid \exists x \vp  \text{~if~}x\in EVar\\[1mm]
& \mid \mu X\vp  \text{~if~}x\in SVar.
\end{align*}
\efact

\bdfn[Alternative definition for $\AMLsig$-patterns]\label{def-pats-equiv}\, \\
The set of \defnterm{$\AMLsig$-patterns}  is the intersection of all 
sets $\Gamma$ of $\AMLsig$-expressions that have the following properties:
\be
\item $\Gamma$ contains all atomic $\AMLsig$-patterns.
\item  $\Gamma$ is closed to $\Appl$,   $\to$,  $\exists x$  (for any element variable $x$), 
and  $\mu X$ (for any set variable $X$), that is:
\bce
if $\vp,\psi\in\Gamma$, then $\Appl\vp\psi$, $\to\vp\psi$, $\exists x\vp$, $\mu X\vp\in\Gamma$.
\ece
\ee
\edfn

When the signature $\AMLsig$ is clear from the context, we shall write simply expression(s), pattern(s) 
and we shall denote the set of expressions by $Expr$, the set of 
patterns by $\setpat$, the set of atomic patterns by $\setatomicpat$, etc.. 

\begin{proposition}[Induction principle on patterns]\label{induction-pat}\, \\
Let $\Gamma$  be a set of patterns  satisfying the following properties:
\be
\item $\Gamma$ contains all atomic patterns.
\item  $\Gamma$ is closed  to  $\Appl$, $\to$,  $\exists x$  (for any element variable $x$), and  $\mu X$ (for any set variable $X$).
\ee
Then $\Gamma=\setpat$.
\end{proposition}
\solution{By hypothesis, $\Gamma \se \setpat$. By Definition~\ref{def-pats-equiv}, we get that 
$ \setpat\se \Gamma$.
}

Induction principle on patterns is used to prove that all patterns have a property $\cP$: we define 
$\Gamma$ as the set of all patterns satisfying  $\cP$ and apply 
induction on patterns to obtain that $\Gamma=\setpat$.

%% file: AML-language-unique-read.tex
\subsection{Unique readability}

\bprop[Unique readability of patterns]\label{unique-read-patterns}\,
\be
\item\label{pattern-positive} Any pattern has a positive length.
\item\label{pattern-read} If $\vp$ is a pattern, then one of the following hold:
\be
\item $\vp = x$, where $x\in EVar$.
\item $\vp = X$, where $X\in SVar$.
\item $\vp = \sigma$, where $\sigma\in \Sigma$.
\item $\vp  = \Appl\psi\chi$, where $\psi,\chi$ are patterns.
\item $\vp  = \to\psi\chi$, where $\psi,\chi$ are patterns.
\item $\vp  = \exists x\psi$, where $x$ is an element variable and $\psi$ is a pattern. 
\item $\vp  = \mu X\psi$, where $X$ is a set variable and $\psi$ is a pattern. 
\ee
\item\label{pattern-initial-segment} Any proper initial segment of a pattern is not a pattern.
\item\label{pattern-unique-read} If  $\vp$ is a pattern, then exactly one of the cases from \eqref{pattern-read}
holds. Moreover, $\vp$  can be written in a unique way in one of these forms. 
\ee 
\eprop
\solution{
\be
\item Let $\Gamma$ be the set of patterns of positive length. We prove that $\Gamma=\setpat$ using 
the Induction principle on patterns (Proposition~\ref{induction-pat}). 
\be
\item If $\vp$ is an atomic pattern, then its length is $1$, so $\vp\in \Gamma$.
\item If $\vp,\psi \in \Gamma$, hence they have positive length, then obviously 
the patterns $\Appl\vp\psi$, $\to\vp\psi$, $\exists x\vp$, $\mu X\vp\in\Gamma$ have positive 
length, hence they are in $\Gamma$. 
\ee

\item Let $\Gamma_1=\{\Appl\psi\chi \mid \psi,\chi \in \setpat\}$, $\Gamma_2=\{\to\psi\chi \mid \psi,\chi \in \setpat\}$, 
$\Gamma_3=\{\exists x\psi \mid x\in EVar \text{~ and~} \psi\in \setpat\}$ and 
$\Gamma_4=\{\mu X\psi \mid X\in SVar \text{~ and~} \psi\in \setpat\}$.
Define 
\begin{align*}
\Gamma & = EVar \cup SVar \cup \Sigma \cup \Gamma_1 \cup \Gamma_2 \cup \Gamma_3 \cup \Gamma_4.
\end{align*}
Then obviously, $\Gamma\se \setpat$. We prove that $\Gamma=\setpat$ using the 
Induction principle on patterns (Proposition~\ref{induction-pat}).
\be
\item As  $EVar \cup SVar \cup \Sigma \se \Gamma$, we have that $\Gamma$ contains all atomic patterns.
\item  Let $\vp, \psi\in \Gamma$, $x\in EVar$ and $X\in SVar$. Then 
$\Appl\vp\psi\in \Gamma_1\se \Gamma$, $\to\vp\psi\in \Gamma_2\se \Gamma$, $\exists x\vp\in \Gamma_3\se \Gamma$, and
 $\mu X\vp\in \Gamma_4\se \Gamma$. Thus, $\Gamma$ is closed to  $\Appl$, $\to$,  $\exists x$, and  $\mu X$.
\ee
\item As, by \eqref{pattern-positive}, patterns have positive length, it follows that we have to prove that for all $n\geq 1$,

\bt{ll}
(P) & if $\vp=\vp_0\ldots \vp_{n-1}$ is a pattern of length $n$, then for any  $0\le i<n-1$, \\[1mm]
& $\vp=\vp_0\ldots \vp_i$ is not a pattern.
\et

The proof is by induction on $n$.

$n=1$:  Then one cannot have $0\leq i < 0$, hence (P) holds.

Assume that $n>1$ and that (P) holds for any pattern of length $<n$. Let $\vp=\vp_0\ldots \vp_{n-1}$ be a pattern of length $n$.
By \eqref{pattern-read}, we have the following cases:
\be
\item $\vp=\star\psi\chi$, where $\star\in\{\Appl, \to\}$ and $\psi$, $\chi$ are patterns. Thus, $\vp_0=\star$,  
$\psi=\vp_1\ldots \vp_{k-1}$ and 
$\chi=\vp_k\ldots \vp_{n-1}$, where $2\leq k \leq n-1$.  

Let $0\le i<n-1$ and assume, by contradiction, that $\vp_0\ldots \vp_i$ is a pattern.
Applying again \eqref{pattern-read}, it follows that $\vp_0\ldots \vp_i=\star\psi^1\chi^1$, where $\psi^1$, $\chi^1$ are patterns.
Thus, $\psi^1=\vp_1\ldots \vp_{p-1}$, $\chi^1=\vp_p\ldots \vp_i$, where $2\leq p \leq i$. We have the following cases:
\be
\item $p<k$. Then $\psi^1$ is a proper initial segment of $\psi$. As the length of $\psi$ is $<n$, we can apply  
the induction hypothesis to get that $\psi^1$ is not a pattern. We have obtained a contradiction.
\item $p=k$. Then $\psi^1=\psi$ and $\chi^1$ is a proper initial segment of $\chi$. As the length of $\chi$ is $<n$, we can apply  
the induction hypothesis  to get that $\chi^1$ is not a pattern. We have obtained a contradiction.
\item $p>k$. Then $\psi$ is a proper initial segment of $\psi^1$. As the length of $\psi^1$ is $<n$, we can apply  
the induction hypothesis to get that $\psi$ is not a pattern. We have obtained a contradiction.
\ee
\item $\vp=\theta\psi$, where $\theta \in \{\exists x, \mu X\}$ for some $x\in EVar$ and $X\in SVar$ and $\psi$ is a pattern.
Then $\vp_0\vp_1=\theta$ and $\psi=\vp_2\ldots \vp_{n-1}$. Let $0\le i<n-1$ and assume, by contradiction, 
that $\vp_0\ldots \vp_i$ is a pattern. Applying again \eqref{pattern-read}, it follows that $\vp_0\ldots \vp_i=\theta\psi^1$, 
where $\psi^1=\vp_2\ldots \vp_i$ is a pattern. Then $\psi^1$ is a proper initial segment of $\psi$. 
As the length of $\psi$ is $<n$, we can apply  the induction hypothesis to get that $\psi^1$ is 
not a pattern. We have obtained a contradiction.
\ee
\item is an immediate consequece of \eqref{pattern-read} and \eqref{pattern-initial-segment}.
\ee
}

The following results will be useful in Subsections~\ref{language-free-bound} and \ref{language-positive}.

\bprop\label{unique-subpat-exists-mu}
Let $\vp=\vp_0\vp_1\ldots\vp_{n-1}$ be a pattern and suppose that $\vp_i\in\{\exists,\mu\}$ for some $i=0,\ldots,n-1$.
Then there exists a unique $j$ such that $i<j\leq n-1$ and
$\vp_i\ldots \vp_j$ is a pattern.
\eprop
\solution{Let us prove first the uniqueness. Assume, by contradiction, that 
$i<j<k\leq n-1$ are such that $\vp_i\ldots \vp_j$  and $\vp_i\ldots \vp_k$ are both patterns. As $\vp_i\ldots \vp_j$  is a
proper initial segment of $\vp_i\ldots \vp_k$, it follows, by Proposition~\ref{unique-read-patterns}.\eqref{pattern-initial-segment}
that $\vp_i\ldots \vp_j$ is not a pattern. We have obtained a contradiction.

Let us prove in the sequel the existence. 

As, by Proposition~\ref{unique-read-patterns}.\eqref{pattern-positive}, patterns have positive length, it follows that we have to prove that for all $n\geq 1$,

\bt{ll}
(P) & if $\vp=\vp_0\ldots \vp_{n-1}$ is a pattern of length $n$ and $\vp_i\in\{\exists,\mu\}$ for some $i=0,\ldots,n-1$, \\[1mm]
& then there exists $j$ such that $i<j\leq n-1$ and
$\vp_i\ldots \vp_j$ is a pattern. \\[1mm]
\et

The proof is by induction on $n$. 

$n=1$. Then $\vp=\vp_0$ is an atomic pattern, so there exists no $i$ satisfying the premise in (P), hence (P) holds.

Assume that $n>1$ and that (P) holds for any pattern of length $<n$. Let $\vp=\vp_0\ldots \vp_{n-1}$ be a pattern of length $n$ 
such that $\vp_i\in\{\exists,\mu\}$ for some $i=0,\ldots,n-1$. 
By Proposition~\ref{unique-read-patterns}.\eqref{pattern-read}, we have the following cases:
\be
\item $\vp=\applton\psi\chi$, where $\applton\in\{\Appl, \to\}$ and $\psi$, $\chi$ are patterns. 
Thus, $\vp_0=\applton$,  
$\psi=\vp_1\ldots \vp_{k-1}$ and 
$\chi=\vp_k\ldots \vp_{n-1}$, where $2\leq k \leq n-1$.  We have the following cases:
\be 
\item $i\leq k-1$. Then $\vp_i$ occurs in $\psi$. As the length of $\psi$ is $<n$, we can apply  
the induction hypothesis to get the existence of $j$ such that $i<j\leq k-1$ and $\vp_i\ldots \vp_j$ is a pattern.
\item $i\geq k$. Then $\vp_i$ occurs in $\chi$. As the length of $\chi$ is $<n$, we can apply  
the induction hypothesis to get the existence of $j$ such that $i<j\leq n-1$ and $\vp_i\ldots \vp_j$ is a pattern.
\ee
\item $\vp=\theta\psi$, where $\theta \in \{\exists x, \mu X\}$ for some $x\in EVar$ or $X\in SVar$ and $\psi$ is a pattern.
Then $\vp_0\vp_1=\theta$ and $\psi=\vp_2\ldots \vp_{n-1}$. We have the following cases:
\be
\item $i=0$. Then $j=n-1$ and $\vp_i\ldots \vp_j=\vp$ is a pattern.
\item $2\le i\le n-1$. Then $\vp_i$ occurs in  $\psi$. As the length of $\psi$ is $<n$, we can apply  
the induction hypothesis to get the existence of $j$ such that $i<j\leq n-1$ and $\vp_i\ldots \vp_j$ is a pattern.
\ee
\ee
} 

\bprop\label{unique-subpat-to-appl}
Let $\vp=\vp_0\vp_1\ldots\vp_{n-1}$ be a pattern and suppose that $\vp_i\in\{\Appl,\to\}$ for some $i=0,\ldots,n-1$. 
Then there exist unique $j$, $l$ such that $i<j < l \leq n-1$ and 
$\vp_{i+1}\ldots \vp_j$,  $\vp_{j+1}\ldots \vp_l$ are patterns. 
\eprop
\solution{Let us prove first the uniqueness. Assume, by contradiction, that $i<j < l \leq n-1$ and $i<j_1 < l_1 \leq n-1$ are such that
$\psi=\vp_{i+1}\ldots \vp_j$,  $\chi=\vp_{j+1}\ldots \vp_l$,  $\psi^1=\vp_{i+1}\ldots \vp_{j_1}$,  
$\chi=\vp_{j+1}\ldots \vp_{l_1}$ are patterns. If $j\ne j_1$, then either $j<j_1$ or $j_1<j$, hence one of  
$\psi$, $\psi^1$  is a proper initial segment of the other one. By 
Proposition~\ref{unique-read-patterns}.\eqref{pattern-initial-segment}, we get that one of  
$\psi$, $\psi^1$  is not a pattern. We have obtained a contradiction. Thus, we must have $j=j_1$. 
We prove similarly that we must have $l=l_1$. 

Let us prove in the sequel the existence. 

As, by Proposition~\ref{unique-read-patterns}.\eqref{pattern-positive}, patterns have positive length, 
 we have to prove that for all $n\geq 1$,\\

\bt{ll}
(P) & if $\vp=\vp_0\ldots \vp_{n-1}$ is a pattern of length $n$ and $\vp_i\in\{\Appl,\to\}$ for some $i=0,\ldots,n-1$,\\[1mm]
& then there exist $j$, $l$ such that $i<j < l \leq n-1$ and 
$\vp_{i+1}\ldots \vp_j$,  $\vp_{j+1}\ldots \vp_l$ are patterns.  \\[1mm]
\et

The proof is by induction on $n$. 

$n=1$. Then $\vp=\vp_0$ is an atomic pattern, so there exists no $i$ satisfying the premise in (P), 
so (P) holds.

Assume that $n>1$ and that (P) holds for any pattern of length $<n$. Let $\vp=\vp_0\ldots \vp_{n-1}$ be a pattern of length $n$ 
such that $\vp_i\in\{\Appl,\to\}$ for some $i=0,\ldots,n-1$. 
By Proposition~\ref{unique-read-patterns}.\eqref{pattern-read}, we have the following cases:
\be
\item $\vp=\applton\psi\chi$, where $\applton\in\{\Appl, \to\}$ and $\psi$, $\chi$ are patterns. 
Thus, $\vp_0=\applton$,  
$\psi=\vp_1\ldots \vp_{k-1}$ and 
$\chi=\vp_k\ldots \vp_{n-1}$, where $2\leq k \leq n-1$.  We have the following cases:
\be 
\item $i=0$. Then we can take $j=k-1$ and $l=n-1$. 
\item $i\in \{1, \ldots, k-1\}$. Then $\vp_i$ occurs in $\psi$. As the length of $\psi$ is $<n$, we can apply  
the induction hypothesis to get the conclusion. 
\item $i\geq k$. Then $\vp_i$ occurs in $\chi$. As the length of $\chi$ is $<n$, we can apply  
the induction hypothesis to get the conclusion
\ee 
\item $\vp=\theta\psi$, where $\theta \in \{\exists x, \mu X\}$ for some $x\in EVar$ and $X\in SVar$ and $\psi$ is a pattern.
Then $\vp_0\vp_1=\theta$ and $\psi=\vp_2\ldots \vp_{n-1}$. As $\vp_i$ does not occur in $\theta$, it follows that 
$\vp_i$ occurs in  $\psi$. As the length of $\psi$ is $<n$, we can apply  
the induction hypothesis to get the conclusion.
\ee
}

%% file: AML-language-recursion.tex
\subsection{Recursion principle on patterns}

\bprop[Recursion principle on patterns]\label{recursion-principle}
Let $D$  be a set and the mappings
\begin{align} G_0:\setatomicpat\to D, & \quad G_{\Appl}, \, G_\to:D^2 \times \setpat^2 \to D, \\ 
G_\exists: D\times EVar \times \setpat \to D, & \quad G_\mu: D\times SVar \times  \setpat \to D.
\end{align}
Then there exists a unique mapping 
\[F:\setpat \to D\]
that satisfies the following properties:
\be
\item $F(\vp)=G_0(\vp)$ for any atomic pattern $\vp$.
\item  $F(\Appl\vp\psi)=G_{\Appl}(F(\vp),F(\psi), \vp, \psi)$ for any patterns $\vp$, $\psi$.
\item  $F(\to\vp\psi)=G_\to(F(\vp),F(\psi), \vp, \psi)$ for any patterns $\vp$, $\psi$.
\item $F(\exists x \vp)=G_\exists(F(\vp), x, \vp)$ for any pattern $\vp$ and any element variable $x$.
\item $F(\mu X\vp)=G_\mu(F(\vp), X, \vp)$ for any pattern $\vp$ and any set variable $X$.
\ee
\eprop
\solution{Apply Proposition~\ref{unique-read-patterns}. 
}

%% file: AML-language-subpatterns.tex
\subsection{Subpatterns}

\bdfn
Let $\vp$ be a pattern. A subpattern of $\vp$ is a pattern $\psi$ that occurs in $\vp$.
\edfn

\bntn
We denote by $\setsubpats(\vp)$  the set of  subpatterns of $\vp$.
\entn

\bfact[Alternative definition]\,\\
The mapping
$$\setsubpats:\setpat \to 2^{\setpat}, \quad \vp\mapsto \setsubpats(\vp)$$  
can be defined by recursion on patterns as follows: \\[1mm]
\begin{tabular}{lll}
$\setsubpats(\vp)$ &  = & $\{\vp\}$ \quad if $\vp$ is an atomic pattern,\\[1mm]
$\setsubpats(\Appl\vp\psi)$ & = & $\setsubpats(\vp)\cup\setsubpats(\psi)\cup\{Appl\vp \psi\}$, \\[1mm]
$\setsubpats(\to\vp\psi)$ & = & $\setsubpats(\vp)\cup \setsubpats(\psi)\cup\{\to\vp \psi\}$, \\[1mm]
$\setsubpats(\exists x\vp)$ & = & $\setsubpats(\vp) \cup \{\exists x \vp\}$, \\[1mm]
$\setsubpats(\mu X\vp)$ & = & $\setsubpats(\vp) \cup \{\mu X\vp\}$.
\end{tabular}
\efact
\solution{Apply Recursion principle on patterns (Proposition~\ref{recursion-principle}) with 
$D=2^{\setpat}$ and 
\begin{align*}
G_0(\vp) =\{\vp\}, & \quad  G_\applton(\Gamma, \Delta, \vp, \psi) = \Gamma \cup \Delta \cup \{\applton\vp \psi\} 
\text{~for~} \applton\in \{\Appl, \to\},\\
G_\exists(\Gamma, x, \vp) = \Gamma \cup \{\exists x \vp\} & \quad G_\mu(\Gamma, X, \vp) = \Gamma \cup \{\mu X \vp\}.
\end{align*} 
Then 
\be
\item $\setsubpats(\vp)=\{\vp\}=G_0(\vp)$ if $\vp$ is an atomic pattern.
\item For $\applton\in\{\Appl, \to\}$, we have that 
\begin{align*} 
\setsubpats(\applton\vp \psi) & =\setsubpats(\vp)\cup\setsubpats(\psi)\cup\{\applton\vp \psi\}\\
& = G_\applton(\setsubpats(\vp), \setsubpats(\psi), \vp, \psi).
\end{align*} 
\item $\setsubpats(\exists x\vp) = \setsubpats(\vp) \cup \{\exists x \vp\} = G_\exists(\setsubpats(\vp), x, \vp)$. 
\item $\setsubpats(\mu X\vp) =\setsubpats(\vp) \cup \{\mu X\vp\}=G_\mu(\setsubpats(\vp), X, \vp)$. 
\ee
Thus, $\setsubpats:\setpat \to 2^{\setpat}$ is the unique mapping given by Proposition~\ref{recursion-principle}.
 }

%% file: AML-language-free-bound.tex
\subsection{Free and bound variables}\label{language-free-bound}

\bdfn 
Let $\vp=\vp_0\vp_1\ldots\vp_{n-1}$ be a pattern and $x$ be an element variable.

\be
\item We say that $\exists$ is \defnterm{a quantifier on $x$ at the $i$th place with scope $\psi$} if
$\vp_i=\exists$, $\vp_{i+1}=x$ and $\psi=\vp_i\ldots \vp_j$ is the unique pattern given by Proposition \ref{unique-subpat-exists-mu}.
\item We say that  $x$  \defnterm{occurs bound at the $k$th place of $\vp$} if 
$\vp_k=x$ and 
there exist $0\leq i, j \le n-1$ such that $i <k \le j$ and 
$\exists$ is a quantifier on $x$ at the $i$th place with scope $\psi=\vp_i\ldots \vp_j$.
\item If $\vp_k=x$ but $x$ does not occur bound at the $k$th place of $\vp$, we say that 
$x$  \defnterm{occurs free at the $k$th place of $\vp$}.
\item $x$ is a  \defnterm{bound variable} of $\vp$ if there exists 
$k$ such that  $x$ occurs bound at the $k$th place of $\vp$.
\item $x$ is a  \defnterm{free variable} of $\vp$ if there exists $k$ such that 
$x$ occurs free at the $k$th place of $\vp$.
\ee
\edfn

\bdfn
Let $\vp=\vp_0\vp_1\ldots\vp_{n-1}$ be a pattern and $X$ be a set variable.
\be
\item We say that $\mu$ is \defnterm{a binder on $X$ at the $i$th place with scope $\psi$} if
$\vp_i=\mu$, $\vp_{i+1}=X$ and $\psi=\vp_i\ldots \vp_j$ is the unique pattern given 
by Proposition \ref{unique-subpat-exists-mu}.
\item We say that   $X$  \defnterm{occurs bound at the $k$th place of $\vp$} if 
$\vp_k=X$ and there exist $0\leq i, j \le n-1$ such that $i <k \le j$ and 
$\mu$ is a binder on $X$ at the $i$th place with scope $\psi=\vp_i\ldots \vp_j$.
\item If $\vp_k=X$ but $X$ does not occur bound at the $k$th place of $\vp$, we say that 
$X$  \defnterm{occurs free at the $k$th place of $\vp$}.
\item $X$ is a  \defnterm{bound variable} of $\vp$ if there exists 
$k$ such that  $X$ occurs bound at the $k$th place of $\vp$.
\item $X$ is a  \defnterm{free variable} of $\vp$ if there exists $k$ such that 
$X$ occurs free at the $k$th place of $\vp$.
\ee
\edfn

\bntn 
$FV(\vp)=$ the set of free element  and set variables of $\vp$.
\entn 

\bfact[Alternative definition]\,\\
The mapping
$$FV:\setpat \to 2^{EVar \cup SVar}, \quad \vp\mapsto FV(\vp)$$
 can be defined by recursion on patterns as follows:\\

\begin{tabular}{llll}
$FV(\vp)$ &  = & $EVar(\vp)\cup SVar(\vp)$  & if $\vp$ is an atomic pattern,\\[1mm]
$FV(\applton\vp \psi)$ & = & $FV(\vp)\cup FV(\psi)$ & for $\applton\in\{Appl, \to\}$, \\[1mm]
$FV(\exists x\vp)$ & = & $FV(\vp) \setminus \{x\}$, \\[1mm]
$FV(\mu X\vp)$ & = & $FV(\vp)\setminus \{X\}$.
\end{tabular}
\efact
\solution{Apply Recursion principle on patterns (Proposition~\ref{recursion-principle}) with 
$D=2^{EVar \cup SVar}$ and 
\begin{align*}
G_0(\vp)=EVar(\vp)\cup SVar(\vp),  & \quad G_\applton(V_1, V_2, \vp, \psi)=V_1\cup V_2,\\
G_\exists(V, x, \vp)=V\setminus \{x\}, & \quad  G_\mu(V, X, \vp)=V\setminus \{X\}.
\end{align*}
Then
\be
\item $FV(\vp)=EVar(\vp)\cup SVar(\vp)=G_0(\vp)$ if $\vp$ is an atomic pattern.
\item $FV(\applton\vp \psi)=FV(\vp)\cup FV(\psi)=G_\applton(FV(\vp), FV(\psi), \vp, \psi)$ for $\applton\in\{Appl, \to\}$.
\item $FV(\exists x\vp)=FV(\vp) \setminus \{x\}=G_\exists(FV(\vp), x, \vp)$.
\item $FV(\mu X\vp)=FV(\vp) \setminus \{X\}=G_\mu(FV(\vp), X, \vp)$.
\ee
Thus, $FV:\setpat \to 2^{EVar \cup SVar}$ is the unique mapping given by Proposition~\ref{recursion-principle}.
}

%% file: AML-language-positive.tex
\subsection{Positive and negative occurences of set variables}\label{language-positive}

\bdfn\label{def-implication-i-left-right-scope-X-occurs-left-at-k-vp}
Let $\vp=\vp_0\vp_1\ldots\vp_{n-1}$ be a pattern and $X$ be a set variable.
\be
\item\label{def-implication-i-left-right-scope} We say that $\to$ is \defnterm{an implication at the $i$th place of $\vp$ with left scope $\psi$ and right scope $\chi$} if
$\vp_i=\to$ and  $\psi=\vp_{i+1}\ldots \vp_j$,   $\chi=\vp_{j+1}\ldots \vp_l$ are the unique patterns  given 
by Proposition \ref{unique-subpat-to-appl}.
\item\label{X-occurs-left-at-k-vp} $X$  \defnterm{\occleft~  at the $k$th place of $\vp$} if $X$ occurs free at the $k$th place of $\vp$
and there exist $0\leq i < k \le j \leq n-1$  such that $\psi=\vp_{i+1}\ldots \vp_j$ is the left scope of an implication $\to$ at the $i$th place of $\vp$.
\ee
\edfn

\bdfn
Let $X$ be a set variable. We define the mapping 
$$N_{X,L}:\setpat \to Fun(\N, \N)$$
by recursion on patterns as follows:
\be 
\item $\vp$ is an atomic pattern. Then $N_{X,L}(\vp)(k)=0$ for every $k\in \N$.
\item $\vp=\Appl\psi\chi$, that is $\vp=\vp_0\vp_1\ldots\vp_{n-1}$ with $\vp_0=\Appl$, $\psi=\vp_1\ldots \vp_j$ and $\chi=\vp_{j+1}\ldots \vp_{n-1}$ for some
$1\le j <n-1$. We have the following cases:
\be
\item $k=0$ or $k\ge n$. Then $N_{X,L}(\vp)(k)=0$.
\item $1\le k \le j$. Then $N_{X,L}(\vp)(k)=N_{X,L}(\psi)(k-1)$.
\item $j+1 \le k \le n-1$. Then $N_{X,L}(\vp)(k)=N_{X,L}(\chi)(k-j-1)$.
\ee
\item $\vp=\to\psi\chi$, that is $\vp=\vp_0\vp_1\ldots\vp_{n-1}$ with $\vp_0=\to$, $\psi=\vp_1\ldots \vp_j$ and $\chi=\vp_{j+1}\ldots \vp_{n-1}$ for some
$1\le j <n-1$. We have the following cases:
\be
\item $k=0$ or $k\ge n$. Then $N_{X,L}(\vp)(k)=0$.
\item $1\le k \le j$. Then $N_{X,L}(\vp)(k)=N_{X,L}(\psi)(k-1)+1$.
\item $j+1 \le k \le n-1$. Then $N_{X,L}(\vp)(k)=N_{X,L}(\chi)(k-j-1)$.
\ee
\item $\vp=\exists x\psi$, that is $\vp=\vp_0\vp_1\ldots\vp_{n-1}$ with $\vp_0=\exists$, $\vp_1=x$ and $\psi=\vp_2\ldots \vp_{n-1}$.
We have the following cases:
\be
\item $k\in \{0,1\}$ or $k\ge n$. Then $N_{X,L}(\vp)(k)=0$.
\item $2\le k \le n-1$. Then $N_{X,L}(\vp)(k)=N_{X,L}(\psi)(k-2)$.
\ee
\item $\vp=\mu Z\psi$, that is $\vp=\vp_0\vp_1\ldots\vp_{n-1}$ with $\vp_0=\mu$, $\vp_1=Z$ and $\psi=\vp_2\ldots \vp_{n-1}$. If
$Z=X$, then $N_{X,L}(\vp)(k)=0$ for every $k\in \N$. If $Z\ne X$, we have the following cases:
\be
\item $k\in \{0,1\}$ or $k\ge n$. Then $N_{X,L}(\vp)(k)=0$.
\item $2\le k \le n-1$. Then $N_{X,L}(\vp)(k)=N_{X,L}(\psi)(k-2)$.
\ee
\ee
\edfn
\solution{Apply
Recursion principle on patterns (Proposition~\ref{recursion-principle}) with $D=Fun(\N, \N)$ and
\begin{align*}
G_0(\vp)(k) & = 0 \qquad  \text{~if~} \vp \text{~is an atomic pattern},\\
G_{\Appl}(f,g, \psi, \chi)(k) &= \begin{cases} 0  & \text{~if~}  k=0 \text{~or~} k\ge n,\\
f(k-1) & \text{~if~}  1\le k \le j,\\
g(k-j-1)  & \text{~if~}  j+1 \le k \le n-1,\\
\end{cases},\\
G_\to(f,g, \psi, \chi)(k) &= \begin{cases} 0  & \text{~if~}  k=0 \text{~or~} k\ge n,\\
f(k-1)+1 & \text{~if~}  1\le k \le j,\\
g(k-j-1)  & \text{~if~}  j+1 \le k \le n-1,\\
\end{cases},\\
G_\exists(f, x,\psi)(k)& =\begin{cases} 0  & \text{~if~}  k\in \{0,1\} \text{~or~} k\ge n,\\
f(k-2) & \text{~if~}  2\le k \le n-1,\\
\end{cases},\\
G_\mu(f, Z,\psi)(k)& =\begin{cases} 0  & \text{~if~}  X=Z,\\
0  & \text{~if~}  X\ne Z \text{~and~} (k\in \{0,1\} \text{~or~} k\ge n),\\
f(k-2)  & \text{~if~}  X\ne Z \text{~and~} 2\le k \le n-1.
\end{cases}
\end{align*}
Then
\be
\item $N_{X,L}(\vp)(k)=0=G_0(\vp)(k)$ for every $k\in \N$ if $\vp$ is an atomic pattern.
\item $\vp=\Appl\psi\chi$. Then  
\begin{align*} N_{X,L}(\to\psi\chi)(k) & =\begin{cases} 0 & \text{~if~} k=0 \text{~or~} k\ge n,\\
N_{X,L}(\psi)(k-1) &  \text{~if~}  1\le k \le j,\\
N_{X,L}(\chi)(k-j-1) & \text{~if~}  j+1 \le k \le n-1 
\end{cases}\\[1mm]
& = G_{\Appl}(N_{X,L}(\psi),N_{X,L}(\chi), \psi, \chi)(k).
\end{align*}
\item $\vp=\to\psi\chi$. Then  
\begin{align*} N_{X,L}(\to\psi\chi)(k) & =\begin{cases} 0 & \text{~if~} k=0 \text{~or~} k\ge n,\\
N_{X,L}(\psi)(k-1)+1 &  \text{~if~}  1\le k \le j,\\
N_{X,L}(\chi)(k-j-1) & \text{~if~}  j+1 \le k \le n-1 
\end{cases}\\[1mm]
& = G_{\Appl}(N_{X,L}(\psi),N_{X,L}(\chi), \psi, \chi)(k).
\end{align*}
\item $\vp=\exists x\psi$. Then 
\begin{align*} N_{X,L}(\exists x\psi)(k) & =\begin{cases} 0 & \text{~if~} k\in \{0,1\} \text{~or~} k\ge n,\\
N_{X,L}(\psi)(k-2) &  \text{~if~} 2\le k \le n-1
\end{cases}\\[1mm]
& = G_{\exists}(N_{X,L}(\psi),x, \psi)(k).
\end{align*}
\item $\vp=\mu Z\psi$. If $X=Z$, then $N_{X,L}(\mu Z\psi)(k)=0= G_\mu(N_{X,L}(\psi), Z,\psi)(k)$. \\ 
Assume that $X\ne Z$. Then 
\begin{align*} N_{X,L}(\mu Z\psi)(k) & =\begin{cases} 0 & \text{~if~} k\in \{0,1\} \text{~or~} k\ge n,\\
N_{X,L}(\psi)(k-2) &  \text{~if~} 2\le k \le n-1
\end{cases}\\[1mm]
& = G_{\mu}(N_{X,L}(\psi),Z, \psi)(k).
\end{align*}
\ee
}

\bntn
Let $X$ be a set variable, $k\in\N$ and $\vp$ be a pattern. We denote $N_{X,L}(\vp)(k)$ $N_L(\vp, X,k)$.
\entn

\bdfn
Let $X$ be a set variable, $k\in\N$ and $\vp$ be a pattern such that $X$ occurs free at the $k$th place of $\vp$.
\be
\item We say that $X$ \defnterm{occurs positively at the $k$th place of $\vp$} (or that $X$ 
\defnterm{has a positive occurence at the $k$th place of $\vp$}) if $N_L(\vp, X,k)=0$  or $N_L(\vp, X,k)$ is an even natural number. 
\item We say that $X$ \defnterm{occurs negatively at the $k$th place of $\vp$} (or that $X$ 
\defnterm{has a negative occurence at the $k$th place of $\vp$}) if $N_L(\vp, X,k)$ is an odd natural number. 
\ee
\edfn

\bdfn\label{def-vp-positive-in-X}
We say that $\vp$ is \defnterm{positive in $X$} if one of the following is true:
\be
\item $X$ does not occur free in $\vp$.
\item For every $k\in\N$, if $X$ occurs free at the $k$th place of $\vp$, then  $X$ occurs positively at the $k$th place of $\vp$.
\ee
\edfn

\bdfn\label{def-vp-negative-in-X}
We say that $\vp$ is \defnterm{negative in $X$} if one of the following is true:
\be
\item $X$ does not occur free in $\vp$.
\item For every $k\in\N$, if $X$ occurs free at the $k$th place of $\vp$, then  $X$ occurs negatively at the $k$th place of $\vp$.
\ee
\edfn

\bex
\be
\item $\vp=\to X\to X\bot$. Then $N_L(X,\vp,1)=N_L(X,X,0)+1=0+1=1$. Furthermore, 
$N_L(X,\vp,3)=N_L(X, \to X\bot,1)=N_L(X,X,0)+1=1$. Thus, $\vp$ is negative in $X$.
\item $\vp =\to\to X\bot\bot$. Then $N_L(X,\vp,2) = N_L(X,\to X\bot,1)+1=N_L(X,X,0)+2=2$. Thus,
$\vp$ is positive in $X$.
\item $\vp=\Appl{\to X\to X\bot}{\to\to X\bot\bot}$. Then $\vp$ is neither positive nor negative in $X$.
\ee
\eex

\bfact[Alternative definition]\label{def-positive-X-recursion}
The property that $\vp$ is positive in $X$ can be defined by recursion on patterns as follows:
\be
\item If $\vp$ is atomic,  then $\vp$ is positive in $X$.
\item If $\vp=\Appl\psi\chi$, then $\vp$ is positive in $X$ iff both $\psi$, $\chi$ are positive in $X$.
\item If $\vp=\to\psi\chi$, then $\vp$ is positive in $X$ iff $\psi$ is negative in $X$ and  $\chi$ is positive in $X$.
\item If $\vp=\exists x\psi$, then $\vp$ is positive in $X$ iff $\psi$ is positive in $X$. 
\item If $\vp=\mu X\psi$, then $\vp$ is positive in $X$.
\item If $\vp=\mu Z\psi$ with $Z\ne X$, then $\vp$ is positive in $X$ iff $\psi$ is positive in $X$. 
\ee
\efact

\bfact[Alternative definition]\label{def-negative-X-recursion}
The property that $\vp$ is negative in $X$ can be defined by recursion on patterns as follows:
\be
\item If $\vp$ is atomic, then $\vp$ is negative in $X$ iff $\vp\ne X$.
\item If $\vp=\Appl\psi\chi$, then $\vp$ is negative in $X$ iff both $\psi$, $\chi$ are negative in $X$.
\item If $\vp=\to\psi\chi$, then $\vp$ is negative in $X$ iff $\psi$ is positive in $X$ and  $\chi$ is negative in $X$.
\item If $\vp=\exists x\psi$, then $\vp$ is negative in $X$ iff $\psi$ is negative in $X$. 
\item If $\vp=\mu X\psi$, then $\vp$ is negative in $X$.
\item If $\vp=\mu Z\psi$ with $Z\ne X$, then $\vp$ is negative in $X$ iff $\psi$ is negative in $X$.
\ee
\efact

%% file: AML-substitution.tex
\subsection{Substitution of element variables}

Let $x$ be an element variable and $\vp$, $\delta$ be patterns.

\bdfn
We define \defnterm{$\subfx{\delta}{\vp}$} to be the expression 
obtained from $\vp$  by replacing every free occurence of $x$ in $\vp$ with $\delta$.
\edfn

\bfact[Alternative definition]\label{def-subf-recursion}\,\\
The mapping $$\subfxdeltaf: \setpat \to Expr, \quad \subfxdeltaf(\vp)=\subfxdelta{\vp}$$ can be 
defined by recursion on patterns as follows: \\

\bt{llll}
$\subfxdeltaf(z)$ & = & $\begin{cases} \delta \text{ if } x=z\\
 z\text{ if } x \ne z\\
\end{cases}$ \quad 
& if $z\in EVar$, \\[3mm]
$\subfxdeltaf(\vp)$ & = & $\vp$  & if $\vp\in SVar \cup \Sigma$, \\[2mm]
$\subfxdeltaf(\applton\psi\chi)$ & = & $\applton \subfxdeltaf(\psi)\subfxdeltaf(\chi)$ & for $\applton\in \{\Appl, \to\}$, \\[2mm]
$\subfxdeltaf(\exists z\psi) $ & = & $ \begin{cases} \exists z\psi  & \text{ if } x=z\\
\exists z \subfxdeltaf(\psi)   & \text{ if } x \ne z\\
\end{cases}$,  \\[5mm] 
$\subfxdeltaf(\mu X\psi)$ & = & $ \mu X \subfxdeltaf(\psi)$.
\et
\efact
\solution{Apply
Recursion principle on patterns (Proposition~\ref{recursion-principle}) with $D=Expr$ and
\begin{align*}
G_0(\vp)& =\begin{cases}
\delta & \text{~if~} \vp=x \\
\vp & \text{~if~} \vp \in (EVar\setminus \{x\}) \cup SVar \cup \Sigma
\end{cases} \\
G_\applton(\theta, \tau, \psi, \chi) &=\applton \theta \tau \text{~for~} \applton\in \{\Appl, \to\},\\
G_\exists(\theta, z, \psi)& =\begin{cases} \exists z\psi  & \text{~if~}  x=z\\
\exists z \theta  & \text{~if~} x \ne z\\
\end{cases},\\
G_\mu(\theta, X, \psi) &= \mu X \theta.
\end{align*}
Then
\be
\item If $\vp$ is an atomic pattern, we have the following cases:
\be
\item $\vp=x$. Then $\subfxdeltaf(\vp)=\subfxdeltaf(x)=\delta =G_0(\vp)$.
\item $\vp \in (EVar\setminus \{x\}) \cup SVar \cup \Sigma$. Then $\subfxdeltaf(\vp)=\vp=G_0(\vp)$. 
\ee
\item For $\applton\in \{\Appl, \to\}$, we have that 
$$\subfxdeltaf(\applton\psi\chi)=\applton \subfxdeltaf(\psi)\subfxdeltaf(\chi)=
G_\applton(\subfxdeltaf(\psi), \subfxdeltaf(\chi), \psi, \chi).$$
\item $\subfxdeltaf(\exists z\psi) = \begin{cases} \exists z\psi  & \text{ if } x=z\\
\exists z \subfxdeltaf(\psi)  & \text{ if } x \ne z\\
\end{cases}= G_\exists(\subfxdeltaf(\psi), z, \psi)$
\item $\subfxdeltaf(\mu X\psi)=\mu X \subfxdeltaf(\psi) = G_\mu(\subfxdeltaf(\psi), X, \psi)$. 
\ee
Thus, $\subfxdeltaf$ is the unique mapping given by Proposition~\ref{recursion-principle}.
}

\bprop
$\subfx{\delta}{\vp}$ is a pattern. 
\eprop
\solution{The proof is immediate  by induction on $\vp$, using the alternative definition 
of $\subfx{\delta}{\vp}$.}

\bex[TRIVIAL CASES]
\be
\item $\subfx{x}{\vp}=\vp$. 
\item If $x$ does not occur free in $\vp$, then $\subfxdelta{\vp}=\vp$. 
\ee
\eex

\bdfn
We say that $x$ is \defnterm{free for $\delta$} in $\vp$ or that $\delta$  
is \defnterm{substitutable for $x$} in $\vp$ if the following hold:
\be
\item if $z$ is an element variable occuring free in $\delta$ and $\exists$ is a quantifier on $z$ in $\vp$ with scope $\theta$, then 
$x$ does not occur free in $\theta$.
\item if $Z$ is a set variable occuring free in $\delta$ and $\mu$ is a binder on $Z$ in $\vp$ 
with scope $\theta$, then $x$ does not occur free in $\theta$.
\ee
\edfn

\bex
$x$ is free for $\delta$ in  $\vp$ in any of the following cases:
\be
\item $x$ does not occur free in $\vp$.
\item Any element or set variable of $\delta$ does not occur bound in $\vp$.
\item $EVar(\delta)=SVar(\delta)=\emptyset$.
\ee
\eex

\bfact
Let $y\ne x$ be an element variable. 
\be
\item $x$ is free for $y$ in $\vp$ if the following holds: if $\exists$ is a quantifier on $y$ in $\vp$ with scope $\theta$, then 
$x$ does not occur free in $\theta$.
\item $x$ is not free for $y$ in $\vp$ if $\exists$ is a quantifier on $y$ in $\vp$ with scope $\theta$ and $x$ occurs free in $\theta$.
\ee
\efact

\subsubsection{Bounded substitution}

Let $\vp$ be a pattern and $x$, $y$ be element variables.

\bdfn
We define \defnterm{$\subbvpxy$} to be the expression 
obtained from $\vp$  by replacing every bound occurence of $x$ in $\vp$ with $y$.
\edfn

\bfact[Alternative definition]\label{subb-inductive-x}\,\\
If $x=y$, then obviously $\subbvpxy=\vp$. Assume that $x\ne y$. Then the mapping 
$$\subb{x}{y}: \setpat \to Expr, \quad \subb{x}{y}(\vp)=\subb{x}{y}{\vp}$$
 can be defined by recursion on patterns as follows: \\
 
\bt{llll}
$\subb{x}{y}(\vp)$ &=& $\vp$  & if $\vp$ is an atomic pattern,\\[2mm]
$\subb{x}{y}(\applton\psi\chi)$ &=& $ \applton \subb{x}{y}(\psi)\subb{x}{y}(\chi)$ & for $\applton\in \{\Appl, \to\}$, \\[2mm]
$\subb{x}{y}(\exists z\psi)$ &=& $ \begin{cases} \exists y\subfx{y}{\left(\subb{x}{y}(\psi)\right)} & \text{ if } x=z\\
\exists z\subb{x}{y}(\psi) & \text{ if } x \ne z\\
\end{cases}$, \\[4mm]
$\subb{x}{y}(\mu X\psi)$ &=& $\mu X \subb{x}{y}(\psi)$.
\et
\efact
\solution{Apply
Recursion principle on patterns (Proposition~\ref{recursion-principle}) with $D=Expr$ and
\begin{align*}
G_0(\vp) = \vp, & \quad G_\applton(\theta, \tau, \psi, \chi)=\applton \theta \tau \text{~for~} \applton\in \{\Appl, \to\},\\
G_\exists(\theta, z, \psi)= \begin{cases} \exists y\subfx{y}{(\theta)} & \text{ if } x=z\\
\exists z\theta & \text{ if } x \ne z\\
\end{cases}, & \quad  G_\mu(\theta, X, \psi) = \mu X \theta.
\end{align*}
Then 
\be
\item If $\vp$ is an atomic pattern, $\subb{x}{y}(\vp)=\vp=G_0(\vp)$. 
\item For $\applton\in \{\Appl, \to\}$, we have that $$\subb{x}{y}(\applton\psi\chi)=\applton \subb{x}{y}(\psi)\subb{x}{y}(\chi)=
G_\applton(\subb{x}{y}(\psi), \subb{x}{y}(\chi), \psi, \chi).$$ 
\item $\subb{x}{y}(\exists z\psi) = \begin{cases} \exists y\subfx{y}{\left(\subb{x}{y}(\psi)\right)} & \text{ if } x=z\\
\exists z\subb{x}{y}(\psi) & \text{ if } x \ne z\\
\end{cases}=G_\exists(\subb{x}{y}(\psi), z, \psi)$
\item $\subb{x}{y}(\mu X\psi)=\mu X \subb{x}{y}(\psi) = G_\mu(\subb{x}{y}(\psi), X, \psi)$. 
\ee
Thus, $\subb{x}{y}$ is the unique mapping given by Proposition~\ref{recursion-principle}.
}

\bprop
$\subbvpxy$ is a pattern. 
\eprop
\solution{The proof is immediate  by induction on $\vp$, using the alternative definition of $\subbvpxy$.}

\bprop\label{subb-useful-1}
Assume that $x\ne y$  and $y$ does not occur in $\vp$. Then $x$ is free for $y$ in $\subb{x}{y}{\vp}$.
\eprop
\solution{The proof is by induction on $\vp$.
\be
\item $\vp$ is an atomic pattern. Then $\subbvpxy = \vp$ and $x$ is free for $y$ in $\vp$, as $y$ does not occur in $\vp$.

\item $\vp=\applton\psi\chi$ for $\applton\in \{\Appl, \to\}$. Then $\subbvpxy=\applton \subb{x}{y}(\psi)\subb{x}{y}(\chi)$. 
As $y$ does not occur in $\psi$, $\chi$,
we can apply the induction hypothesis to get that $x$ is free for $y$ in $\subb{x}{y}{\psi}$, $\subb{x}{y}{\chi}$.
It follows that $x$ is free for $y$ in $\subb{x}{y}{\vp}$.
\item $\vp=\exists z\psi$. We have two cases:
\be
\item $x=z$. Then $\subbvpxy=\exists y\subfx{y}{\left(\subb{x}{y}(\psi)\right)}$. It is obvious that $x$ 
does not occur in 
$\subbvpxy$, so $x$ is free for $y$ in $\subb{x}{y}{\vp}$.
\item  $x \ne z$. Then $\subbvpxy=\exists z\subb{x}{y}(\psi)$. As $y$ does not occur in $\psi$, we 
can apply the induction hypothesis to get that $x$ is free for $y$ in $\subb{x}{y}{\psi}$. As $y$ does not occur in $\vp$, we must 
have that $y\ne z$. Then, obviously $x$ is free for $y$ in $\subbvpxy$. 
\ee
\item $\vp=\mu X\psi$. Then $\subbvpxy=\mu X \subb{x}{y}(\psi)$. As $y$ does not occur in $\psi$, we 
can apply the induction hypothesis to get that $x$ is free for $y$ in $\subb{x}{y}{\psi}$. 
Then, obviously $x$ is free for $y$ in $\subbvpxy$. 
\ee
}

%% file: AML-substitution-svar.tex
\subsection{Substitution of set variables}

Let $X$ be a set variable and $\vp$, $\delta$ be patterns.

\bdfn
We define \defnterm{$\subfvpXdelta$} to be the expression 
obtained from $\vp$  by replacing every free occurence of $X$ in $\vp$ with $\delta$.
\edfn

\bfact[Alternative definition]\label{def-recursion-subfvpX}\,\\
The mapping $$\subfXdeltaf: \setpat \to Expr, \quad \subfXdeltaf(\vp)=\subfXdelta{\vp}$$ can be 
defined by recursion on patterns as follows: \\

\bt{llll}
$\subfXdeltaf(Z)$ & = & $\begin{cases} \delta \text{ if } X=Z\\
 Z\text{ if } X \ne Z\\
\end{cases}$ & if $Z\in SVar$, \\[2mm]
$\subfXdeltaf(\vp)$ & = & $\vp$ &  if $\vp\in EVar \cup \Sigma$, \\[2mm]
$\subfXdeltaf(\applton\psi\chi)$ & = & $\applton \subfXdeltaf(\psi)\subfXdeltaf(\chi)$ & for $\applton\in \{\Appl, \to\}$, \\[2mm]
$\subfXdeltaf(\exists x\psi)$ & = & $\exists x \subfXdeltaf(\psi)$, \\[2mm]
$\subfXdeltaf(\mu Z\psi)$ & = & $\begin{cases} \mu Z\psi & \text{ if } X=Z\\
\mu Z\subfXdeltaf(\psi) & \text{ if } X \ne Z.
\end{cases}$
\et 
\efact
\solution{Apply
Recursion principle on patterns (Proposition~\ref{recursion-principle}) with $D=Expr$ and
\begin{align*}
G_0(\vp) & =\begin{cases}
\delta & \text{~if~} \vp=X \\
\vp & \text{~if~} \vp \in EVar \cup (SVar\setminus \{X\}) \cup \Sigma
\end{cases} \\
G_\applton(\theta, \tau, \psi, \chi) &=\applton \theta \tau \text{~for~} \applton\in \{\Appl, \to\},\\
G_\exists(\theta, x, \psi)& =\exists x \theta, \\
G_\mu(\theta, Z, \psi) &= \begin{cases} \mu Z\psi & \text{ if } X=Z\\
\mu Z\theta & \text{ if } X \ne Z.
\end{cases}
\end{align*}
Then
\be
\item If $\vp$ is an atomic pattern,  we have the following cases:
\be
\item $\vp=X$. Then $\subfXdeltaf(\vp)=\subfXdeltaf(X)=\delta =G_0(\vp)$.
\item $\vp \in EVar \cup (SVar\setminus \{X\}) \cup \Sigma$. Then $\subfXdeltaf(\vp)=\vp=G_0(\vp)$. 
\ee
\item  For $\applton\in \{\Appl, \to\}$, we have that 
$$\subfXdeltaf(\applton\psi\chi)=\applton \subfXdeltaf(\psi)\subfXdeltaf(\chi)=
G_\applton(\subfXdeltaf(\psi), \subfXdeltaf(\chi), \psi, \chi).$$
\item  $\subfXdeltaf(\exists x\psi)=\exists x \subfXdeltaf(\psi)=G_\exists(\subfXdeltaf(\psi), x, \psi).$
\item $\subfXdeltaf(\mu Z\psi) = \begin{cases} \mu Z\psi & \text{ if } X=Z\\
\mu Z\subfXdeltaf(\psi) & \text{ if } X \ne Z.
\end{cases}=G_\mu(\subfXdeltaf(\psi), Z, \psi)$
\ee 
Thus, $\subfXdeltaf$ is the unique mapping given by Proposition~\ref{recursion-principle}.
}

\bprop
$\subfXdelta{\vp}$ is a pattern. 
\eprop
\solution{The proof is immediate  by induction on $\vp$, using the alternative definition 
of $\subfXdelta{\vp}$.}

\bdfn
We say that $X$ is \defnterm{free for $\delta$} in $\vp$ or that $\delta$  
is \defnterm{substitutable for $X$} in $\vp$ if  the following hold:
\be
\item if $z$ is an element variable occuring free in $\delta$ and $\exists$ is a quantifier on $z$ in $\vp$ with scope $\theta$, then 
$X$ does not occur free in $\theta$.
\item if $Z$ is a set variable occuring free in $\delta$ and $\mu$ is a binder on $Z$ in $\vp$ with scope $\theta$, then 
$X$ does not occur free in $\theta$.
\ee
\edfn

\bex
$X$ is free for $\delta$ in  $\vp$ in any of the following cases:
\be
\item $X$ does not occur free in $\vp$.
\item Any element or set variable of $\delta$ does not occur bound in $\vp$.
\item $EVar(\delta)=SVar(\delta)=\emptyset$.
\ee
\eex

\subsubsection{Bounded substitution}

Let $\vp$ be a pattern   and $X$, $Y$  be set variables.

\bdfn
We define \defnterm{$\subbvpXY$} to be the expression 
obtained from $\vp$  by replacing every bound occurence of $X$ in $\vp$ with $Y$.
\edfn

\bfact[Alternative definition]\label{subb-inductive-X}\,\\
If $X=Y$, then obviously $\subbvpXY=\vp$. Assume that $X\ne Y$. Then the mapping 
$$\subb{X}{Y}: \setpat \to Expr, \quad \subb{X}{Y}(\vp)=\subbvpXY$$
 can be defined by recursion on patterns as follows: \\
 
\bt{llll}
$\subb{X}{Y}(\vp)$ &=& $\vp$  & if $\vp$ is an atomic pattern,\\[2mm]
$\subb{X}{Y}(\applton\psi\chi)$ &=& $\applton \subb{X}{Y}(\psi)\subb{X}{Y}(\chi)$ & for $\applton\in \{\Appl, \to\}$, \\[2mm]
$\subb{X}{Y}(\exists x\psi)$ &=& $\exists x \subb{X}{Y}(\psi)$, \\[2mm]
$\subb{X}{Y}(\mu Z\psi)$ &=& $\begin{cases} \mu Y\subfX{Y}{\left(\subb{X}{Y}(\psi)\right)} & \text{ if } X=Z\\
\mu Z\subb{X}{Y}(\psi) & \text{ if } X \ne Z\\
\end{cases}$.
\et
\efact
\solution{Apply
Recursion principle on patterns (Proposition~\ref{recursion-principle}) with $D=Expr$ and
\begin{align*}
G_0(\vp) = \vp, & \quad G_\applton(\theta, \tau, \psi, \chi)=\applton \theta \tau \text{~for~} \applton\in \{\Appl, \to\},\\
G_\exists(\theta, x, \psi)= \exists x \theta, & \quad  G_\mu(\theta, Z, \psi) =
\begin{cases} \mu Y\subfX{Y}{(\theta)} & \text{ if } X=Z\\
\mu Z \theta & \text{ if } X\ne Z.
\end{cases}
\end{align*}
Then 
\be
\item If $\vp$ is an atomic pattern, then $\subb{X}{Y}(\vp)=\vp=G_0(\vp)$. 
\item For $\applton\in \{\Appl, \to\}$, we have that $$\subb{X}{Y}(\applton\psi\chi)=\applton \subb{X}{Y}(\psi)\subb{X}{Y}(\chi)=
G_\applton(\subb{X}{Y}(\psi), \subb{X}{Y}(\chi), \psi, \chi).$$ 
\item $\subb{X}{Y}(\exists x\psi) = \exists x \subb{X}{Y}(\psi)=G_\exists(\subb{X}{Y}(\psi), x, \psi)$. 
\item $\subb{X}{Y}(\mu Z\psi)=\begin{cases} \mu Y\subfX{Y}{\left(\subb{X}{Y}(\psi)\right)} & \text{ if } X=Z\\
\mu Z\subb{X}{Y}(\psi) & \text{ if } X \ne Z\end{cases} = G_\mu(\subb{X}{Y}(\psi), Z, \psi)$. 
\ee
Thus, $\subb{X}{Y}$ is the unique mapping given by Proposition~\ref{recursion-principle}.
}

\bprop
$\subbvpXY$ is a pattern. 
\eprop
\solution{The proof is immediate  by induction on $\vp$, using the alternative definition of $\subbvpXY$.}

\bprop\label{subb-X-useful-1}
Assume that $X\ne Y$  and $Y$ does not occur in $\vp$. Then $X$ is free for $Y$ in $\subb{X}{Y}{\vp}$.
\eprop
\solution{The proof is by induction on $\vp$.
\be
\item $\vp$ is an atomic pattern. Then $\subb{X}{Y}{\vp} = \vp$ and $X$ is free for $Y$ in $\vp$, as $Y$ does not occur in $\vp$.
\item $\vp=\applton\psi\chi$ for $\applton\in \{\Appl, \to\}$. Then 
$\subb{X}{Y}{\vp}=\Appl\subb{X}{Y}{\psi}\subb{X}{Y}{\chi}$. As $Y$ does not occur in $\psi$, $\chi$,
we can apply the induction hypothesis to get that $X$ is free for $Y$ in $\subb{X}{Y}{\psi}$, $\subb{X}{Y}{\psi}$.
It follows that $X$ is free for $Y$ in $\subb{X}{Y}{\vp}$.
\item $\vp=\exists x\psi$. Then $\subb{X}{Y}{\vp}=\exists x \subb{X}{Y}{\psi}$. As $Y$ does not occur in $\psi$, we 
can apply the induction hypothesis to get that $X$ is free for $Y$ in $\subb{X}{Y}{\psi}$. 
Then, obviously $X$ is free for $Y$ in $\subb{X}{Y}{\vp}$. 
\item $\vp=\mu Z\psi$. 
 We have two cases:
\be
\item $X=Z$. Then $\subb{X}{Y}{\vp}=\mu Y\subfX{Y}{\left(\subb{X}{Y}{\psi}\right)}$. It is obvious that $X$ does not occur in 
$\subb{X}{Y}{\vp}$, so $X$ is free for $Y$ in $\subb{X}{Y}{\vp}$.
\item  $X \ne Z$. Then $\subb{X}{Y}{\vp}=\mu Z\subb{X}{Y}{\psi}$. As $Y$ does not occur in $\psi$, we 
can apply the induction hypothesis to get that $X$ is free for $Y$ in $\subb{X}{Y}{\psi} $. As $Y$ does not occur in $\vp$, we must 
have that $Y\ne Z$. Then, obviously $X$ is free for $Y$ in $\subb{X}{Y}{\vp}$. 
\ee

\ee
}

%% file: AML-substitution-free.tex
\subsection{Free substitution}

\bdfn\label{def-free-substitution-element}
Let $x$ be an element variable and $\vp$, $\delta$ be patterns.
We define $\fsubvpxdelta$ as follows:
\be
\item If $x$ is free for $\delta$ in $\vp$, then $\fsubvpxdelta=\subfvpxdelta$.
\item Assume that $x$ is not free for $\delta$ in $\vp$ and let $u_1,\ldots, u_k$ be the element 
variables and $U_1,\ldots, U_p$ be the set variables that occur bound in $\vp$ and also 
occur in $\delta$. Let $z_1, \ldots, z_k$ be new element variables and $Z_1, \ldots, Z_p$  be 
new set variables, that do not occur in $\vp$ or $\delta$. Consider the pattern 
$$\theta=\subb{U_1}{Z_1}\subb{U_2}{Z_2}\ldots \subb{U_p}{Z_p}\subb{u_1}{z_1}
\subb{u_2}{z_2}\ldots \subb{u_k}{z_k}\vp.$$
Then  $\fsubvpxdelta=\subfxdelta{\theta}$.
\ee
\edfn

\bdfn\label{def-free-substitution-set}
Let $X$ be a set variable and $\vp$, $\delta$ be patterns.
We define $\fsubvpXdelta$ as follows:
\be
\item If $X$ is free for $\delta$ in $\vp$, then $\fsubvpXdelta=\subfvpXdelta$.
\item Assume that $X$ is not free for $\delta$ in $\vp$ and let $u_1,\ldots, u_k$ be the element 
variables and $U_1,\ldots, U_p$ be the set variables that occur bound in $\vp$ and also 
occur in $\delta$. Let $z_1, \ldots, z_k$ be new element variables and $Z_1, \ldots, Z_p$  be 
new set variables, that do not occur in $\vp$ or $\delta$. Consider the pattern 
$$\theta=\subb{U_1}{Z_1}\subb{U_2}{Z_2}\ldots \subb{U_p}{Z_p}\subb{u_1}{z_1}
\subb{u_2}{z_2}\ldots \subb{u_k}{z_k}\vp.$$
Then  $\fsubvpXdelta=\subfXdelta{\theta}$.
\ee
\edfn

\bfact
In the above definitions, we could define $z_1, \ldots, z_k$ and $Z_1,\ldots, Z_p$  as follows: 
\be 
\item Let $l\in\N$ be maximum such that $v_l$ occurs in $\vp$ or  $\delta$.
Then define $z_i=v_{l+i}$ for all $i=1, \ldots, k$. 
\item Let $m\in\N$ be maximum such that $V_m$ occurs in $\vp$ or  $\delta$.
Then define $Z_i=V_{m+i}$ for all $i=1, \ldots, p$.
\ee
\efact

%% file: AML-language-contexts.tex
\subsection{Contexts}

Let $\Box$ be a new symbol and let us denote
$$\langcontext=Sym_{\AMLsig}\cup\{\Box\}\cup\{\ApplC\}.$$

\bdfn
The \defnterm{$\AMLsig$-contexts}  are the expressions over $\langcontext$ inductively defined 
as follows:
\be
\item $\Box$ is a $\AMLsig$-context.
\item If $C$ is a $\AMLsig$-context and $\vp$ is a $\AMLsig$-pattern, then 
$\ApplC C\vp$ and $\ApplC \vp C$ are $\AMLsig$-contexts.
\item Only the expressions over $\langcontext$ obtained by applying the above rules  are $\AMLsig$-contexts.
\ee
\edfn

The set of $\AMLsig$-contexts is denoted by $\setcontextsAML$ and $\AMLsig$-contexts are denoted by  $C$, $C_1$, $C_2$, \ldots.

\bfact
The definition of $\AMLsig$-contexts can be written using the BNF notation:
\begin{align*}
C ::= & \Box \mid \ApplC C\vp \mid \ApplC \vp C.
\end{align*}
\efact

\bdfn[Alternative definition for $\AMLsig$-contexts]\label{def-contexts-equiv}\,\\
The set of \defnterm{$\AMLsig$-contexts}  is the intersection of all 
sets $\Gamma$ of expressions over $\langcontext$ that have the following properties:
\be
\item $\Box\in \Gamma$.
\item If $C\in \Gamma$ and $\vp$ is a $\AMLsig$-pattern, then $\ApplC C\vp, \ApplC \vp C\in \Gamma$.
\ee
\edfn

\begin{proposition}\label{induction-contexts}[Induction principle on contexts]\, \\
Let $\Gamma$  be a set of $\AMLsig$-contexts satisfying the following properties:
\be
\item $\Box\in \Gamma$.
\item If $C\in \Gamma$ and $\vp$ is a pattern, then $\ApplC C\vp, \ApplC \vp C\in \Gamma$.
\ee
Then $\Gamma=\setcontexts$.
\end{proposition}
\solution{By hypothesis, $\Gamma \se \setcontexts$. By Definition~\ref{def-contexts-equiv}, we get that 
$ \setcontexts\se \Gamma$.
}

When the signature $\AMLsig$ is clear from the context, we shall say simply context(s) and 
we shall denote the set of contexts by $\setcontexts$. 

\bprop
$\Box$ occurs exactly once in every context $C$.
\eprop
\solution{The proof is by induction on the context $C$:
\be
\item $C=\Box$. Obviously.
\item $C=\ApplC C_1\psi$ and, by the induction hypothesis, $\Box$ occurs exactly once in $C_1$. As $\psi$
is a pattern, $\Box$ does not occur in $\psi$. Thus, $\Box$ occurs exactly once in $C$.
\item $C=\ApplC \psi C_1$ and, by the induction hypothesis, $\Box$ occurs exactly once in $C_1$. As $\psi$
is a pattern, $\Box$ does not occur in $\psi$. Thus, $\Box$ occurs exactly once in $C$.
\ee
}

\subsubsection{Unique readability and recursion principle}

\bprop[Unique readability of contexts]\label{unique-read-contexts}\,
\be
\item\label{context-positive} Any context has a positive length.
\item\label{context-read} If $C$ is a context,  then one of the following hold:
\be
\item $C = \Box$.
\item $C=\ApplC C_1\vp$, where $C_1$ is a context and $\vp$ is a pattern.
\item $C=\ApplC \vp C_1$, where $C_1$ is a context and $\vp$ is a pattern. 
\ee
\item\label{context-initial-segment} Any proper initial segment of a context is not a context.
\item\label{context-unique-read} If $C$ is a context,  then exactly one of the cases from \eqref{context-read}
holds. Moreover, $C$   can be written in a unique way in one of these forms. 
\ee 
\eprop
\solution{Similarly with the proof of Proposition~\ref{unique-read-patterns}.}

\bprop[Recursion principle on contexts]\label{recursion-principle-contexts}\,\\
Let $A$  be a set, $\Box^A\in A$ and the mappings
\begin{align*} 
G_1, G_2:A \times \setpat \to A.
\end{align*}
Then there exists a unique mapping 
\[F:\setcontexts \to A\]
that satisfies the following properties:
\be
\item $F(\Box)=\Box^A$ for any atomic pattern $\vp$.
\item  $F(\ApplC C_1\vp)=G_1(F(C_1), \vp)$ for any context $C_1$ and any pattern $\vp$.
\item  $F(\ApplC \vp C_1)=G_2(F(C_1), \vp)$ for any context $C_1$ and any pattern $\vp$.
\ee
\eprop
\solution{Apply Proposition~\ref{unique-read-contexts}. 
}

\subsubsection{Context substitution}

\bdfn
Let $C$ be a context and $\delta$ be a pattern. We denote by $C[\delta]$ the $\AMLsig$-expression  
obtained by replacing all the occurences of $\Box$ with $\delta$ and all occurences of 
$\ApplC$ with $\Appl$.
\edfn

\bfact[Alternative definition]\, \\
$C[\delta]$ can be defined by recursion on contexts as follows:\\

\begin{tabular}{lll}
$\Box[\delta]$ & = & $\delta$, \\[1mm]
$(\ApplC C_1\vp)[\delta]$ &=& $\Appl C_1[\delta]\vp$, \\[1mm]
$(\ApplC \vp C_1)[\delta]$ &=& $\Appl \vp C_1[\delta]$.
\et
\efact
\solution{Let $F:\setcontexts \to Fun(\setpat, Expr_{\AMLsig})$ be defined by $F(C)(\delta)=C[\delta]$. 
We show in the sequel that $F$ can be defined by recursion on patterns. The above definitions
become 
$$
F(\Box)(\delta)=\delta, \quad F(\ApplC C_1\vp)(\delta)=\Appl F(C_1)(\delta)\vp, \quad 
F(\ApplC \vp C_1)(\delta) =\Appl \vp F(C_1)(\delta).
$$
Apply Proposition~\ref{recursion-principle-contexts} with 
\bce 
$A=Fun(\setpat, Expr_{\AMLsig})$, 
$\Box^A(\delta)=\delta$, $G_1(f,\vp)(\delta)=\Appl f(\delta)\vp$ and $G_2(f,\vp)(\delta)=\Appl \vp f(\delta)$
\ece
for every $f:\setpat\to Expr_{\AMLsig}$ and every patterns $\delta$, $\vp$.
Then 
\be
\item $F(\Box)(\delta)=\delta=\Box^A(\delta)$ for every pattern $\delta$. It follows that $F(\Box)=\Box^A$.
\item $F(\ApplC C_1\vp)(\delta)=\Appl F(C_1)(\delta)\vp=G_1(F(C_1), \vp)(\delta)$ for every pattern $\delta$. 
 It follows that  $F(\ApplC C_1\vp)=G_1(F(C_1), \vp)$.
\item $F(\ApplC \vp C_1)(\delta)=\Appl \vp F(C_1)(\delta)=G_2(F(C_1), \vp)(\delta)$ for every pattern $\delta$. 
 It follows that  $F(\ApplC \vp C_1)=G_2(F(C_1), \vp)$.
\ee
Thus, $F$ is the unique mapping given by Proposition~\ref{recursion-principle-contexts}.
}

\bprop
$C[\delta]$ is a pattern. 
\eprop
\solution{The proof is by induction on the context $C$, using the alternative definition 
of $C[\delta]$.
\be
\item $C=\Box$. Then $C[\delta]=\delta$ is a pattern.
\item $C=\ApplC C_1\vp$ and, by the induction hypothesis,  $C_1[\delta]$ is a pattern. As
 $C[\delta]=\Appl C_1[\delta]\vp$, it follows that $C[\delta]$ is a pattern.
\item $C=\ApplC \vp C_1$ and, by the induction hypothesis,  $C_1[\delta]$ is a pattern. 
As $C[\delta]=\Appl \vp C_1[\delta]$, it follows that $C[\delta]$ is a pattern.
\ee
}

\subsubsection{Free variables of contexts}

\bdfn\label{def-free-var-contexts}
The set $FV(C)$ of \defnterm{free variables} of $C$ is defined by recursion on contexts as follows:\\

\bt{lll}
$FV(\Box)$ &  = & $\emptyset$, \\[1mm]
$FV(\ApplC C_1 \vp)$ & = & $FV(C_1)\cup FV(\vp)$, \\[1mm]
$FV(\ApplC \vp C_1)$ & = & $FV(C_1)\cup FV(\vp)$.
\et
\edfn
\solution{Apply Proposition~\ref{recursion-principle-contexts} with  
$$A=2^{EVar \cup SVar}, \quad \Box^A=\emptyset, \quad G_1(V, \vp)=G_2(V, \vp)=V\cup FV(\vp).$$
Then
\be
\item $FV(\Box)=\emptyset=\Box^A$.
\item $FV(\ApplC C_1 \vp)=FV(C_1)\cup FV(\vp)=G_1(FV(C_1), \vp)$.
\item $FV(\ApplC \vp C_1)=FV(C_1)\cup FV(\vp)=G_2(FV(C_1), \vp)$.
\ee
Thus, $FV:\setcontexts \to 2^{EVar \cup SVar}$ is the unique mapping given by Proposition~\ref{recursion-principle-contexts}.
}

%% file: AML-language-derived-conventions.tex
\subsection{Derived connectives. Conventions}

We shall  write in the sequel $\appln{\vp}{\psi}$ instead of $Appl\vp\psi$, $\vp\to\psi$ instead of $\to\vp\psi$, 
$\exists x.\vp$ instead of $\exists x\vp$, and $\mu X.\vp$ instead of $\mu X\vp$.
We shall also use parentheses when it is necessary to make clear the 
structure of a pattern. For example, we write $\appln{(\vp\to \psi)}{\chi}$. 

Furthermore, from now on, $\vp_1, \vp_2, \ldots$ will denote patterns 

The pattern $\appln{\vp}{\psi}$ is called an \defnterm{application}. As a convention, application is 
associative to the left. Thus, if $\vp_1, \ldots, \vp_n$ are patterns, we write $\appln{\appln{\appln{\vp_1}{\vp_2}}{\vp_3}\ldots}{\vp_n}$ instead of 
$(\ldots \appln{(\appln{(\appln{\vp_1}{\vp_2})}{\vp_3)} \ldots}{\vp_n})$.

We introduce derived  connectives by the following abbreviations:
$$
\ba{lll}
\bot = \mu X.X & \neg \vp = \vp\to \bot, & \top =\neg\bot, \\[1mm]
\vp\sau\psi = \neg\vp\to \psi, & \vp\si\psi = \neg(\neg\vp \sau \neg\psi),  & \varphi\dnd\psi =  (\varphi\to\psi)\si(\psi\to\varphi), \\[1mm]
\forall x.\vp = \neg\exists x.\neg\vp, & \nu X.\vp = \neg\mu X.\neg \subfX{\neg X}{\vp}.	
\ea
$$

The scope of $\exists$, $\forall$, $\mu$, $\nu$  goes as far as possible to the right unless it 
is limited by parentheses. For example, $\forall x.\vp\ra \psi$ is $\forall x.(\vp\ra \psi)$, 
$\mu X.\vp\si \psi$ is $\mu X.(\vp\si \psi)$, and 
$\exists x. \psi \ra ((\exists y.\vp)\ra z)$ is $\exists x. \big(\psi \ra ((\exists y.\vp)\ra z)\big)$.

To reduce the use of parentheses, we assume that 
\be
\item application has the highest precedence.
\item $\mu$, $\nu$ have higher precedence than the quantifiers $\exists$, $\forall$.
\item quantifiers $\exists$, $\forall$ have higher precedence than the propositional connectives.
\item $\neg$ has higher precedence than $\ra$, $\si$, $\sau$, $\dnd$;
\item $\si, \sau$ have higher precedence than $\ra$, $\dnd$.
\ee

We define in the sequel \defnterm{finite conjunctions and disjunctions}. Let $\vp_1, \ldots, \vp_n (n\geq 1)$ be patterns. 
Then $\bigwedge_{i=1}^n \vp_i$ and $\bigvee_{i=1}^n \vp_i$ are defined inductively as follows:
\begin{align}
\bigwedge_{i=1}^1 \vp_i=\vp_1, \,\, \bigwedge_{i=1}^2 \vp_i= \vp_1\si\vp_2, \quad 
\bigwedge_{i=1}^{n+1} \vp_i = \left(\bigwedge_{i=1}^n \vp_i\right)\si \vp_{n+1}, \label{def-finite-si}\\
\bigvee_{i=1}^1 \vp_i=\vp_1, \,\, \bigvee_{i=1}^2 \vp_i= \vp_1\sau\vp_2, \quad 
\bigvee_{i=1}^{n+1} \vp_i = \left(\bigvee_{i=1}^n \vp_i\right)\sau \vp_{n+1}. \label{def-finite-sau}
\end{align}

We also write $\vp_1\si \vp_2 \si \ldots \si \vp_n$ instead of $\bigwedge_{i=1}^n \vp_i$ and 
$\vp_1\sau \vp_2 \sau \ldots \sau \vp_n$ instead of $\bigvee_{i=1}^n \vp_i$.

%% file: AML-semantics.tex
\section{Semantics}

Let $\AMLsig=(EVar, SVar, \Sigma)$ be a signature.

\bdfn
A $\AMLsig$-structure is a pair 
$$\cA=(A, \_\appls\_\,, \Sigma^{\cal A}),$$
where
\be
\item $A$ is a nonempty set.
\item  $\_\appls\_:A\times A\to 2^A$ is a binary \defnterm{application} function;
\item  $\Sigma^{\cal A}=\{\sigma^{\cal A}\se A \mid \sigma\in \Sigma\}$.
\ee
\edfn
Let $\cA$ be a $\AMLsig$-structure. $A$ is called  the \defnterm{universe} or the \defnterm{domain} of the structure $\cA$. 
We use the notation $A=|\cA|$. If $a\in A$, we also write $a\in \cA$. Furthermore, $\sigma^\cA$  is called the  
\defnterm{interpretation} or the \defnterm{denotation} of the constant $\sigma$ in $\cA$.
We extend the application function $\_\appls\_$  as follows:
$$\_\appls\_: 2^A \times 2^A\to 2^A, \quad 
B \appls C =\bigcup_{b\in B, c\in C} b \appls c.
$$

\bdfn
An \defnterm{$\cA$-valuation}  is a function  $\eval:EVar\cup SVar\to A\cup 2^A$
that satisfies: 
\bce
$\eval(x)\in A$ for all $x\in EVar$ and $\eval(X)\se A$ for all $X\in SVar$.
\ece 
\edfn
\defnterm{$\cA$-valuations} are also called  \defnterm{$\cA$-assignments}.

\mbox{}

In the following, $\cA$ is a $\AMLsig$-structure and $\eval$ is an $\cA$-valuation.

\bntn
For any variable $x\in EVar$ and any $a\in A$, we define  a new $\cA$-valuation $\evalxa$
as follows: for all $v\in EVar$ and all  $V\in SVar$, 
\bce
 $\evalxa(V)=\eval(V)$  and $\evalxa(v) = \begin{cases}
\eval(v) & \text{ if } v\neq x\\
a & \text{ if } v = x.
\end{cases}$
\ece
For any variable $X\in SVar$ and any $B\se A$, we define  a new $\cA$-valuation 
$\evalXB$ as follows: for all $v\in EVar$ and all  $V\in SVar$, 
\bce
 $\evalXB(v) =\eval(v)$ and $\evalXB(V) =\begin{cases}
\eval(V) & \text{ if } V\neq X\\
B & \text{ if } V = X.
\end{cases}$
\ece
\entn

\bdfn\label{def-val-patterns}
The \defnterm{patterns  $\cA$-valuation} 
$$ \evalp: \setpat \to 2^A$$
is defined by recursion on patterns  as follows:
\be
\item $\evalp(x)=\{\eval(x)\}$ if $x\in EVar$.
\item $\evalp(X)=\eval(X)$ if $X\in SVar$.
\item $\evalp(\sigma)=\sigma^\cA$ for every $\sigma \in \Sigma$.
\item $\evalp(\appln{\vp}{\psi}) = \evalp(\vp)\appls \evalp(\psi)$.
\item $\evalp(\vp \to \psi)= C_A(\evalp(\vp) \setminus\evalp(\psi))$.
\item $\evalp(\exists x.\vp)=\bigcup_{a\in A} \evalpxa(\vp)$.
\item\label{def-val-patterns-mu} $\evalp(\mu X.\vp)=\bigcap \left\{ B \subseteq A \mid \evalpXB(\vp) 
 \subseteq B \right\}.$
\ee
\edfn

We say that \defnterm{$\vp$ evaluates to $B\se A$} (with $\eval$) if $\evalp(\vp)=B$. If $a\in \evalp(\vp)$, 
we say that \defnterm{$a$ matches $\vp$} (with witness $\eval$).  
 
\bprop\label{basic-val}
For all patterns  $\vp,\psi$,
\be
\item\label{basic-val-imp} $\evalp(\vp \to \psi)= C_A\evalp(\vp)\cup  \evalp(\psi)$.
\item\label{basic-val-bot} $\evalp(\bot)=\emptyset$.
\item\label{basic-val-neg} $\evalp(\neg\vp) = C_A\evalp(\vp)$.
\item\label{basic-val-top} $\evalp(\top)=A$. 
\item\label{basic-val-sau} $\evalp(\vp \sau \psi)= \evalp(\vp) \cup \evalp(\psi)$.
\item\label{basic-val-si} $\evalp(\vp \si \psi)= \evalp(\vp) \cap \evalp(\psi)$.
\item\label{basic-val-dnd} $\evalp(\vp \dnd \psi)=A\setminus(\evalp(\vp) \diffsym \evalp(\psi))$.
\item\label{basic-val-forall} $\evalp(\forall x.\vp)=\bigcap_{a\in A} \evalpxa(\vp)$.
\ee
\eprop
\solution{
\be
\item We have that  
\begin{align*}
\evalp(\vp \to \psi)= 
C_A(\evalp(\vp) \setminus\evalp(\psi))= C_A\evalp(\vp) \cup \evalp(\psi).
\end{align*}  
\item We have that  
\begin{align*}
\evalp(\bot) &= \evalp(\mu X.X)= \bigcap \left\{ B \subseteq A \mid \evalpXB(X)  \subseteq B \right\} =
\bigcap \left\{ B \subseteq A \mid \evalXB(X)  \subseteq B \right\} \\
& = \bigcap \left\{ B \subseteq A \mid B  \subseteq B \right\}  =\bigcap \left\{ B \subseteq A \right\}=\emptyset.
\end{align*}  
\item $\evalp(\neg\vp) =\evalp(\vp\to\bot) = C_A\evalp(\vp)\cup \evalp(\bot)=C_A\evalp(\vp)\cup \emptyset 
= C_A\evalp(\vp)$.
\item $\evalp(\top) = \evalp(\neg\bot)=C_A\evalp(\bot)=C_A \emptyset = A$.
\item We have that
\begin{align*}
\evalp(\vp \sau \psi) &= \evalp(\neg\vp\to \psi)= C_A\evalp(\neg\vp) \cup \evalp(\psi)=
C_A(C_A\evalp(\vp)) \cup \evalp(\psi)\\
& = \evalp(\vp) \cup \evalp(\psi).
\end{align*}  
\item We have that
\begin{align*}
\evalp(\vp \si \psi) &= \evalp(\neg(\neg\vp\sau\neg\psi))= C_A\evalp(\neg\vp\sau\neg\psi)
=C_A(\evalp(\neg\vp)\cup\evalp(\neg\psi))\\
&=C_A(C_A\evalp(\vp)\cup C_A\evalp(\psi)) = 
\evalp(\vp)\cap \evalp(\psi).
\end{align*}  
\item We have that 
\begin{align*}
\evalp(\vp \dnd \psi) &= \evalp\big((\varphi\to\psi)\si(\psi\to\varphi)\big)=
\evalp(\varphi\to\psi) \cap \evalp(\psi\to\varphi)\\
& = C_A(\evalp(\vp) \setminus\evalp(\psi)) \cap C_A(\evalp(\psi) \setminus\evalp(\vp))
= C_A(\evalp(\vp) \diffsym \evalp(\psi)).
\end{align*}
\item We have that 
\begin{align*}
\evalp(\forall x.\vp) &= \evalp(\neg \exists x.\neg\vp)=C_A \evalp(\exists x.\neg\vp)=
C_A\left(\bigcup_{a\in A} \evalpxa(\neg\vp)\right) \\
& = C_A\left(\bigcup_{a\in A} \left(C_A\evalpxa(\vp)\right)\right) 
= \bigcap_{a\in A} C_A\left(C_A\evalpxa(\vp)\right)= \bigcap_{a\in A} \evalpxa(\vp).
\end{align*}
\ee
}

\bdfn
Let $\vp$ be a pattern. We say that
\be
\item $\eval$ \defnterm{satisfies} $\vp$ in $\cA$ 
if $\evalp(\vp)=A$.  \\
Notation:  $\cA\models \vp[\eval]$.
\item $\eval$ \defnterm{does not satisfy} $\vp$ in $\cA$ 
if $\evalp(\vp) \ne A$. \\
 Notation:  $\cA\not\models \vp[\eval]$.
\ee
\edfn

\bprop\label{A-e-models}
\be
\item\label{A-e-models-x} For every element variable $x$, $\cA\models x[\eval]$ iff $|A|=1$.
\item\label{A-e-models-X} For every set variable $X$, $\cA\models X[\eval]$ iff $\eval(X)=A$.
\item For every constant $\sigma$, $\cA\models \sigma[\eval]$ iff $\sigma^\cA=A$.
\item $\cA\not\models \bot[\eval]$ and $\cA\models \top[\eval]$.
\item For every pattern  $\vp$, $\cA\models \neg\vp[\eval]$ iff $\evalp(\vp)=\emptyset$.
\item For every patterns  $\vp$, $\psi$, $\cA\models (\appln{\vp}{\psi})[\eval]$ iff 
$\bigcup_{b\in \evalp(\vp), c\in \evalp(\psi)} b \appls c=A$.
\item\label{A-e-models-si}  For every patterns $\vp$, $\psi$, $\cA\models (\vp\si\psi)[\eval]$ iff ($\cA\models \vp[\eval]$ and $\cA\models \psi[\eval]$).
\item For every patterns  $\vp$, $\psi$, $\cA\models (\vp\sau\psi)[\eval]$ iff $\evalp(\vp)\cup \evalp(\psi)=A$.
\item\label{A-e-models-to} For every patterns  $\vp$, $\psi$, $\cA\models (\vp\to\psi)[\eval]$ iff $\evalp(\vp) \se \evalp(\psi)$. 
\item\label{A-e-models-dnd} For every patterns  $\vp$, $\psi$, $\cA\models (\vp\dnd\psi)[\eval]$ iff $\evalp(\vp) = \evalp(\psi)$.
\item\label{A-e-models-dnd-2} For every patterns  $\vp$, $\psi$, $\cA\models (\vp\dnd\psi)[\eval]$ iff ($\cA\models (\vp\to\psi)[\eval]$
and $\cA\models (\psi\to\vp)[\eval])$.
\item\label{A-e-models-exists-iff} For every pattern  $\vp$ and every element variable $x$, $\cA\models (\exists x.\vp)[\eval]$ 
iff $\bigcup_{a\in A} \evalpxa(\vp)=A$.
\item\label{A-e-models-exists-imp} For every pattern  $\vp$ and every element variable $x$, if there exists $b\in A$ such that 
$\cA\models \vp[\evalxb]$, then 
$\cA\models (\exists x.\vp)[\eval]$. 
\item For every pattern $\vp$ and every element variable $x$, $\cA\models (\forall x.\vp)[\eval]$ iff 
$\cA\models \vp[\evalxa]$ for every $a\in A$.
\item For every pattern $\vp$ and every set variable $X$, if $\cA\models \vp[\evalXB]$ for every $B\se A$, then 
$\cA\models (\mu X.\vp)[\eval]$.
\ee
\eprop
\solution{
\be
\item $\cA\models x[\eval]$ iff $\evalp(x)=A$ iff $\{\eval(x)\}=A$ iff $|A|=1$.
\item  $\cA\models X[\eval]$ iff $\evalp(X)=A$ iff $\eval(X)=A$.
\item  $\cA\models \sigma[\eval]$ iff $\evalp(\sigma)=A$ iff $\sigma^\cA=A$.
\item Obviously, since $\evalp(\bot)=\emptyset$ and $\evalp(\top)=A$.
\item $\cA\models \neg\vp[\eval]$ iff  $\evalp(\neg\vp)=A$ iff $A\setminus \evalp(\vp)=A$ iff 
$\evalp(\vp)=\emptyset$.
\item $\cA\models (\appln{\vp}{\psi})[\eval]$ iff $\evalp(\appln{\vp}{\psi})=A$ iff $\evalp(\vp)\appls \evalp(\psi)=A$
iff $\bigcup_{b\in \evalp(\vp), c\in \evalp(\psi)} b \appls c=A$.
\item $\cA\models (\vp\si\psi)[\eval]$ iff $\evalp(\vp\si\psi)=A$ iff $\evalp(\vp)\cap \evalp(\psi)=A$ iff
$\evalp(\vp)=\evalp(\psi)=A$ (since $\evalp(\vp),\evalp(\psi) \se A$) iff ($\cA\models \vp[\eval]$ and $\cA\models \psi[\eval]$).
\item $\cA\models (\vp\sau\psi)[\eval]$ iff $\evalp(\vp\sau\psi)=A$ iff $\evalp(\vp)\cup \evalp(\psi)=A$.
\item $\cA\models (\vp\to\psi)[\eval]$ iff  $\evalp(\vp\to\psi)=A$ iff $C_A(\evalp(\vp) \setminus\evalp(\psi))=A$ iff
$\evalp(\vp) \setminus\evalp(\psi)=\emptyset$ iff $\evalp(\vp) \se \evalp(\psi)$.
\item $\cA\models (\vp\dnd\psi)[\eval]$ iff $\evalp(\vp\dnd\psi)=A$ iff 
$A\setminus (\evalp(\vp) \diffsym \evalp(\psi))=A$ iff $\evalp(\vp) \diffsym \evalp(\psi)=\emptyset$ 
iff $\evalp(\vp) = \evalp(\psi)$.
\item Apply \eqref{A-e-models-si}.
\item $\cA\models (\exists x.\vp)[\eval]$ iff $\evalp(\exists x. \vp)=A$ iff $\bigcup_{a\in A} \evalpxa(\vp)=A$.
\item Assume that  $b\in A$ is such that $\cA\models \vp[\evalxb]$, so $\evalpxb(\vp)=A$. Since  $\evalpxb(\vp)\se 
\bigcup_{a\in A} \evalpxa(\vp)=\evalp(\exists x.\vp)$, we must have $\evalp(\exists x.\vp)=A$, hence 
$\cA\models (\exists x.\vp)[\eval]$.
\item $\cA\models (\forall x\vp)[\eval]$ iff $\evalp(\forall x.\vp)=A$ iff $\bigcap_{a\in A} \evalpxa(\vp)=A$ iff
for every $a\in A$, $\evalpxa(\vp)=A$ iff for every $a\in A$, $\cA\models \vp[\evalxa]$.
\item Let us denote $\cB=\left\{ B \subseteq A \mid \evalpXB(\vp)\subseteq B \right\}$. Obviously, $A\in \cB$.
Assume that for every $B\se A$ we have that $\cA\models \vp[\evalXB]$, hence $\evalpXB(\vp)=A$. It follows that $\cB=\{A\}$, 
since if $B\se A, B\ne A$, then
$\evalpXB(\vp)=A\not \subseteq B$. Thus, $\evalp(\mu X.\vp)=\bigcap \cB =A$, so $\cA\models (\mu X.\vp)[\eval]$.
\ee
}

\bprop\label{evaluation-patterns-free-variables}
Let $\cA$ be a $\AMLsig$-structure. Then for every pattern $\vp$, the following holds:
\bce
$(\star)$ \quad for every $\cA$-valuations $\evalu$, $\evald$, if $\evalu|_{FV(\vp)}=\evald|_{FV(\vp)}$, then
$\evalup(\vp)=\evaldp(\vp)$.
\ece
\eprop
\solution{We prove $(\star)$  by induction on $\vp$.
\be
\item $\vp=x\in EVar$. Then $FV(\vp)=\{x\}$. Let $\evalu$, $\evald$ be two $\cA$-valuations such that 
$\evalu(x)=\evald(x)$. We get that
$\evalup(\vp)=\{\evalu(x)\}=\{\evald(x)\}=\evaldp(\vp)$.
\item $\vp=X\in SVar$. Then $FV(\vp)=\{X\}$. Let $\evalu$, $\evald$ be two $\cA$-valuations such that 
$\evalu(X)=\evald(X)$. We get that
$\evalup(\vp)=\evalu(X)=\evald(X)=\evaldp(\vp)$.
\item $\vp=\sigma\in \Sigma$. Then $FV(\vp)=\emptyset$. Let $\evalu$, $\evald$ be two $\cA$-valuations. Then 
$\evalup(\vp)=\sigma^\cA=\evaldp(\vp)$. 
\item $\vp=\appln{\psi}{\chi}$. Then $FV(\vp)=FV(\psi)\cup FV(\chi)$. Let $\evalu$, $\evald$ be two $\cA$-valuations such that 
$\evalu|_{FV(\vp)}=\evald|_{FV(\vp)}$. Then $\evalu|_{FV(\psi)}=\evald|_{FV(\psi)}$ and 
$\evalu|_{FV(\chi)}=\evald|_{FV(\chi)}$. Applying the induction hypothesis  for $\psi$, $\chi$, 
it follows that $\evalup(\psi)=\evaldp(\psi)$ and $\evalup(\chi)=\evaldp(\chi)$. We get that
$\evalup(\vp)=\evalup(\psi)\appls\evalup(\chi)=\evaldp(\psi)\appls\evaldp(\chi)=\evaldp(\vp$.)
\item $\vp=\psi\to\chi$. Then $FV(\vp)=FV(\psi)\cup FV(\chi)$. Let $\evalu$, $\evald$ be two $\cA$-valuations such that 
$\evalu|_{FV(\vp)}=\evald|_{FV(\vp)}$. Then $\evalu|_{FV(\psi)}=\evald|_{FV(\psi)}$ and 
$\evalu|_{FV(\chi)}=\evald|_{FV(\chi)}$. Applying the induction hypothesis  for $\psi$, $\chi$, 
it follows that $\evalup(\psi)=\evaldp(\psi)$ and $\evalup(\chi)=\evaldp(\chi)$. We get that 
$\evalup(\vp)=C_A(\evalup(\psi)\setminus \evalup(\chi))= C_A(\evaldp(\psi)\setminus \evaldp(\chi))
=\evaldp(\vp)$.
\item $\vp=\exists x.\psi$. Then $FV(\vp)=FV(\psi)\setminus \{x\}$. Let $\evalu$, $\evald$ be two $\cA$-valuations such that 
$\evalu|_{FV(\vp)}=\evald|_{FV(\vp)}$. For every $a\in A$, we have that 
$\evaluxa|_{FV(\psi)}=\evaldxa|_{FV(\psi)}$, hence,  applying the induction hypothesis  for $\psi$, that 
$\evalupxa(\psi)=\evaldpxa(\psi)$. We get that 
$\evalup(\vp)=\bigcup_{a\in A} \evalupxa(\psi)=\bigcup_{a\in A} \evaldpxa(\psi)=\evaldp(\vp)$.
\item $\vp=\mu X.\psi$. Then $FV(\vp)=FV(\psi)\setminus \{X\}$. Let $\evalu$, $\evald$ be two $\cA$-valuations such that 
$\evalu|_{FV(\vp)}=\evald|_{FV(\vp)}$. For every $B\se A$, we have that 
$\evaluXB|_{FV(\psi)}=\evaldXB|_{FV(\psi)}$,  hence,  applying the induction hypothesis  for $\psi$, that 
$\evalupXB(\psi)=\evaldpXB(\psi)$. We get that 
$\evalup(\vp)=\bigcap \{B \se A \mid \evalupXB(\psi) \se B\}=
\bigcap \{B \se A \mid \evaldpXB(\psi)\se B\}=\evaldp(\vp)$. 
\ee
}

Hence, the $\cA$-valuation of a pattern depends only on the $\cA$-valuations of the free variables of the pattern. As an immediate 
application of Proposition \ref{evaluation-patterns-free-variables}, we get 

\bcor
Let $\cA$ be a $\AMLsig$-structure, $\vp$ be a pattern and $\evalu$, $\evald$ be two $\cA$-valuations such that 
$\evalu|_{FV(\vp)}=\evald|_{FV(\vp)}$.
Then $\cA\models \vp[\evalu]$  iff $\cA\models \vp[\evald]$.
\ecor

%% file: AML-semantics-models-valid-log-equiv.tex
\subsection{Model, validity, satisfiability}

\bdfn
Let $\vp$ be a pattern and $\cA$ be a $\AMLsig$-structure. We say that 
$\cA$ \defnterm{satisfies} $\vp$ or that $\cA$ is a \defnterm{model} of $\vp$ if $\cA\models \vp[\eval]$ 
for every $\cA$-valuation $\eval$. \\
Notation: $\cA\models \vp$.
\edfn

\bprop\label{A-models}
Let $\cA$ be a $\AMLsig$-structure. 
\be
\item\label{A-models-1} For every element variable $x$, $\cA\models x$ iff $|A|=1$.
\item\label{A-models-2} For every constant symbol $\sigma$, $\cA\models \sigma$ iff $\sigma^\cA=A$.
\item\label{A-models-3} $\cA\not\models \bot$ and $\cA\models \top$.
\item\label{A-models-4} For every patterns $\vp$, $\psi$,  $\cA\models \vp\si\psi$ iff ($\cA\models \vp$ and $\cA\models \psi$).
\item\label{A-models-5} For every patterns $\vp$, $\psi$,  $\vp$, $\cA\models \vp\dnd\psi$ iff ($\cA\models \vp\to\psi$ and $\cA\models \psi\to\vp$).
\item\label{A-models-6} For every patterns $\vp$, $\psi$, if $\cA\models \vp\to\psi$ and $\cA\models \vp$, then $\cA\models \psi$.
\item\label{A-models-7} For every patterns $\vp$, $\psi$, if $\cA\models \vp\dnd\psi$, then  ($\cA\models \vp$ iff $\cA\models \psi$).
\item\label{A-models-forall} For every pattern $\vp$ and every element variable $x$, $\cA \models \forall x.\vp$ iff $\cA\models \vp$.
\ee
\eprop
\solution{
\eqref{A-models-1}-\eqref{A-models-7} are obtained as an immediate consequence of Proposition \ref{A-e-models}.

\eqref{A-models-forall} $\Ra$ Let $\eval$ be an arbitrary $\cA$-valuation. Since $\cA \models \forall x.\vp$,  we have that $\cA\models \vp[\evalxa]$ 
for all $a\in A$. By letting $a:=\eval(x)$, we get that $\eval=\evalxa$. Thus, $\cA\models \vp[\eval]$.

$\La$ Let $\eval$ be an arbitrary $\cA$-valuation and $a\in A$. We have to 
prove that $\cA \models \vp[\evalxa]$. This follows immediately from the fact that $\cA\models \vp$.
}

\bdfn
Let $\vp$ be a pattern. We say that 
$\vp$ is \defnterm{valid} if $\cA\models \vp$ for every $\AMLsig$-structure $\cA$.\\
Notation: $\models \vp$.
\edfn

As an application of Propositions~\ref{A-models}, we get

\bfact\label{valid-basic}
\be
\item $\top$ is valid.
\item For every patterns  $\vp$, $\psi$, $\models\vp\si\psi$   iff (\,$\models\vp$ and $\models\psi$).
\item\label{valid-basic-dnd} For every patterns  $\vp$, $\psi$, $\models\vp\dnd\psi$ iff 
(\,$\models\vp\to\psi$ and $\models\psi\to\vp$).
\item\label{valid-mp} For every patterns $\vp$, $\psi$, if $\models \vp\to\psi$ and $\models \vp$, then $\models \psi$.
\item\label{valid-mp-dnd} For every patterns $\vp$, $\psi$, if $\models \vp\dnd\psi$, then  ($\models \vp$ iff $\models \psi$).
\ee
\efact

\mbox{}

In the sequel, we write  ``Let $(\cA, \eval)$'' instead of ``Let $\cA$ be a $\AMLsig$-structure 
and $\eval$ be an $\cA$-valuation'', 
``there exists $(\cA, \eval)$'' instead of ``there exists a $\AMLsig$-structure 
$\cA$ and an $\cA$-valuation $\eval$'' and  ``for all $(\cA, \eval)$'' instead of ``for every 
$\AMLsig$-structure $\cA$ and every $\cA$-valuation $\eval$".

%% file: AML-semantics-finite-conj-disj.tex
\subsubsection{Finite disjunctions and conjunctions}

Let $\vp_1, \ldots, \vp_n (n\geq 1)$ be patterns.

\bprop
Let $(\cA, \eval)$. Then 
\begin{align}
\evalp(\vp_1\si \vp_2 \si \ldots \si \vp_n)=\bigcap_{i=1}^n \evalp(\vp_i) \quad \text{and} \quad 
\evalp(\vp_1\sau \vp_2 \sau \ldots \sau \vp_n)=\bigcup_{i=1}^n \evalp(\vp_i)
\end{align}
\eprop
\solution{By induction on $n$. The case $n=1$ is obvious. For $n=2$  apply Proposition~\ref{basic-val}.\eqref{basic-val-si}, \eqref{basic-val-sau}.

$n\Ra n+1$: We have that 
\begin{align*}
\evalp(\vp_1\si \vp_2 \si \ldots \si \vp_n\si \vp_{n+1}) & = \evalp((\vp_1\si \vp_2 \si \ldots \si \vp_n)\si \vp_{n+1})\\
& =\evalp(\vp_1\si \vp_2 \si \ldots \si \vp_n) \cap \evalp(\vp_{n+1}) =\bigcap_{i=1}^n \evalp(\vp_i)\cap \evalp(\vp_{n+1})\\
&= \bigcap_{i=1}^{n+1} \evalp(\vp_i), \\
\evalp(\vp_1\sau \vp_2 \sau \ldots \sau \vp_n\sau \vp_{n+1}) & = \evalp((\vp_1\sau \vp_2 \sau \ldots \sau \vp_n)\sau \vp_{n+1})\\
& =\evalp(\vp_1\sau \vp_2 \sau \ldots \sau \vp_n) \cup \evalp(\vp_{n+1}) =\bigcup_{i=1}^n \evalp(\vp_i)\cup \evalp(\vp_{n+1})\\
&= \bigcup_{i=1}^{n+1} \evalp(\vp_i).
\end{align*}
}

It follows immediately that 

\bprop\label{A-e-models-finite-disj-conj}
The following hold: 
\be
\item\label{A-e-models-finite-disj-conj-A-e} For every $(\cA, \eval)$,
\be
\item $\cA\models (\vp_1\si \ldots \si \vp_n)[\eval]$ iff $\cA\models \vp_i[\eval]$ for every $i=1, \ldots, n$.
\item $\cA\models (\vp_1\sau \ldots \sau \vp_n)[\eval]$ iff $\bigcup_{i=1}^n \evalp(\vp_i)=A$.
\ee
\item\label{A-e-models-finite-disj-conj-A} For every $\cA$, $\cA \models\vp_1\si \ldots \si \vp_n$ iff $\cA \models \vp_i$ for every $i=1, \ldots, n$.
\item\label{A-e-models-finite-disj-conj-valid} $\models\vp_1\si \ldots \si \vp_n$ iff $\models \vp_i$ for every $i=1, \ldots, n$.
\ee
\eprop

%% file: AML-semantics-consequence.tex
\subsection{Semantic consequence and equivalence}\label{semantic-cons-equiv-section}

We define in the sequel three notions of semantic consequence and logic equivalence.

\subsubsection{Notion 1}

\bdfn\label{global-conseq-logic-equiv-vp-psi}
Let $\vp$, $\psi$ be patterns. We say that 
\be
\item\label{gsc} $\psi$ is a \defnterm{\gscn} of  $\vp$ if for every $\cA$, 
\bce 
$\cA \models \vp$ implies $\cA \models \psi$.
\ece
Notation: $\vp \gsc \psi$.
\item\label{gleq} $\vp$ and $\psi$ are \defnterm{\gleqn}  if for every $\cA$, 
$\cA \models \vp$ iff $\cA \models \psi$. \\[1mm]
Notation: $\vp \gleq \psi$.
\ee
\edfn

\bfact\label{gleq-two-gsc}
$\vp \gleq \psi$ iff ($\vp \gsc  \psi$ and $\psi \gsc  \vp$).
\efact
\solution{Obviously.}

\bfact\label{gsc-gleq-models}
\be
\item\label{gsc-gleq-models-to} If $\vp \gsc \psi$, then $\models\vp$ implies $\models\psi$.
\item\label{gsc-gleq-models-dnd} If $\vp \gleq \psi$, then $\models\vp$ iff $\models\psi$.
\ee
\efact
\solution{
\be
\item Assume that $\vp \gsc \psi$ and that $\models\vp$. We have to prove that $\models\psi$, hence that for every $\cA$, 
$\cA\models \psi$. Let $\cA$ be arbitrary. Since $\models\vp$, we have that $\cA\models\vp$. Since $\vp \gsc \psi$, 
it follows that $\cA\models \psi$.
\item By \eqref{gsc-gleq-models-to}.
\ee
}

\subsubsection{Notion 2}

\bdfn\label{local-conseq-logic-equiv-vp-psi}
Let $\vp$, $\psi$ be patterns. We say that 
\be
\item\label{lsc} $\psi$ is a \defnterm{\lscn} of  $\vp$ if for every $(\cA, \eval)$, 
\bce 
$\cA \models \vp[e]$ implies $\cA \models \psi[e]$.
\ece
Notation: $\vp \lsc \psi$.
\item\label{lleq}  $\vp$ and $\psi$ are \defnterm{\lleqn}  if for every $(\cA, \eval)$, 
$\cA \models \vp[e]$ iff $\cA \models \psi[e]$. \\[1mm]
Notation: $\vp \lleq \psi$.
\ee
\edfn

\bfact\label{lleq-two-lsc}
$\vp \lleq \psi$ iff ($\vp \lsc \psi$ and $\psi \lsc \vp$).
\efact

\bfact\label{lsc-lleq-models}
\be
\item\label{lsc-lleq-models-to} If $\vp \lsc \psi$, then $\models\vp$ implies $\models\psi$.
\item\label{lsc-lleq-models-dnd} If $\vp \lleq \psi$, then $\models\vp$ iff $\models\psi$.
\ee
\efact
\solution{
\be
\item Assume that $\vp \lsc \psi$ and that $\models\vp$. We have to prove that $\models\psi$, hence that for every $(\cA, \eval)$, 
$\cA\models \psi[\eval]$. Let $(\cA, \eval)$ be arbitrary. Since $\models\vp$, we have that $\cA\models\vp[\eval]$. 
Since $\vp \lsc \psi$, 
it follows that $\cA\models \psi[\eval]$.
\item By \eqref{gsc-gleq-models-to}.
\ee
}

\subsubsection{Notion 3} 

\bdfn\label{strong-conseq-logic-equiv-vp-psi}
Let $\vp$, $\psi$ be patterns. We say that 
\be
\item $\psi$ is a \defnterm{\sscn} of  $\vp$ if for every $(\cA, \eval)$, 
$\evalp(\vp)\se \evalp(\psi)$. \\[1mm]
Notation: $\vp \ssc \psi$.
\item $\vp$ and $\psi$ are \defnterm{\sleqsn}  if for every $(\cA, \eval)$, 
$\evalp(\vp)=\evalp(\psi)$. \\[1mm] 
Notation: $\vp \sleqs \psi$.
\ee
\edfn

\bfact\label{sleqs-two-ssc}
$\vp \sleqs \psi$ iff ($\vp \ssc \psi$ and $\psi \ssc \vp$).
\efact

\bfact\label{ssc-sleqs-to-dnd}
\be
\item\label{ssc-sleqs-to-dnd-ssc} $\vp \ssc  \psi$ iff $\models \vp\to \psi$.
\item\label{ssc-sleqs-to-dnd-sleqs} $\vp \sleqs \psi$ iff $\models \vp\dnd \psi$.
\ee
\efact
\solution{
\be
\item $\vp \ssc  \psi$ iff for every $\eval$, $\evalp(\vp)\se \evalp(\psi)$ iff for every $\eval$, 
$\models (\vp\to \psi)[\eval]$ (by Proposition \ref{A-e-models}.\eqref{A-e-models-to} iff 
$\models \vp\to \psi$.
\item $\vp \sleqs \psi$ iff for every $\eval$, $\evalp(\vp) = \evalp(\psi)$ iff for every $\eval$, 
$\models (\vp\dnd \psi)[\eval]$ 
(by Proposition \ref{A-e-models}.\eqref{A-e-models-dnd}) iff $\models \vp\dnd \psi$.
\ee
}

\bfact\label{ssc-sleqs-models}
\be
\item\label{ssc-sleqs-models-to} If $\vp \ssc \psi$, then $\models\vp$ implies $\models\psi$.
\item\label{ssc-sleqs-models-dnd} If $\vp \sleqs \psi$, then $\models\vp$ iff $\models\psi$.
\ee
\efact
\solution{
By Remark~\ref{ssc-sleqs-to-dnd} and Remark~\ref{valid-basic},\eqref{valid-mp},\eqref{valid-mp-dnd}.
}

\subsubsection{Relations between the three notions}

\bprop\label{remark-conseq-equiv-to-dnd}
\be
\item\label{remark-conseq-equiv-to-imp-dnd} $\vp \ssc \psi$ implies $\vp \lsc \psi$ implies $\vp \gsc \psi$.
\item\label{remark-conseq-equiv-to-dnd-dnd} $\vp \sleqs \psi$ implies $\vp \lleq \psi$ implies 
$\vp \gleq \psi$.
\ee
\eprop
\solution{
\be
\item 
$\vp \ssc \psi$ implies $\vp \lsc \psi$: Let $(\cA,\eval)$ be such that  $\cA\models \vp[\eval]$.
We have to prove that $\cA\models \psi[\eval]$. As $\cA\models \vp[\eval]$, we have that 
$\evalp(\vp)=A$. Since $\evalp(\vp)\se\evalp(\psi)$, it follows that 
$\evalp(\psi)=A$, hence $\cA\models \psi[\eval]$.\\[1mm]
$\vp \lsc \psi$ implies $\vp \gsc \psi$: Let $\cA\models \vp$ and $\eval$  be an arbitrary 
$\cA$-valuation. We have to prove that  $\cA\models \psi[\eval]$.
Since $\cA\models \vp$, we have that  $\cA\models \vp[\eval]$. As $\vp \lsc \psi$, it follows that 
$\cA\models \psi[\eval]$.
\item Follows immediately from \eqref{remark-conseq-equiv-to-imp-dnd}.
\ee
}

\bfact\label{x-y-lleq-gleq}
Let  $x$, $y$ be element variables. Then $x \lleq y$, hence  $x \gleq y$. 
\efact
\solution{ Let $(A,\eval)$ be arbitrary. Applying Proposition \ref{A-e-models}.\eqref{A-e-models-x}, we get that 
$(A,\eval)$ is a model of $x$ iff $\cA\models x[\eval]$ iff $|A|=1$
iff $\cA\models y[\eval]$ iff $(A,\eval)$ is a model of $y$.
Thus, $x \lleq y$. By Proposition \ref{remark-conseq-equiv-to-dnd}.\eqref{remark-conseq-equiv-to-dnd-dnd}, it follows that
$x \gleq y$. 
}

\bfact\label{does-not-imply-gsc-lsc-ssc}
\be
\item $\vp \gsc \psi$ does not imply $\vp \lsc \psi$ and $\vp \gleq \psi$ does not imply $\vp \lleq \psi$.
\item $\vp \lsc \psi$ does not imply $\vp \ssc \psi$ and $\vp \lleq \psi$ does not imply $\vp \sleqs \psi$.
\ee
\efact
\solution{
Let $x$ and $y$ be two different element variables. We give the following counterexamples.
\be
\item  Let us prove that $x\vee y \gleq x\si y$. Let $\cA$ be arbitrary. 
Then $\cA\models x\sau y$ iff (for every $e$,  $\cA\models (x\sau y)[e]$) iff 
(for every $e$, $\evalp(x\sau y)=A$) iff (for every $e$, $\{e(x)\}\cup \{e(y)\}=A$) iff $|A|=1$, 
as we can take $e$ such that $e(x)=e(y)$. 
Furthermore, $\cA\models x\si y$ iff (for every $e$,  $\cA\models (x\si y)[e]$) iff 
(for every $e$, $\evalp(x\si y)=A$) iff (for every $e$, $\{e(x)\}\cap \{e(y)\}=A$) iff $|A|=1$. 

Thus, $\cA\models x\sau y$ iff $\cA\models x\si y$ iff $|A|=1$.  Thus, $x\vee y \gleq x\si y$.

Let us prove now that $x\sau y \not\lsc x \si y$. Let $A=\{1,2\}$ and $e$ be such that $e(x)=1$ and $e(y)=2$. Then 
$\evalp(x\sau y)=\{e(x)\}\cup \{e(y)\}=A$. Thus, $\cA\models (x\sau y)[e]$. On the other hand,
$\evalp(x\si y)=\{e(x)\}\cap \{e(y)\}=\emptyset$, hence $\cA\not \models (x\si y)[e]$.

\item By Proposition \ref{x-y-lleq-gleq}, $x\lleq y$. On the other hand, $x \not\ssc y$: take an $(A,\eval)$ such that 
$|A|\geq 2$ and $\eval(x)\ne \eval(y)$. Then $\evalp(x)=\{\eval(x)\}\not\se \{\eval(y)\}=\evalp(y)$.
\ee
}

%% file: AML-semantics-sets-patterns.tex
\subsection{Sets of patterns}

Let $\Gamma$ be a set of patterns. 

\bdfn
Let $(\cA, \eval)$.  We say that
$\eval$ \defnterm{satisfies} $\Gamma$ in $\cA$ if  $\cA\models \gamma[\eval]$ for every $\gamma\in \Gamma$.
Notation: $\cA\models \Gamma[\eval]$.
\edfn

\bdfn
Let $\cA$ be a $\AMLsig$-structure. We say that
$\cA$ \defnterm{satisfies} $\Gamma$  or that $\cA$ is a \defnterm{model} of $\Gamma$ if  every $\cA$-valuation  satisfies $\Gamma$ in $\cA$.\\
Notation:   $\cA\models \Gamma$.
\edfn

%% file: AML-semantics-sets-consequence.tex
\subsection{Semantic consequence and equivalence for sets of patterns}\label{semantic-consequence-vp-Gamma}

We extend to sets of patterns the notions introduced in Subsection~\ref{semantic-cons-equiv-section}.

\subsubsection{Notion 1}

\bdfn\label{def-gsc-Gamma}
Let $\vp$ be a pattern. We say that 
$\vp$ is a \defnterm{\gscn} of $\Gamma$ if for every $\cA$, 
\bce
$\cA\models\Gamma$ \, implies \, $\cA \models\vp$.
\ece
Notation: $\Gamma\gsc \vp$.
\edfn

This is the definition of semantic consequence from \cite{Ros17}. 
By letting $\Gamma=\{\psi\}$, we get the notion introduced in Definition 
\ref{global-conseq-logic-equiv-vp-psi}.\eqref{gsc}. Hence, we write $\psi\gsc \vp$ instead of $\{\psi\}\gsc \vp$.

\bfact\label{basic-gsc-gamma}
\be
\item $\emptyset\gsc \vp$ iff $\models \vp$. 
\item If $\vp$ is valid, then $\Gamma \gsc \vp$.
\ee
\efact
\solution{ 
\be
\item $\emptyset\gsc \vp$ iff (for every $\cA$, we have that $\cA \models \vp$) iff $\models \vp$.
\item Obviously, as $\cA \models\vp$ for every $\cA$.
\ee
}

\blem\label{gsc-gleq-Gamma}
For all patterns  $\vp$, $\psi$, 
\be
\item\label{gsc-gleq-Gamma-gsc}  If $\vp  \gsc  \psi$, then  $\Gamma  \gsc  \vp$ implies $\Gamma  \gsc \psi$.
\item\label{gsc-gleq-Gamma-gleq}  If $\vp \gleq \psi$, then $\Gamma \gsc \vp$ iff $\Gamma \gsc   \psi$.
\item\label{gsc-gleq-Gamma-si} $\Gamma  \gsc  \vp\si \psi$ iff ($\Gamma  \gsc  \vp$ and $\Gamma  \gsc  \psi)$.
\item\label{gsc-gleq-Gamma-dnd} $\Gamma  \gsc  \vp\dnd \psi$ iff ($\Gamma  \gsc  \vp\to \psi$ and $\Gamma  \gsc  \psi\to \vp)$.
\ee
\elem
\solution{
\be
\item Assume that $\vp  \gsc  \psi$ and $\Gamma  \gsc  \vp$. Let $\cA$ be a model of $\Gamma$.
As $\Gamma \gsc  \vp$, we have that $\cA\models \vp$. Furthermore, $\cA\models \psi$ as $\vp  \gsc  \psi$.
\item It follows from (i) and Remark \ref{gleq-two-gsc}.
\item We have that \\[1mm]
\bt{lll}
$\Gamma  \gsc  \vp\si \psi$ & iff & for every model $\cA $ of $\Gamma$, $\cA\models \vp\si \psi$\\
& iff & for every model $\cA $ of $\Gamma$ and every $\cA$-valuation $\eval$, $\cA\models (\vp\si\psi)[\eval]$ \\[1mm]
& iff & for every model $\cA $ of $\Gamma$ and every $\cA$-valuation $\eval$,  $\cA\models \vp[\eval]$ and $\cA\models \psi[\eval]$ \\[1mm]
&& (by Proposition~\ref{A-e-models}.\eqref{A-e-models-si})\\[1mm]
& iff &  (for every model $\cA $ of $\Gamma$ and every $\cA$-valuation $\eval$,  $\cA\models \vp[\eval]$) and \\
&& (for every model $\cA $ of $\Gamma$ and every $\cA$-valuation $\eval$,  $\cA\models \psi[\eval]$)\\
& iff &  (for every model $\cA $ of $\Gamma$,  $\cA\models \vp$) and (for every model $\cA $ of $\Gamma$,  $\cA\models \psi$)\\
& iff & $\Gamma  \gsc  \vp$ and $\Gamma \gsc \psi$.
\et
\item Apply \eqref{gsc-gleq-Gamma-si}.
\ee
}

\bdfn\label{def-gsc-gleq-Gamma-Delta}
Let $\Delta$ be a set of patterns. We say that 
\be
\item $\Delta$ is a \defnterm{semantic consequence} of $\Gamma$ if for every $\cA$, 
\bce
$\cA\models\Gamma$ \, implies \, $\cA \models\Delta$.
\ece
\blue{Notation:} $\Gamma \gsc \Delta$.
\item $\Gamma$ and $\Delta$ are \defnterm{logic equivalent}  if for every $\cA$, 
\bce
$\cA\models\Gamma$ \, iff \, $\cA \models\Delta$.
\ece
\blue{Notation:} $\Gamma \gleq \Delta$.
\ee
\edfn

Obviously, by letting $\Delta=\{\vp\}$, we obtain the notion introduced in  Definition~\ref{def-gsc-Gamma}. Furthermore, 
for  $\Gamma=\{\psi\}$ and $\Delta=\{\vp\}$, we get the notions introduced in 
Definition \ref{global-conseq-logic-equiv-vp-psi}.

\bfact\label{Gamma-gleqn-two-gsc}
$\Gamma \gleq \Delta$ iff ($\Gamma \gsc \Delta$ and $\Delta \gsc \Gamma$).
\efact

\blem\label{Gamma-Delta-vp-psi-gleq}
\be
\item\label{Gamma-Delta-vp-psi-gleq-1} Assume that $\Gamma \gsc \Delta$ and that $\vp \gleq \psi$.
 Then  $\Delta  \gsc  \vp$ implies $\Gamma \gsc \psi$.
\item\label{Gamma-Delta-vp-psi-gleq-2} Assume that $\Gamma \gleq \Delta$ and that $\vp \gleq \psi$. 
Then  $\Gamma \gsc \vp$ iff $\Delta \gsc \psi$. 
\ee 
\elem
\solution{ 
\be
\item We have that 

\bt{lll}
$\Delta \gsc \vp$ & iff & for every model $\cA$ of $\Delta$, $\cA\models \vp$ \\
& iff & for every model $\cA$ of $\Delta$, $\cA\models \psi$  (as $\vp \gleq \psi$)\\
& implies & for every model $\cA$ of $\Gamma$, $\cA\models \psi$  (as $\Gamma \gsc \Delta$)\\
& iff & $\Gamma \gsc \psi$.
\et 

\item Apply \eqref{Gamma-Delta-vp-psi-gleq-1}.
\ee
}

\bprop[Finite sets]
Let $\Gamma=\{\vp_1, \ldots, \vp_n\}$ be a finite set of patterns. Then $\Gamma \gleq \{\vp_1\si\ldots \si \vp_n\}$. 
\eprop
\solution{Applying Proposition~\ref{A-e-models-finite-disj-conj}.\eqref{A-e-models-finite-disj-conj-A}, we get that for 
every $\cA$, $\cA\models\Gamma$ iff $\cA\models\vp_i$ for every $i=1,\ldots, n$ iff
$\cA\models \vp_1\si\ldots \si \vp_n$.
}

\subsubsection{Notion 2}

\bdfn\label{local-conseq-logic-equiv-Gamma-vp}
Let $\vp$ be a pattern. We say that $\vp$ is a \defnterm{\lscn} of $\Gamma$ if for every $(\cA, \eval)$, 
\bce
$\cA\models \Gamma[\eval]$\, implies \, $\cA\models \vp[\eval]$.
\ece
Notation: $\Gamma \lsc \vp$.
\edfn

By letting $\Gamma=\{\psi\}$, we get the notion introduced in Definition \ref{local-conseq-logic-equiv-vp-psi}. 
Hence, we write $\psi\lsc \vp$ instead of $\{\psi\}\lsc \vp$.

\bfact\label{basic-lsc-gamma}
\be
\item $\emptyset\lsc \vp$ iff $\models \vp$. 
\item If $\vp$ is valid, then $\Gamma \lsc  \vp$.
\ee
\efact
\solution{ 
\be
\item $\emptyset\lsc \vp$ iff (for every $(\cA, \eval)$, we have that $\cA\models \vp[\eval]$) iff (for every $\cA$, we have that 
$\cA\models \vp$) iff $\models \vp$. 
\item Obviously, as $\cA\models \vp[\eval]$ for every $(\cA,\eval)$.
\ee
}

\blem\label{lsc-lleq-Gamma}
For all patterns  $\vp$, $\psi$, 
\be
\item\label{lsc-lleq-Gamma-lsc}  If $\vp  \lsc  \psi$, then  $\Gamma  \lsc  \vp$ implies $\Gamma  \lsc  \psi$.
\item\label{lsc-lleq-Gamma-lleq} If $\vp\lleq \psi$, then $\Gamma \lsc \vp$ iff 
$\Gamma \lsc  \psi$.
\item\label{lsc-lleq-Gamma-si} $\Gamma  \lsc  \vp\si \psi$ iff ($\Gamma  \lsc  \vp$ and $\Gamma  \lsc  \psi)$.
\item\label{lsc-lleq-Gamma-dnd} $\Gamma  \lsc  \vp\dnd \psi$ iff ($\Gamma  \lsc  \vp\to \psi$ and $\Gamma  \lsc  \psi\to \vp)$.
\ee
\elem
\solution{
\be
\item Assume that $\vp  \lsc  \psi$ and $\Gamma  \lsc  \vp$.
Let $(\cA, \eval)$ be such that $\cA\models \Gamma[\eval]$. Since $\Gamma  \lsc  \vp$, we get that  
$\cA\models \vp[\eval]$. Since $\vp \lsc \psi$, it follows that $\cA\models \psi[\eval]$.
\item It follows from (i) and Remark \ref{lleq-two-lsc}.
\item We have that \\[1mm]
\bt{lll}
$\Gamma  \lsc  \vp\si \psi$ & iff & for every  $(\cA, \eval)$ such that  $\cA\models \Gamma[\eval]$ we have that 
$\cA\models (\vp\si\psi)[\eval]$ \\[1mm]
& iff & for every  $(\cA, \eval)$ such that  $\cA\models \Gamma[\eval]$ we have that  $\cA\models \vp[\eval]$ and 
$\cA\models \psi[\eval]$ \\[1mm]
&& (by Proposition~\ref{A-e-models}.\eqref{A-e-models-si})\\[1mm]
& iff &  (for every  $(\cA, \eval)$ such that  $\cA\models \Gamma[\eval]$ we have that  $\cA\models \vp[\eval]$) and \\
&& (for every  $(\cA, \eval)$ such that  $\cA\models \Gamma[\eval]$ we have that  $\cA\models \psi[\eval]$)\\[1mm]
& iff & $\Gamma  \lsc  \vp$ and $\Gamma \lsc \psi$. 
\et
\item Apply \eqref{lsc-lleq-Gamma-si}.
\ee
}

\bdfn\label{def-lsc-Gamma}
Let $\Delta$ be a set of patterns. We say that 
\be
\item $\Delta$ is a \defnterm{\lscn} of $\Gamma$ if for every $(\cA, \eval)$, 
\bce
$\cA\models\Gamma[\eval]$ \, implies \, $\cA \models\Delta[\eval]$.
\ece
Notation: $\Gamma \lsc \Delta$.
\item $\Gamma$ and $\Delta$ are \defnterm{\lleqn}  if for every $(\cA, \eval)$, 
\bce
$\cA\models\Gamma[\eval]$ \, iff \, $\cA \models\Delta[\eval]$.
\ece
Notation: $\Gamma \lleq \Delta$.
\ee
\edfn

Obviously, by letting $\Delta=\{\vp\}$, we obtain the notion introduced in  Definition~\ref{local-conseq-logic-equiv-Gamma-vp}. Furthermore, 
for  $\Gamma=\{\psi\}$ and $\Delta=\{\vp\}$, we get the notions introduced in 
Definition \ref{local-conseq-logic-equiv-vp-psi}.

\bfact\label{Gamma-lleq-two-lsc}
$\Gamma \lleq \Delta$ iff ($\Gamma \lsc \Delta$ and $\Delta \lsc \Gamma$).
\efact

\blem\label{Gamma-Delta-vp-psi-lleq}
\be
\item\label{Gamma-Delta-vp-psi-lleq-1} Assume that $\Gamma \lsc \Delta$ and that $\vp \lleq \psi$.
 Then  $\Delta  \lsc  \vp$ implies $\Gamma \lsc \psi$.
\item\label{Gamma-Delta-vp-psi-lleq-2} Assume that $\Gamma \lleq \Delta$ and that $\vp \lleq \psi$. 
Then  $\Gamma \lsc \vp$ iff $\Delta \lsc \psi$. 
\ee 
\elem
\solution{ 
\be
\item We have that 

\bt{lll}
$\Delta \lsc \vp$ & iff & for every  $(\cA, \eval)$ such that  $\cA\models \Delta[\eval]$, $\cA\models \vp[\eval]$ \\
& iff  & for every  $(\cA, \eval)$ such that  $\cA\models \Delta[\eval]$, $\cA\models \psi[\eval]$ (as $\vp \lleq \psi$)\\
& iff  & for every $(\cA, \eval)$ such that  $\cA\models \Gamma[\eval]$, $\cA\models \psi[\eval]$ (as $\Gamma \lsc \Delta$)\\
& iff & $\Gamma\lsc \psi$.
\et 

\item Apply \eqref{Gamma-Delta-vp-psi-lleq-1}.
\ee
}

\bprop[Finite sets]
Let $\Gamma=\{\vp_1, \ldots, \vp_n\}$ be a set of patterns. Then $\Gamma \lleq \{\vp_1\si\ldots \si \vp_n\}$. 
\eprop
\solution{Applying Proposition~\ref{A-e-models-finite-disj-conj}.\eqref{A-e-models-finite-disj-conj-A-e}, we get that for 
every $(\cA,\eval)$, $\cA\models\Gamma[\eval]$ iff $\cA\models\vp_i[\eval]$ for every $i=1,\ldots, n$ iff
$\cA\models (\vp_1\si\ldots \si \vp_n)[\eval]$.
}

\subsubsection{Notion 3}

\bdfn\label{def-ssc-Gamma}
Let $\vp$ be a pattern. We say that $\vp$ is a \defnterm{\sscn} of $\Gamma$ if for every $(\cA, \eval)$, 
\bce
$\bigcap_{\gamma \in \Gamma} \evalp(\gamma) \se \evalp(\vp)$.
\ece
Notation: $\Gamma \ssc \vp$.
\edfn

By letting $\Gamma=\{\psi\}$, we get the notion introduced in Definition \ref{strong-conseq-logic-equiv-vp-psi}.

\bfact\label{basic-ssc-gamma}
\be
\item\label{basic-ssc-gamma-empty} $\emptyset\ssc \vp$ iff $\models \vp$. 
\item\label{basic-ssc-gamma-valid} If $\vp$ is valid, then $\Gamma \ssc \vp$.
\ee
\efact
\solution{ 
\be
\item $\emptyset\ssc \vp$ iff (for every $(\cA,\eval)$, we have that $\bigcap_{\gamma \in \emptyset} \evalp(\gamma) \se \evalp(\vp)$)
iff  (for every $(\cA,\eval)$, we have that $A\se  \evalp(\vp)$) iff (for every $(\cA,\eval)$, we have that $\evalp(\vp)=A$) iff
$\models \vp$. 
\item Obviously, as $\evalp(\vp)=A$ for every $(\cA,\eval)$.
\ee
}

\bfact\label{ssc-sleqs-Gamma}
For all patterns  $\vp$, $\psi$, 
\be
\item\label{ssc-sleqs-Gamma-ssc}  If $\vp  \ssc  \psi$, then  $\Gamma  \ssc  \vp$ implies $\Gamma  \ssc  \psi$.
\item\label{ssc-sleqs-Gamma-sleqs}  If $\vp\sleqs \psi$, then $\Gamma \ssc \vp$ iff $\Gamma \ssc  \psi$.
\item\label{ssc-sleqs-Gamma-si} $\Gamma  \ssc  \vp\si \psi$ iff ($\Gamma  \ssc  \vp$ and $\Gamma  \ssc  \psi)$.
\item\label{ssc-sleqs-Gamma-dnd} $\Gamma  \ssc  \vp\dnd \psi$ iff ($\Gamma  \ssc  \vp\to \psi$ and $\Gamma  \ssc  \psi\to \vp)$.
\ee
\efact
\solution{
\be
\item Assume that $\vp  \ssc  \psi$ and that $\Gamma  \ssc  \vp$. Let $(\cA, \eval)$. 
Then $\bigcap_{\gamma \in \Gamma} \evalp(\gamma) \se \evalp(\vp)\se \evalp(\psi)$, hence $\Gamma  \ssc  \psi$.
\item It follows from (i) and Remark \ref{sleqs-two-ssc}.

\item $\,$ \\
\bt{lll}
$\Gamma \ssc \vp\si \psi$ & iff & for every  $(\cA, \eval)$,  $\bigcap_{\gamma \in \Gamma} \evalp(\gamma) \se \evalp(\vp\si\psi)$\\
& iff & for every  $(\cA, \eval)$,  $\bigcap_{\gamma \in \Gamma} \evalp(\gamma) \se \evalp(\vp)\cap \evalp(\psi)$\\
& iff & for every  $(\cA, \eval)$,  $\bigcap_{\gamma \in \Gamma} \evalp(\gamma) \se \evalp(\vp)$ and 
 $\bigcap_{\gamma \in \Gamma} \evalp(\gamma) \se \evalp(\psi)$\\
 & iff & for every  $(\cA, \eval)$,  $\bigcap_{\gamma \in \Gamma} \evalp(\gamma) \se \evalp(\vp)$ and \\
 && for every  $(\cA, \eval)$,  $\bigcap_{\gamma \in \Gamma} \evalp(\gamma) \se \evalp(\psi)$\\
 & iff & $\Gamma  \ssc  \vp$ and $\Gamma  \ssc  \psi$.
\et 

 \item Apply \eqref{ssc-sleqs-Gamma-si}.
\ee
}

\bdfn
Let $\Delta$ be a set of patterns. We say that 
\be
\item $\Delta$ is a \defnterm{\sscn} of $\Gamma$ if for every $(\cA, \eval)$, 
$$\bigcap_{\gamma \in \Gamma} \evalp(\gamma) \se \bigcap_{\delta\in \Delta}\evalp(\delta).$$
Notation: $\Gamma \ssc \Delta$.
\item $\Gamma$ and $\Delta$ are \defnterm{\sleqsn}  if for every $(\cA, \eval)$, 
\bce
$\bigcap_{\gamma \in \Gamma} \evalp(\gamma) = \bigcap_{\delta\in \Delta}\evalp(\delta)$.
\ece
Notation: $\Gamma \sleqs \Delta$.
\ee
\edfn

Obviously, by letting $\Delta=\{\vp\}$, we obtain the notion introduced in  Definition~\ref{def-ssc-Gamma}. Furthermore, 
for  $\Gamma=\{\psi\}$ and $\Delta=\{\vp\}$, we get the notions introduced in 
Definition \ref{strong-conseq-logic-equiv-vp-psi}.

\bfact\label{Gamma-sleqs-two-ssc}
$\Gamma \sleqs \Delta$ iff ($\Gamma \ssc \Delta$ and $\Delta \ssc \Gamma$).
\efact

\blem\label{Gamma-Delta-vp-psi-sleqs}
\be
\item\label{Gamma-Delta-vp-psi-sleqs-1} Assume that $\Gamma \ssc \Delta$ and that $\vp \sleqs \psi$.
Then  $\Delta  \ssc  \vp$ implies $\Gamma \ssc \psi$.
\item\label{Gamma-Delta-vp-psi-sleqs-2} Assume that $\Gamma \sleqs \Delta$ and that $\vp \sleqs \psi$. 
Then  $\Gamma \ssc \vp$ iff $\Delta \ssc \psi$. 
\ee 
\elem
\solution{ 
\be
\item We have that 

\bt{lll}
$\Delta \ssc \vp$ & iff & for every  $(\cA, \eval)$ $\bigcap_{\gamma \in \Delta} \evalp(\gamma) \se \evalp(\psi)$ \\
& iff  & for every  $(\cA, \eval)$ $\bigcap_{\gamma \in \Delta} \evalp(\gamma) \se \evalp(\psi)$  (as $\vp \sleqs \psi$)\\
& iff  & for every  $(\cA, \eval)$ $\bigcap_{\gamma \in \Gamma} \evalp(\gamma) \se \evalp(\psi)$  (as $\Gamma \ssc \Delta$)\\
& iff & $\Gamma\ssc \psi$.
\et 

\item Apply \eqref{Gamma-Delta-vp-psi-sleqs-1}.
\ee
}

\bprop
Let $\Gamma=\{\vp_1, \ldots, \vp_n\}$ be a finite set of patterns. Then $\Gamma \sleqs \{\vp_1\si\ldots \si \vp_n\}$. 
\eprop
\solution{For every $(\cA,\eval)$,
\begin{align*}
\bigcap_{\gamma \in \Gamma} \evalp(\gamma)=\bigcap_{i=1}^n\evalp(\vp_i)=\evalp(\vp_1\si\ldots \si \vp_n).
\end{align*}
}

\subsubsection{Relations between the three notions}

\bprop\label{Gamma-ssc-imp-lsc-imp-gsc}
$\Gamma \ssc \vp$ implies $\Gamma \lsc \vp$ implies $\Gamma\gsc \vp$.
\eprop
\solution{
$\Gamma \ssc \vp$ implies $\Gamma\lsc \vp$: 
Let $(\cA,\eval)$ be such that $\cA\models \Gamma[\eval]$, so $\evalp(\gamma)=A$ for every $\gamma\in\Gamma$. 
Hence, $\bigcap_{\gamma \in \Gamma} \evalp(\gamma)=A$. As $\Gamma \ssc \vp$,  it follows that $\evalp(\vp)=A$, that is
 $\cA\models \vp[\eval]$.

$\Gamma \lsc \vp$ implies $\Gamma \gsc \vp$:  Assume that $\cA\models \Gamma$ and let $\eval$ be arbitrary. Then 
$\cA\models \Gamma[\eval]$. Applying now the fact that $\Gamma \lsc \vp$, we get that $\cA\models \vp[\eval]$.

}

%% file: AML-semantics-closed-patterns.tex
\subsection{Closed  patterns}

\bdfn	A  pattern $\vp$  is said to be a \defnterm{closed  pattern} if $FV(\vp)=\emptyset$, 
that is $\vp$ does not have free variables. 
\edfn

Let $\cA$  be a $\AMLsig$-structure.

\bprop
Let $\vp$ be a closed  pattern.  For every $\cA$-valuations $\evalu$, $\evald$, we have that $\evalup(\vp)=\evaldp(\vp)$. In particular, 
$\cA\models \vp[e_1]$ iff $\cA\models \vp[e_2]$.
\eprop
\solution{Apply Proposition \ref{evaluation-patterns-free-variables} and the fact that $FV(\vp)=\emptyset$. 
}

\bcor\label{closed-eval-all}
Let $\vp$ be a closed  pattern.  Then either $\cA\models \vp[\eval]$ for all $\eval$ or $\cA\not\models \vp[\eval]$ for all $\eval$.
Hence, $\cA\models \vp$ iff  $\cA\models \vp[e]$ for one $\cA$-valuation $e$. 
\ecor

\bprop\label{closed-gsc-lsc}
Let $\Gamma$ be a set of closed  patterns. Then for every pattern $\vp$,  $\Gamma \lsc \vp$ iff $\Gamma \gsc \vp$.
\eprop
\solution{
 $\Ra$ By Proposition \ref{Gamma-ssc-imp-lsc-imp-gsc}.

$\La$ Let $(\cA, \eval)$ be  such that  $\cA\models \Gamma[\eval]$, that is $\cA\models \gamma[\eval]$ for all
$\gamma\in\Gamma$. By Corollary \ref{closed-eval-all}, we get that
$\cA\models \gamma$ for all $\gamma\in\Gamma$, that is $\cA\models \Gamma$. As $\Gamma \gsc \vp$, it follows that 
$\cA\models \vp$. In particular, $\cA\models \vp[\eval]$. 
}

%% file: AML-semantics-predicates.tex
\subsection{Predicates}

\bdfn
Let $\vp$ be a pattern. 
\be
\item If $\cA$ is a $\tau$-structure, we say that $\vp$ is an \defnterm{$\cA$-predicate} or a \defnterm{predicate in $A$} if for any 
$\cA$-valuation $\eval$, $\evalp(\vp)\in \{\emptyset, A\}$.
\item $\vp$ is a \defnterm{predicate} iff it is an $\cA$-predicate for every $\tau$-structure $\cA$.
\ee
\edfn
We shall denote by \defnterm{$\cA$-Predicates} the set of $\cA$-predicates and by \defnterm{Predicates} 
the set of  predicates.

\bprop\label{predicates-closed}
\be
\item\label{predicates-closed-1} For every  $\tau$-structure $\cA$, the set $\cA$-Predicates contains $\bot$, $\top$ and is 
closed to $\neg, \to, \sau, \si, \dnd$ and to $\exists x$, $\forall x$ (for any element variable $x$). Furthermore, the set $\cA$-Predicates 
is closed to finite conjunctions and disjunctions.
\item\label{predicates-closed-2} The set Predicates contains $\bot$, $\top$ and is closed to $\neg, \to, \sau, \si, \dnd$ and 
to $\exists x$, $\forall x$ (for any variable $x$). Furthermore, the set Predicates 
is closed to finite conjunctions and disjunctions.
\ee
\eprop
\solution{$\,$
\be
\item Let $\eval$ be an arbitrary $\cA$-valuation. 
\be
\item Since $\evalp(\bot)=\emptyset$ and $\evalp(\top)=A$, we have that 
$\bot, \top \in \cA$-Predicates. 
\item If $\vp$ is an $\cA$-predicate, then $\evalp(\vp)\in \{\emptyset, A\}$, hence  
$\evalp(\neg\vp)=A\setminus \evalp(\vp)\in \{\emptyset, A\}$, that is $\neg\vp$ is an $\cA$-predicate.
\item\label{pred-finite-conj-disj} Assume that $\vp_1, \ldots, \vp_n (n\geq 1)$ are $\cA$-predicates, so $\evalp(\vp_i)\in \{\emptyset, A\}$ for every $i=1, \ldots, n$.
Then 
\begin{align*}
\evalp(\vp_1\si \ldots \si \vp_n) & =\bigcap_{i=1}^n \evalp(\vp_i) = 
\begin{cases}
\emptyset & \text{if there exists~} i=1, \ldots, n \text{~with~} \evalp(\vp_i)=\emptyset\\
A & \text{otherwise}, \\
\end{cases}\\
\evalp(\vp_1 \sau \ldots \sau \vp_n) & =\bigcup_{i=1}^n \evalp(\vp_i) = 
\begin{cases}
A & \text{if there exists~} i=1, \ldots, n \text{~with~} \evalp(\vp_i)=A, \\
\emptyset & \text{otherwise}.
\end{cases}
\end{align*}
Thus, $\vp_1\si \ldots \si \vp_n$ and $\vp_1 \sau \ldots \sau \vp_n$ are $\cA$-predicates.
\item Assume that $\vp$, $\psi$ are $\cA$-predicates, so $\evalp(\vp), \evalp(\psi)\in \{\emptyset, A\}$. 
Then  $\vp \si \psi$ and $\vp \sau \psi$. 
Furthermore, 
\begin{align*}
\evalp(\vp \ra \psi) & = \begin{cases} \emptyset & \text{if~} \evalp(\vp)=A \text{~and~}\evalp(\psi)=\emptyset \\
A & \text{otherwise} 
\end{cases},\\
\evalp(\vp \dnd \psi) & = \begin{cases} A & \text{if~} \evalp(\vp)=\evalp(\psi) \\
 & \emptyset \text{otherwise}.
 \end{cases} 
\end{align*}
Thus, $\vp \ra \psi, \vp \dnd \psi$ are $\cA$-predicates.
\item Assume that  $\vp$ is an $\cA$-predicate. Then, for every $a\in A$, $\evalpxa(\vp)\in \{\emptyset, A\}$,
hence  $\evalp(\exists x.\vp)=\bigcup_{a\in A} \evalpxa(\vp)\in \{\emptyset, A\}$ and 
$\evalp(\forall x.\vp)=\bigcap_{a\in A} \evalpxa(\vp)\in \{\emptyset, A\}$. Thus, $\exists x.\vp$ and $\forall x.\vp$ are
 $\cA$-predicates.
\ee
\item Obviously, by \eqref{predicates-closed-1}. 
\ee
}

\bprop\label{patterns-gsc-lsc-ssc}
Let $\Gamma$ be a set of closed predicates (that is, closed patterns that are predicates). 
Then  for every pattern $\vp$, $\Gamma \gsc \vp$ iff $\Gamma \lsc \vp$ iff $\Gamma \ssc \vp$.
\eprop
\solution{
$\Gamma \gsc \vp$ iff $\Gamma \lsc \vp$: By Proposition~\ref{closed-gsc-lsc}.

$\Gamma \lsc \vp$ iff $\Gamma \ssc \vp$:  $\La$ By Proposition~\ref{Gamma-ssc-imp-lsc-imp-gsc}.

$\Ra$ Let $(\cA,\eval)$. We have the following cases:
\be
\item There exists $\gamma_0\in \Gamma$ such that $\evalp(\gamma_0)\ne A$. Then $\evalp(\gamma_0)=\emptyset$, so 
$\bigcap_{\gamma \in \Gamma} \evalp(\gamma)=\emptyset\se \evalp(\vp)$. Thus, $\Gamma \ssc \vp$.
\item $\evalp(\gamma)= A$ for all $\gamma\in \Gamma$, hence $\cA\models \Gamma[\eval]$. Since $\Gamma \lsc \vp$, we get that 
$\cA \models \vp[\eval]$, that is $\evalp(\vp)=A$. We have got that $\bigcap_{\gamma \in \Gamma} \evalp(\gamma)=A=\evalp(\vp)$. 
Thus, $\Gamma \ssc \vp$.
\ee
}

\bprop\label{predicate-imp-rules-lsc-gsc-Gamma}
Let $\Gamma$ be a set of patterns, $\sigma$ be a predicate pattern and  $\vp$, $\psi$ be patterns such 
that 
\bce
$\vp \lsc \psi$.
\ece
Then 
\be
\item\label{predicate-imp-rules-lsc-gsc-1} $\sigma \to \vp \lsc \sigma \to \psi$;
\item\label{predicate-imp-rules-lsc-gsc-2} $\sigma \to \vp \gsc \sigma \to \psi$;
\item\label{predicate-imp-rules-lsc-gsc-1-Gamma} $\Gamma \lsc \sigma \to \vp$ implies 
$\Gamma \lsc \sigma \to \psi$;
\item\label{predicate-imp-rules-lsc-gsc-2-Gamma} $\Gamma \gsc \sigma \to \vp$ implies 
$\Gamma \gsc \sigma \to \psi$. 
\ee
\eprop
\solution{
\be
\item  Let  $(\cA, \eval)$ be such that 
 $\cA \models (\sigma \to \vp)[\eval]$, hence 
$$(*)\qquad \evalp(\sigma)\se \evalp(\vp).$$
 We have to prove that  $\cA \models (\sigma \to \psi)[\eval]$, hence that 
$$(**) \qquad \evalp(\sigma)\se \evalp(\psi).$$
We have two cases:
\be
\item $\evalp(\sigma)=\emptyset$. Then obviously, $(**)$ holds. 
\item $\evalp(\sigma)=A$. Then, by $(*)$, $\evalp(\vp)=A$, so $\cA \models \vp[\eval]$. 
As $\vp \lsc \psi$, we must have that $\cA \models \psi[\eval]$, that is $\evalp(\psi)=A$. Thus, 
$(**)$ holds.
\ee
\item Apply \eqref{predicate-imp-rules-lsc-gsc-1} and the fact that $\sigma \to \vp \lsc \sigma \to \psi$
implies $\sigma \to \vp \gsc \sigma \to \psi$, 
by Proposition \ref{remark-conseq-equiv-to-dnd}.\eqref{remark-conseq-equiv-to-imp-dnd}.
\item Apply \eqref{predicate-imp-rules-lsc-gsc-1} and Lemma \ref{lsc-lleq-Gamma}.\eqref{lsc-lleq-Gamma-lsc}.
\item Apply \eqref{predicate-imp-rules-lsc-gsc-2} and Lemma \ref{gsc-gleq-Gamma}.\eqref{gsc-gleq-Gamma-gsc}.
\ee
}

%% file: AML-semantics-tautologies.tex
\subsection{Tautologies}

Let  $\cA$ be a $\AMLsig$-structure.

\bdfn
A \propeval~ is a mapping  $\evt: \setpat\to 2^A$ 
that satisfies for any patterns  $\vp,\psi$, 
\be
\item $\evt(\bot)=\emptyset$.
\item $\evt(\vp\to\psi)=C_A(\evt(\vp) \setminus \evt(\psi))$.
\ee
\edfn

\blem\label{evt-basic-1}
Let $\evt$ be a \propeval. For any patterns  $\vp$,$\psi$,
\be
\item $\evt(\vp \to \psi)= C_A\evt(\vp)\cup  \evt(\psi)$.
\item $\evt(\top)=A$. 
\item $\evt(\neg\vp) = C_A\evt(\vp)$.
\item $\evt(\vp \sau \psi)= \evt(\vp) \cup \evt(\psi)$.
\item $\evt(\vp \si \psi)= \evt(\vp) \cap \evt(\psi)$.
\item $\evt(\vp \dnd \psi)=C_A(\evt(\vp) \diffsym \evt(\psi))$.
\ee
\elem
\solution{Replace $\evalp$ with $\evt$ in the proof of Proposition \ref{basic-val}.
}

\blem\label{evt-basic-2}
Let $\evt$ be a \propeval. For any patterns  $\vp$,$\psi$,
\be
\item $\evt(\vp\to\psi)=A$ iff $\evt(\vp) \se \evt(\psi)$.
\item $\evt(\vp\dnd\psi)=A$ iff $\evt(\vp) = \evt(\psi)$.
\ee
\elem
\solution{Replace $\evalp$ with $\evt$ in the proof of Proposition~\ref{A-e-models}.\eqref{A-e-models-to}, \eqref{A-e-models-dnd}.
} 

\bprop\label{main-propeval}
Let $\eval$ be an $\cA$-valuation. The mapping 
$$V_{\eval,\cA}: \setpat \to 2^A, \quad V_{e,\cA}(\vp)= \evalp(\vp)
$$
is a \propeval. 
\eprop
\solution{We have that 
\be
\item $V_{e,\cA}(\bot)=\evalp(\bot)=\emptyset$ 
\item $V_{e,\cA}(\vp\to \psi)= \evalp(\vp\to \psi)=C_A(\evalp(\vp) \setminus\evalp(\psi))=
C_A(V_{e,\cA}(\vp) \setminus V_{e,\cA}(\psi)).$
\ee
}

\bdfn\label{tautology-def}
A pattern  $\vp$ is a \defnterm{tautology} if $\evt(\vp)=A$  for any $\AMLsig$-structure $\cA$ and 
any \propeval~$\evt$. \\
Notation: $\tautsign \vp$.
\edfn

\bprop\label{tautology-valid}
Any tautology is valid.
\eprop
\solution{
Assume that $\vp$ is a tautology and  let $(\cA, \eval)$.  Since, by Proposition~\ref{main-propeval}, $V_{e,\cA}$ is a \propeval, we have that 
$V_{e,\cA}(\vp)=A$, that is $\evalp(\vp)=A$, so $\cA\models \vp[e]$.
}

%% file: AML-semantics-tautologies-examples.tex
\subsubsection{Examples of tautologies and tautological equivalences}

\bprop
The following are tautologies, hence valid, for all patterns $\vp$, $\psi$, $\chi$, 
\begin{align}
& \vp \dnd \vp, \label{dnd-vp-vp}\\
& \vp\vee\vp  \dnd \vp,  \label{dnd-contraction-sau}\\
& \vp  \dnd \vp\wedge\vp, \label{dnd-contraction-si}\\
& \vp\vee\psi  \dnd \psi\vee\vp, \label{dnd-permutation-sau}\\
& \vp\wedge\psi \dnd \psi\wedge\vp, \label{dnd-permutation-si}\\
& \vp\to \vp  \vee\psi, \label{weakening-sau}\\
& \vp\wedge\psi \to \vp, \label{weakening-si}\\
& \bot\to \vp, \label{ex-falso}\\
& \vp\sau\neg\vp, \label{G-lem}\\
& (\vp \to (\psi \to \chi))\dnd (\vp \si \psi \to \chi),\label{exp+imp}\\
& (\vp \to (\psi \to \chi))\dnd (\psi\to (\vp\to \chi)), \label{vp-to-psi-chi=vp-psi-to-chi}
\end{align}
\eprop
\solution{
Let $\cA$ be a $\AMLsig$-structure and $\evt$ be an \propeval.   

\eqref{dnd-vp-vp}: We have that  $\evalp(\vp)=\evalp(\vp)$, hence, 
by Lemma \ref{evt-basic-2}, $F(\vp\dnd \vp)=A$.

\eqref{dnd-contraction-sau}: We have that $\evt(\vp\vee\vp) =\evt(\vp)\cup\evt(\vp)=\evt(\vp)$,
hence, by Lemma \ref{evt-basic-2}, $F(\vp\vee\vp\dnd \vp)=A$.

\eqref{dnd-contraction-si}: We have that $\evt(\vp\wedge\vp) =\evt(\vp)\cap\evt(\vp)=\evt(\vp)$,
hence, by Lemma \ref{evt-basic-2}, $F(\vp \dnd \vp\wedge\vp)=A$.

\eqref{dnd-permutation-sau}: We have that $\evt(\vp\vee\psi)  =\evt(\vp)\cup\evt(\psi) = \evt(\psi)\cup\evt(\vp)=
\evt(\psi\vee\vp)$, hence, by Lemma \ref{evt-basic-2}, $F(\vp\vee\psi \dnd \psi\vee\vp)=A$.

\eqref{dnd-permutation-si}: We have that $\evt(\vp\wedge\psi) =\evt(\vp)\cap\evt(\psi) = \evt(\psi)\cap\evt(\vp)=
\evt(\psi\wedge\vp)$, hence, by Lemma \ref{evt-basic-2}, $F(\vp\wedge\psi \dnd \psi\wedge\vp)=A$.

\eqref{weakening-sau}: We have that $\evt(\vp)  \se \evt(\vp)\cup\evt(\psi)=\evt(\vp\vee\psi)$, 
hence,
by Lemma \ref{evt-basic-2}, $F(\vp\to \vp\vee\psi)=A$.

\eqref{weakening-si}: We have that $\evt(\vp\wedge\psi) =\evt(\vp)\cap \evt(\psi) \se \evt(\vp)$,
hence, by Lemma \ref{evt-basic-2}, $F(\vp\wedge\psi \to \vp)=A$.

\eqref{ex-falso}: We have that $\evt(\bot\to \vp)=C_A(\evt(\bot)\setminus \evt(\vp))=
C_A (\emptyset \setminus \evt(\vp))=C_A\emptyset =A$.

\eqref{G-lem}: We have that 
$\evt(\vp\sau\neg\vp) = \evt(\vp)\cup \evt(\neg\vp)= \evt(\vp) \cup C_A \evt(\vp)=A$.

\eqref{exp+imp}: We have that 
\begin{align*}
\evt(\vp \to (\psi \to \chi)) & =C_A\evt(\vp)\cup \evt(\psi \to \chi) =
C_A\evt(\vp)\cup (C_A\evt(\psi)\cup \evt(\chi))\\
& = (C_A\evt(\vp)\cup  C_A\evt(\psi))\cup \evt(\chi) \stackrel{\eqref{de-morgan-2}}{=} C_A(\evt(\vp)\cap \evt(\psi))\cup \evt(\chi)\\
& = C_A \evt(\vp\si\psi)\cup \evt(\chi)=\evt(\vp \si \psi \to \chi),
\end{align*}
hence, by Lemma \ref{evt-basic-2}, $\evt((\vp \to (\psi \to \chi))\dnd (\vp \si \psi \to \chi))=A$.

\eqref{vp-to-psi-chi=vp-psi-to-chi}: We have that 
\begin{align*}
\evt(\vp \to (\psi \to \chi)) &  \stackrel{\eqref{exp+imp}}{=}\evt(\vp \si \psi \to \chi) = C_A\evt(\vp\si\psi)\cup \evt(\chi) \\
& = C_A\evt(\psi\si\vp)\cup \evt(\chi) \quad \text{as } \evt(\vp\si\psi)= \evt(\psi\si\vp) \text{ by \eqref{dnd-permutation-si}}\\
& = \evt(\psi \si \vp \to \chi) \stackrel{\eqref{exp+imp}}{=}\evt(\psi\to (\vp\to \chi)). 
\end{align*}
}

%% file: AML-semantics-change-val.tex
\subsection{Change of $\cA$-valuations}

\bntn 
For any distinct element variables $x,z$ and any $a,b\in A$, we define  a new $\cA$-valuation 
$\evalxzd{a}{b}$ as follows: for all $v\in EVar$ and all $V\in SVar$,
\begin{align}
\evalxzd{a}{b}(V) = \eval(V),  & \qquad \evalxzd{a}{b}(v) =\begin{cases}
\eval(v) & \text{ if } v\neq x, z\\
a 	& 		\text{~if~}  v = x\\
b 	& 		\text{~if~}  v = z
\end{cases}
\end{align}
It is obvious that $\evalxzd{a}{b}=\evalxz{a}{b}=\evalzx{b}{a}$.
\entn

\bntn 
For any distinct set variables $X,Z$ and any $B,C\se A$, we define  a new $\cA$-valuation 
$\evalXZd{B}{C}$ as follows: for all $v\in EVar$ and all $V\in SVar$, 
\begin{align}
\evalXZd{B}{C}(v)=\eval(v), & \qquad 
\evalXZd{B}{C}(V) =\begin{cases}
\eval(V) & 			\text{~if~} V\neq X, Z\\
B & 			\text{~if~} V=X\\
C & 			\text{~if~} V=Z
\end{cases}
\end{align}
It is obvious that $\evalXZd{B}{C}=\evalXZ{B}{C}=\evalZX{C}{B}$.
\entn

\bntn 
For any element variable $x$ and set variable $X$ and any $a\in A, B\se A$, we define  a new $\cA$-valuation 
$\evalxXd{a}{B}$ as follows: for all $v\in EVar$ and all $V\in SVar$,
\begin{align}
\evalxXd{a}{B}(v) = \begin{cases}
\eval(v) & 	\text{~if~} v\neq x\\
a & 		\text{~if~} v = x
\end{cases}
 & \qquad 
\evalxXd{a}{B}(V) =\begin{cases}
\eval(V) & 	\text{~if~} V\neq X\\
B & 		\text{~if~} V = X
\end{cases}
\end{align}
It is obvious that $\evalxXd{a}{b}=\evalxX{a}{B}=\evalXx{B}{a}$.
\entn

%% file: AML-semantics-substitution.tex
\subsection{Substitution}

\subsubsection{Element variables}

\bprop\label{eval-subfvpxdeltagma}
Let $\vp$, $\delta$ be  patterns  and $x$ be an element variable such that $x$ is free for $\delta$ in $\vp$. Assume that 
\bce
 (*) \quad $\cA$ is a $\tau$-structure  such that for every $\cA$-valuation $\eval$, there exists $\aesing\in A$ such that $\evalp(\delta)=\{\aesing\}$. 
\ece
Then for  every $\cA$-valuation $\eval$,
$$\evalp(\subfvpxdelta)=\evalpx{\aesing}(\vp).$$
\eprop
\solution{If $x\notin FV(\vp)$, then $\subfvpxdelta=\vp$ and $\eval=\evalx{\aesing}$, by Proposition~\ref{evaluation-patterns-free-variables}. 
The conclusion is obvious. 

We prove by induction on patterns that for all patterns $\vp$ such that $x\in FV(\vp)$, the following holds:
\bce 
for  every $\cA$-valuation $\eval$, $\evalp(\subfvpxdelta)=\evalpx{\aesing}(\vp)$.
\ece
We use the definition by recursion on paterns of $\subfvpxdelta$, given by Remark~\ref{def-subf-recursion}.
\be
\item $\vp$ is an atomic pattern. Since  $x\in FV(\vp)$, we have that $\vp=x\in EVar$. Then $\subfvpxdelta=\delta$.
For every $\cA$-valuation $\eval$,  we get that $\evalp(\subfvpxdelta)=\evalp(\delta)=\{\aesing\}$. Furthermore,
 $\evalpx{\aesing}(\vp)=\evalpx{\aesing}(x)=\{\aesing\}$.
 
\item  $\vp=\appln{\psi}{\chi}$. Then $x\in FV(\psi)$, $x\in FV(\chi)$ and $x$ is free for 
$\delta$ in $\psi$, $\chi$. Let $\eval$ be an $\cA$-valuation. We can apply the induction hypothesis for $\psi$ and $\chi$ to get that 
$\evalp(\subfx{\delta}{\psi})=\evalpx{\aesing}(\psi)$ and $\evalp(\subfx{\delta}{\chi})=\evalpx{\aesing}(\chi)$. It follows that 
\begin{align*}
\evalp(\subfvpxdelta) & = \evalp(\appln{\subfx{\delta}{\psi}}{\subfx{\delta}{\chi}})
= \evalp(\subfx{\delta}{\psi}) \appls \evalp(\subfx{\delta}{\chi}) \\
& = \evalpx{\aesing}(\psi) \appls \evalpx{\aesing}(\chi) =\evalpx{\aesing}(\appln{\psi}{\chi})=\evalpx{\aesing}(\vp). 
\end{align*}

\item  $\vp=\psi\to\chi$. Then $x\in FV(\psi)$, $x\in FV(\chi)$ and $x$ is free for 
$\delta$ in $\psi$, $\chi$. Let $\eval$ be an $\cA$-valuation. We can apply the induction hypothesis for $\psi$ and $\chi$ to get that 
$\evalp(\subfx{\delta}{\psi})=\evalpx{\aesing}(\psi)$ and $\evalp(\subfx{\delta}{\chi})=\evalpx{\aesing}(\chi)$. It follows that
\begin{align*}
\evalp(\subfvpxdelta) &= \evalp(\subfxdeltaf{\psi}\to \subfxdeltaf{\chi})
=C_A\left(\evalp(\subfxdeltaf{\psi}) \setminus \evalp(\subfxdeltaf{\chi})\right) \\
& = C_A\left(\evalpx{\aesing}(\psi)\setminus \evalpx{\aesing}(\chi)\right)=\evalpx{\aesing}(\psi\to\chi)\\
&=\evalpx{\aesing}(\vp).
\end{align*}

\item $\vp=\exists z.\psi $. Since  $x\in FV(\vp)$, we have that $x\ne z$.
 Then $x\in FV(\psi)$ and $x$ is free for $\delta$ in $\psi$. Let $\eval$ be an $\cA$-valuation. We get that
\begin{align*}
\evalp(\subfvpxdelta) &= \evalp(\exists z.\subfxdeltaf{\psi})=\bigcup_{a\in A} \evalpza(\subfxdeltaf{\psi})\\
&= \evalpzx{a}{\aesingza}(\psi) \text{ by the induction hypothesis for } \psi \text{ applied to } \evalza.
\end{align*}
As $x$ is free for $\delta$ in $\vp$ and $x\in FV(\psi)$, it follows that 
$z\notin FV(\delta)$. Hence, 
by Proposition \ref{evaluation-patterns-free-variables}, $\evalp(\delta)=\evalpza(\delta)$, so 
$\aesing=\aesingza$. It follows that 
\begin{align*}
\evalp(\subfvpxdelta) &=  \evalpzx{a}{\aesing}(\psi) =\bigcup_{a\in A} \evalpzax(\psi)
=\bigcup_{a\in A} \evalpxz{\aesing}{a}(\psi)\\
&= \evalpx{\aesing}(\vp), 
\end{align*}
since $x\ne z$, hence $\evalxz{\aesing}{a}=\evalzx{\aesing}{a}=\eval_{x\mapsto \aesing, z\mapsto a}$.

\item $\vp=\mu X.\psi$. Then $x\in FV(\psi)$ and $x$ is free for $\delta$ in $\psi$. Let $\eval$ be an $\cA$-valuation.
\begin{align*}
\evalp(\subfvpxdelta) & = \evalp(\mu X. \subfx{\delta}{\psi})=\bigcap \left\{ B \subseteq A \mid \evalpXB(\subfx{\delta}{\psi}) 
 \subseteq B \right\}\\
 &=  \bigcap \left\{ B \subseteq A \mid  \evalpXx{B}{\aesingXB}(\psi)\se B \right\} \\
 & \text{ by the induction hypothesis for } \psi \text{ applied to } \evalXB.
\end{align*}
As $x$ is free for $\delta$ in $\vp$ and $x\in FV(\psi)$, it follows that 
$X\notin FV(\delta)$. Hence, 
by Proposition \ref{evaluation-patterns-free-variables}, $\evalp(\delta)=\evalpXB(\delta)$, so 
$\aesing=\aesingXB$. Thus, $\evalpXx{B}{\aesingXB}(\psi)=\evalpXx{B}{\aesing}(\psi)$

It follows that 
\begin{align*}
\evalp(\subfvpxdelta)  &=  \bigcap \left\{ B \subseteq A \mid  \evalpXx{B}{\aesing}(\psi)\se B \right\}\\
&= 
 \bigcap \left\{ B \subseteq A \mid  \evalpxX{\aesing}{B}(\psi)\se B \right\}\\
&=  \evalpx{\aesing}(\vp).
\end{align*}
\ee
}

As an immediate consequence of Proposition \ref{eval-subfvpxdeltagma}, we get the following.

\bcor\label{eval-subfvpxpy}
Let $\vp$ be a pattern  and $x,y$ be variables such that $x$ is free for $y$ in $\vp$. 
Then for every $\tau$-structure $\cA$ and every $\cA$-valuation $\eval$,
$$\evalp(\subfvpxy)=\evalpx{\eval(y)}(\vp).$$
\ecor

\bprop\label{exists-forall-substitution-subfvpxy}
Let $\vp$ be a pattern  and $x$, $y$ be variables such that $x$ is free for $y$ in $\vp$.  Then
\begin{align}
\models \subfvpxy \to \exists x. \varphi, \label{exists-axiom-subfvpxy}\\
\models \forall x.\varphi \to \subfvpxy. \label{forall-axiom-subfvpxy} 
\end{align}
\eprop
\solution{Let  $(\cA, \eval)$. Apply  Corollary \ref{eval-subfvpxpy} to get that 
$\evalp(\subfvpxy)=\evalpx{\eval(y)}(\vp)$. It follows that 

\eqref{exists-axiom}:  $\evalp(\fsubvpxy) = \evalpx{\eval(y)}(\vp) \se \bigcup_{a\in A} \evalpxa(\vp)=\evalp(\exists x. \varphi)$.

\eqref{forall-axiom}: $\evalp(\fsubvpxy) = \evalpx{\eval(y)}(\vp)\supseteq
\bigcap_{a\in A} \evalpxa(\vp)=\evalp(\forall x. \varphi)$.
}

\subsubsection*{Bounded substitution}

\bprop\label{bounded-subst-equiv-x}
Let $\vp$ be a pattern  and $x,y$ variables such that $y$ does not occur in $\vp$. Then
$$\models \vp \dnd \subbvpxy.$$
\eprop
\solution{If $x=y$, then obviously $\subbvpxy=\vp$, hence $\models \vp \dnd \subbvpxy$. Assume that $x\ne y$. 

Let $\cA$ be a $\tau$-structure. We prove by induction on $\vp$ that 
\bce
for every $\cA$-valuation $\eval$, $\evalp(\vp)=\evalp(\subbvpxy)$.
\ece
We use the definition by recursion on paterns of $\subbvpxy$, given by Remark~\ref{subb-inductive-x}. 

\be
\item $\vp$ is an atomic pattern. Then $\subbvpxy = \vp$, so $\evalp(\vp)=\evalp(\subbvpxy)$ for every 
$\cA$-valuation $\eval$. 
\item  $\vp=\appln{\psi}{\chi}$. Let $\eval$ be an $\cA$-valuation. By the induction hypothesis for $\psi$, $\chi$, we have  
that $\evalp(\psi)=\evalp(\subbf{x}{y}{\psi})$ and  $\evalp(\chi)=\evalp(\subbf{x}{y}{\chi})$. It follows that 
\begin{align*}
\evalp(\subbvpxy) &= \evalp(\appln{\subb{x}{y}(\psi)}{\subb{x}{y}(\chi)})=
\evalp(\subbf{x}{y}{\psi}) \appls \evalp(\subbf{x}{y}{\chi})=
\evalp(\psi) \appls \evalp(\chi)\\
& = \evalp(\appln{\psi}{\chi}) =\evalp(\vp).
\end{align*}
\item  $\vp=\psi\to\chi$. Let $\eval$ be an $\cA$-valuation. By the induction hypothesis for $\psi$, $\chi$, we have  
that $\evalp(\psi)=\evalp(\subbf{x}{y}{\psi})$ and  $\evalp(\chi)=\evalp(\subbf{x}{y}{\chi})$. It follows that 
\begin{align*}
\evalp(\subbvpxy) &= \evalp(\subbf{x}{y}{\psi}\to \subbf{x}{y}{\chi})=C_A\evalp(\subbf{x}{y}{\psi}) \cup \evalp(\subbf{x}{y}{\chi})=
C_A\evalp(\psi) \cup \evalp(\chi)\\
&=\evalp(\psi\to\chi)=\evalp(\vp).
\end{align*}
\item $\vp=\exists z.\psi $. We have two cases:

1. $x \ne z$. Then 
\begin{align*}
\evalp(\subbvpxy) &=\evalp(\exists z.\subbf{x}{y}{\psi})=\bigcup_{a\in A} \evalpza(\subbf{x}{y}{\psi})\\
&= \bigcup_{a\in A} \evalpza(\psi) \text{ by the induction hypothesis for } \psi \text{ applied to } \evalza \\
&=\evalp(\vp).
\end{align*}

2. $x=z$. Then 
\begin{align*}\evalp(\subbvpxy) &=\evalp\left(\exists y.\subfx{y}{\left(\subbf{x}{y}{\psi}\right)}\right)=
\bigcup_{a\in A} \evalpya\left(\subfx{y}{\left(\subbf{x}{y}{\psi}\right)}\right).
\end{align*}
As $x\ne y$ and $y$ does not occur in $\psi$,  we have that $x$ is free for 
$y$ in $\subbf{x}{y}{\psi}$, by Proposition \ref{subb-useful-1}. We can apply then Corollary \ref{eval-subfvpxpy} to get that

\begin{align*}
\evalpya\left(\subfx{y}{\left(\subbf{x}{y}{\psi}\right)}\right) & =\evalpyx{a}{\evalya(y)}\left(\subbf{x}{y}{\psi}\right)=
\evalpyx{a}{a}\left(\subbf{x}{y}{\psi}\right)\\
& = \evalpxy{a}{a}\left(\subbf{x}{y}{\psi}\right) = \evalpxa\left(\subbf{x}{y}{\psi}\right),
\end{align*}
since $y$ does not occur in $\psi$, so $y\notin FV\left(\subbf{x}{y}{\psi}\right)$, hence we can apply 
Proposition \ref{evaluation-patterns-free-variables}.
It follows that 
\begin{align*}
\evalp(\subbvpxy) &=\bigcup_{a\in A}\evalpxa\left(\subbf{x}{y}{\psi}\right)\\
&= \bigcup_{a\in A} \evalpxa(\psi) \text{ by the induction hypothesis for } \psi \text{ applied to } \evalxa \\
&=\evalp(\vp).
\end{align*}

\item $\vp=\mu X.\psi$. Then 
\begin{align*}
\evalp(\subbvpxy) & =\evalp(\mu X. \subbf{x}{y}{\psi}) = \bigcap \left\{B \subseteq A \mid \evalpXB(\subbf{x}{y}{\psi}) 
 \subseteq B \right\}\\
 &= \bigcap \left\{B \subseteq A \mid \evalpXB(\psi)  \subseteq B \right\} \\
 &\text{ by the induction hypothesis for } \psi \text{ applied to } \evalXB \\
 &=  \evalp(\vp).
\end{align*}
\ee
}

\subsubsection*{Free substitution}

\bprop
Let $\vp$ be a pattern  and $x$,$y$ be variables.  Then
\begin{align}
\models \fsubvpxy \to \exists x. \vp, \label{exists-axiom}\\
\models \forall x.\vp \to \fsubvpxy. \label{forall-axiom} 
\end{align}
\eprop
\solution{We have the following two cases:
\be
\item $x$ is free for $y$ in $\vp$. Then $\fsubvpxy=\subfvpxy$, so we apply Corollary~\ref{exists-forall-substitution-subfvpxy} 
to get the conclusion.
\item $x$ is not free for $y$ in $\vp$.  Then $\fsubvpxy=\subfx{y}{\theta}$, where $\theta=\subbf{y}{z}{\vp}$, 
with  $z$ a new variable, $z\notin Var(\vp)\cup \{y\}$. Since $x$ is free for $y$ in $\theta$, we can apply 
Corollary~\ref{exists-forall-substitution-subfvpxy} (with $\vp=\theta$)
to get that 
\begin{equation}\label{fs-1}
\models \fsubvpxy \to \exists x. \theta \text{~and~} \models \forall x.\theta \to \fsubvpxy.
\end{equation}
Furthermore, by Proposition \ref{bounded-subst-equiv-x}, we have that 
\begin{equation}\label{fs-2}
\models \theta \dnd\vp.
\end{equation}
Let $(\cA, \eval)$. Then 
\begin{align*} 
\evalp(\fsubvpxy) \stackrel{\eqref{fs-1}}{\se} \evalp(\exists x.\theta)=\bigcup_{a\in A} \evalpxa(\theta) \stackrel{\eqref{fs-2}}{=} 
\bigcup_{a\in A} \evalpxa(\vp)=\evalp(\exists x. \vp),\\
\evalp(\fsubvpxy) \stackrel{\eqref{fs-1}}{\supseteq} \evalp(\forall x.\theta)=\bigcap_{a\in A} \evalpxa(\theta) \stackrel{\eqref{fs-2}}{=} 
\bigcap_{a\in A} \evalpxa(\vp)=\evalp(\forall x.\vp),
\end{align*}
Hence, $\cA\models (\fsubvpxy \to \exists x. \vp)[\eval]$ and $\cA\models (\forall x.\vp \to \fsubvpxy)[\eval]$. 
\ee
}

%% file: AML-semantics-substitution-svar.tex
\subsubsection{Set variables}

\bprop\label{eval-subfvpXdelta}
Let $\vp$, $\delta$ be  patterns  and $X$ be a set variable such that $X$ is free for $\delta$ in $\vp$.
Then for every $\tau$-structure $\cA$ and every $\cA$-valuation $\eval$,
$$\evalp(\subfvpXdelta)=\evalpX{\evalp(\delta)}(\vp).$$
\eprop
\solution{If $X\notin FV(\vp)$, then $\subfvpXdelta=\vp$ and $\eval=\evalX{\evalp(\delta)}$, so the 
conclusion is obvious.

Let $\cA$ be a $\tau$-structure. 
We prove by induction on patterns that for all patterns $\vp$ such that $X\in FV(\vp)$, the following holds:
\bce 
for  every $\cA$-valuation $\eval$, $\evalp(\subfvpXdelta)=\evalpX{\evalp(\delta)}(\vp)$.
\ece
We use the definition by recursion on paterns of $\subfvpXdelta$, given by Remark~\ref{def-recursion-subfvpX}.

Assume in the sequel that $X\in FV(\vp)$. The proof is by induction on $\vp$. 

\be
\item $\vp$ is an atomic pattern. Since  $X\in FV(\vp)$, we have that $\vp=X$. Then $\subfvpXdelta=\delta$. 
For every $\cA$-valuation $\eval$,  we get that $\evalp(\subfvpXdelta)=\evalp(\delta)$. Furthermore, 
 $\evalpX{\evalp(\delta)}(\vp)=\evalpX{\evalp(\delta)}(X)=\evalp(\delta)$.
\item  $\vp=\appln{\psi}{\chi}$. Then $X\in FV(\psi)$, $X\in FV(\chi)$ and $X$ is free for 
$\delta$ in $\psi$, $\chi$. Let $\eval$ be an $\cA$-valuation. We can apply the induction hypothesis for $\psi$ and $\chi$ to get that 
$\evalp(\subfX{\delta}{\psi})=\evalpX{\evalp(\delta)}(\psi)$ and $\evalp(\subfX{\delta}{\chi})=\evalpX{\evalp(\delta)}(\chi)$. It follows that 
\begin{align*}
\evalp(\subfvpXdelta) & = \evalp(\appln{\subfX{\delta}{\psi}}{\subfX{\delta}{\chi}})
= \evalp(\subfX{\delta}{\psi}) \appls \evalp(\subfX{\delta}{\chi}) \\
& = \evalpX{\evalp(\delta)}(\psi) \appls \evalpX{\evalp(\delta)}(\chi) =\evalpX{\evalp(\delta)}(\appln{\psi}{\chi})\\
& =\evalpX{\evalp(\delta)}(\vp). 
\end{align*}

\item  $\vp=\psi\to\chi$. Then $X\in FV(\psi)$, $X\in FV(\chi)$ and $X$ is free for 
$\delta$ in $\psi$, $\chi$. Let $\eval$ be an $\cA$-valuation. We can apply the induction hypothesis for $\psi$ and $\chi$ to get that 
$\evalp(\subfX{\delta}{\psi})=\evalpX{\evalp(\delta)}(\psi)$ and $\evalp(\subfX{\delta}{\chi})=\evalpX{\evalp(\delta)}(\chi)$. It follows that
\begin{align*}
\evalp(\subfvpXdelta) &= \evalp(\subfXdeltaf{\psi}\to \subfXdeltaf{\chi})
=C_A\evalp\subfXdeltaf{\psi} \cup  \evalp(\subfXdeltaf{\chi}) \\
& = C_A\evalpX{\evalp(\delta)}(\psi)\cup \evalpX{\evalp(\delta)}(\chi)=\evalpX{\evalp(\delta)}(\psi\to\chi)\\
&=\evalpX{\evalp(\delta)}(\vp).
\end{align*}

\item $\vp=\exists x.\psi $. Then $X\in FV(\psi)$ and $X$ is free for $\delta$ in $\psi$. Let $\eval$ be an $\cA$-valuation.
\begin{align*}
\evalp(\subfvpXdelta) & =\evalp(\exists x. \subfXdelta{\psi}) = \bigcup_{a\in A} \evalpxa(\subfXdelta{\psi})\\
& = \bigcup_{a\in A} \evalpxX{a}{\evalpxa(\delta)}(\psi)\\
&  \text{ by the induction hypothesis for } \psi \text{ applied to }  \evalxa.
\end{align*}
As $X$ is free for $\delta$ in $\vp$ and $X\in FV(\psi)$, it follows that 
$x\notin FV(\delta)$. Hence, 
by Proposition \ref{evaluation-patterns-free-variables}, $\evalp(\delta)=\evalpxa(\delta)$.

It follows that 
\begin{align*}
\evalp(\subfvpXdelta) & = \bigcup_{a\in A} \evalpxX{a}{\evalp(\delta)}(\psi)=
\bigcup_{a\in A} \evalpXx{\evalp(\delta)}{a}(\psi)\\
& = \evalpX{\evalp(\delta)}(\exists x. \psi)\\
\end{align*}
	
\item $\vp=\mu Z.\psi$. Since  $X\in FV(\vp)$, we have that $X \ne Z$.
 Then $X\in FV(\psi)$ and $X$ is free for $\delta$ in $\psi$. Let $\eval$ be an $\cA$-valuation. We get that 
\begin{align*}
\evalp(\subfvpXdelta) &= \evalp(\mu Z.\subfX{\delta}{\psi})=\bigcap\left\{ B \subseteq A \mid \evalpZB(\subfX{\delta}{\psi}) 
 \subseteq B \right\} \\
&= \bigcap\left\{ B \subseteq A \mid \evalpZX{B}{\evalpZB(\delta)}(\psi) 
 \subseteq B \right\} \\ 
 &  \text{ by the induction hypothesis for } \psi \text{ applied to }  \evalZB.
\end{align*}
As $X$ is free for $\delta$ in $\vp$ and $X\in FV(\vp)$, it follows that $Z\notin FV(\delta)$. Hence, 
by Proposition \ref{evaluation-patterns-free-variables}, $\evalp(\delta)=\evalpZB(\delta)$. 
It follows that 
\begin{align*}
\evalp(\subfvpXdelta) &=  \bigcap\left\{ B \subseteq A \mid \evalpZX{B}{\evalp(\delta)}(\psi) 
 \subseteq B \right\} \\
&= \bigcap\left\{ B \subseteq A \mid \evalpXZ{\evalp(\delta)}{B}(\psi) 
 \subseteq B \right\} \\
& = \evalpX{\evalp(\delta)}(\mu Z.\psi) = \evalpX{\evalp(\delta)}(\vp).
\end{align*}
\ee
}

As an immediate consequence of Proposition \ref{eval-subfvpXdelta}, we get the following.

\bcor\label{eval-subfvpXpY}
Let $\vp$ be a pattern  and $X,Y$ be variables such that $X$ is free for $Y$ in $\vp$. 
Then for every $\tau$-structure $\cA$ and every $\cA$-valuation $\eval$,
$$\evalp(\subfvpXY)=\evalpX{\eval(Y)}(\vp).$$
\ecor

\subsubsection*{Bounded substitution}

\bprop\label{bounded-subst-equiv-X}
Let $\vp$ be a pattern  and $X,Y$ variables such that $Y$ does not occur in $\vp$. Then
$$\models \vp \dnd \subbvpXY.$$
\eprop
\solution{If $X=Y$, then obviously $\subbvpXY=\vp$, hence $\models \vp \dnd \subbvpXY$. Assume that $X\ne Y$. 

Let $\cA$ be a $\tau$-structure. We prove by induction on $\vp$ that 
\bce
for any $\cA$-valuation $\eval$, $\evalp(\vp)=\evalp(\subbvpXY)$.
\ece
We use the definition by recursion on paterns of  $\subbvpXY$ given by Remark~\ref{subb-inductive-X}. 
\be
\item $\vp$ is an atomic pattern. Then $\subbvpXY = \vp$, hence $\evalp(\vp)=\evalp(\subbvpXY)$  for every 
$\cA$-valuation $\eval$. 
\item  $\vp=\appln{\psi}{\chi}$. Let $\eval$ be an $\cA$-valuation. By the induction hypothesis for $\psi$, $\chi$, we have  
that
$\evalp(\psi)=\evalp(\subbf{X}{Y}{\psi})$  and $\evalp(\chi)=\evalp(\subbf{X}{Y}{\chi})$. 
It follows that 
\begin{align*}
\evalp(\subbvpXY) &= \evalp(\appln{\subb{X}{Y}(\psi)}{\subb{X}{Y}(\chi)})=
\evalp(\subbf{X}{Y}{\psi}) \appls \evalp(\subbf{X}{Y}{\psi})=
\evalp(\psi) \appls \evalp(\chi)\\
& = \evalp(\appln{\psi}{\chi}) =\evalp(\vp).
\end{align*}

\item  $\vp=\psi\to\chi$. Let $\eval$ be an $\cA$-valuation. By the induction hypothesis for $\psi$, $\chi$, we have  
that $\evalp(\psi)=\evalp(\subbf{X}{Y}{\psi})$ and  $\evalp(\chi)=\evalp(\subbf{X}{Y}{\chi})$. It follows that 
\begin{align*}
\evalp(\subbvpXY) &= \evalp(\subbf{X}{Y}{\psi}\to \subbf{X}{Y}{\chi})=C_A\evalp(\subbf{X}{Y}{\psi}) \cup \evalp(\subbf{X}{Y}{\chi})\\
& = C_A\evalp(\psi) \cup \evalp(\chi)=\evalp(\psi\to\chi)=\evalp(\vp).
\end{align*}

\item $\vp=\exists x.\psi$. Then 
\begin{align*}
\evalp(\subbvpXY) & =\evalp(\exists x. \subbf{X}{Y}{\psi}) = \bigcap_{a\in A} \evalpxa(\subbf{X}{Y}{\psi})\\
 &= \bigcap_{a\in A} \evalpxa(\vp) \quad \text{ by the induction hypothesis for } \psi \text{ applied to } \evalxa \\
 &=  \evalp(\vp).
\end{align*}
\item $\vp=\mu Z.\psi$. We have two cases:

1. $X \ne Z$, then 
\begin{align*}
\evalp(\subbvpXY) &=\evalp(\mu Z.\subb{X}{Y}{\psi})=\bigcap \left\{ B \subseteq A \mid \evalpZB(\subb{X}{Y}{\psi}) 
 \subseteq B \right\}\\
& = \bigcap \left\{ B \subseteq A \mid \evalpZB(\psi) \subseteq B \right\}\\
&  \text{ by the induction hypothesis for } \psi \text{ applied to } \evalZB \\
&=\evalp(\vp).
\end{align*}
2. $X=Z$. Then 
\begin{align*}\evalp(\subbvpXY) &=\evalp\left(\mu Y.\subfX{Y}{\left(\subb{X}{Y}{\psi}\right)}\right)\\
& = \bigcap \left\{ B \subseteq A \mid \evalpYB(\subfX{Y}{\left(\subb{X}{Y}{\psi}\right)}) 
 \subseteq B \right\}.
\end{align*}
As $Y$ does not occur in $\psi$,  we have that that  $X$ is free for 
$Y$ in $\subb{X}{Y}{\psi}$, by Proposition \ref{subb-X-useful-1}. We can apply then Corollary \ref{eval-subfvpXpY} to get that 
\begin{align*}
\evalpYB\left(\subfX{Y}{\left(\subb{X}{Y}{\psi}\right)}\right) 
& =\evalpYX{B}{\evalYB(Y)}\left(\subb{X}{Y}{\psi}\right)\\
& = \evalpYX{B}{B}\left(\subb{X}{Y}{\psi}\right)=\evalpXB\left(\subb{X}{Y}{\psi}\right)
\\
\end{align*}
since $Y$ does not occur in $\psi$, so $Y\notin FV\left(\subb{X}{Y}{\psi}\right)$, hence we can apply 
Proposition \ref{evaluation-patterns-free-variables}. It follows that 
\begin{align*}
\evalp(\subbvpXY) &= \bigcap \left\{ B \subseteq A \mid \evalpXB\left(\subb{X}{Y}{\psi}\right) 
 \subseteq B \right\} \\
&=  \bigcap \left\{ B \subseteq A \mid \evalpXB\left(\psi\right) \subseteq B \right\} \\
& \text{ by the induction hypothesis for } \psi \text{ applied to }  \evalXB \\
&=\evalp(\vp).
\end{align*}
\ee
}

%% file: AML-semantics-nu.tex
\subsection{Semantics of $\nu$}

 \bprop\label{val-patterns-nu}
 For every pattern $\vp$ and set variable $X$, 
$$\evalp(\nu X.\vp)=\bigcup \left\{ D \subseteq A \mid \evalpXD(\vp) 
 \supseteq D \right\}.$$
\eprop
\solution{As $\nu X.\vp=\neg\mu X.\neg \subfX{\neg X}{\vp}$, we get that
\begin{align*}
\evalp(\nu X.\vp) &= \evalp(\neg\mu X.\neg \subfX{\neg X}{\vp}) = C_A\evalp(\mu X.\neg \subfX{\neg X}{\vp})\\
&= C_A \bigcap \left\{ B \subseteq A \mid \evalpXB(\neg \subfX{\neg X}{\vp}))\subseteq B \right\}  \\
&= C_A \bigcap \left\{ B \subseteq A \mid C_A\evalpXB(\subfX{\neg X}{\vp})) \subseteq B \right\}  \\
& = \bigcup \left\{ C_A B  \mid B \subseteq A \text{~such that~} C_A\evalpXB(\subfX{\neg X}{\vp})) \subseteq B \right\} \\
& = \bigcup \left\{ C_A B \mid B \subseteq A \text{~such that~}  \evalpXB(\subfX{\neg X}{\vp})) \supseteq C_AB \right\}.
\end{align*}
As, obviously, $X$ is free for $\neg X$ in $\vp$, we have by Proposition~\ref{eval-subfvpXdelta} that 
\begin{align*}
\evalpXB(\subfX{\neg X}{\vp}) & = \evalpXX{B}{\evalpXB(\neg X)}(\vp)= \evalpXX{B}{C_AB}(\vp) \\
& = \evalpX{C_AB}(\vp).
\end{align*}
It follows that 
\begin{align*}
\evalp(\nu X.\vp) &=\bigcup \left\{ C_A B  \mid B \subseteq A \text{~such that~} \evalpX{C_AB}(\vp) \supseteq C_AB \right\}\\
& = \bigcup \left\{ D \subseteq A \mid \evalpXD(\vp) \supseteq D \right\}.
\end{align*}
}

%% file: AML-semantics-lfn-lgn.tex
\subsection{Least and greatest fixpoints}

Let $\vp$ be a pattern and $X$ be a set variable. Define, for every $(\cA, \eval)$, the functions
\begin{align}
\fkntar{\vp}{X}: 2^A \to 2^A,  & \quad \fkntar{\vp}{X}(B)=\evalpXB(\vp), \\[1mm]
\gkntar{\vp}{X}: 2^A \to 2^A,  & \quad \gkntar{\vp}{X}(B) = C_A \fkntar{\vp}{X}(B)=C_A\evalpXB(\vp).
\end{align}

\bprop\label{fkntar-monotone}
\be
\item\label{fkntar-monotone-positive} If $\vp$ is positive in $X$, then for all $(\cA, \eval)$, 
$\fkntar{\vp}{X}$ is monotone. 
\item\label{fkntar-monotone-negative} If $\vp$ is negative in $X$, then for all $(\cA, \eval)$, $\gkntar{\vp}{X}$ is monotone. 
\ee
\eprop
\solution{We prove \eqref{fkntar-monotone-positive} and \eqref{fkntar-monotone-negative} simultaneously 
by induction on $\vp$. We use Remarks~\ref{def-positive-X-recursion},~\ref{def-negative-X-recursion}. Let $B,C\se A$ be such that $B\se C$. 
\be 
\item[(a)] $\vp$ is atomic and $\vp\ne X$. Then $\vp$ is both positive and negative in $X$. As $X\notin FV(\vp)$,
we have that $\evalpXB(\vp)=\evalpXC(\vp)=\evalp(\vp)$. It follows that 
$\fkntar{\vp}{X}(B)=\fkntar{\vp}{X}(C)=\evalp(\vp)$ and $\gkntar{\vp}{X}(B)=\gkntar{\vp}{X}(C)=C_A\evalp(\vp)$.
 
\item[(b)]  $\vp=X$. Then $\vp$ is positive in $X$, hence 
\[\fkntar{\vp}{X}(B)=\evalpXB(X)=B\se C=\evalpXC(X)=\fkntar{\vp}{X}(C).\]

\item[(c)]  $\vp=\appln{\psi}{\chi}$. 
\be
\item[(i)] Assume that $\vp$ is positive in $X$. We have that  both $\psi$, $\chi$ are positive in $X$.
We can apply the induction hypothesis to get that $\fkntar{\psi}{X}(B)\se \fkntar{\psi}{X}(C)$ and $\fkntar{\chi}{X}(B)\se \fkntar{\chi}{X}(C)$.
It follows that 
\begin{align*}
\fkntar{\vp}{X}(B) &= \evalpXB(\appln{\psi}{\chi})=\evalpXB(\psi)\appls \evalpXB(\chi)=\fkntar{\psi}{X}(B)\appls \fkntar{\chi}{X}(B)\\
& \stackrel{\eqref{B-se-C-appls-D-se E}}{\se} \fkntar{\psi}{X}(C)\appls \fkntar{\chi}{X}(C) = \evalpXC(\psi)\appls \evalpXC(\chi)= 
\evalpXC(\appln{\psi}{\chi})\\
& =\fkntar{\vp}{X}(C).
\end{align*}
\item[(ii)] Assume that $\vp$ is negative in $X$. We have that  both $\psi$, $\chi$ are negative in $X$.
We can apply the induction hypothesis to get that $\gkntar{\psi}{X}(B)\se \gkntar{\psi}{X}(C)$ and 
$\gkntar{\chi}{X}(B)\se \gkntar{\chi}{X}(C)$, hence  $\fkntar{\psi}{X}(B)\supseteq \fkntar{\psi}{X}(C)$ and 
$\fkntar{\chi}{X}(B)\supseteq \fkntar{\chi}{X}(C)$.
It follows that 
\begin{align*}
\fkntar{\vp}{X}(B) &= \evalpXB(\appln{\psi}{\chi})=\evalpXB(\psi)\appls \evalpXB(\chi)=\fkntar{\psi}{X}(B)\appls \fkntar{\chi}{X}(B)\\
& \stackrel{\eqref{B-se-C-appls-D-se E}}{\supseteq} \fkntar{\psi}{X}(C)\appls \fkntar{\chi}{X}(C) = \evalpXC(\psi)\appls \evalpXC(\chi)= 
\evalpXC(\appln{\psi}{\chi})\\
&=\fkntar{\vp}{X}(C).
\end{align*}
Thus, $\gkntar{\vp}{X}(B)\se \gkntar{\vp}{X}(C)$.
\ee

\item[(d)] $\vp=\psi\to\chi$. 
\be
\item[(i)] Assume that $\vp$ is positive in $X$.  We have that $\psi$ is negative in $X$ and  $\chi$ is positive in $X$.
We can apply  the induction hypothesis for  $\psi$ to get that 
$\gkntar{\psi}{X}(B)\se \gkntar{\psi}{X}(C)$ and the induction hypothesis for $\chi$ to get that 
$\fkntar{\chi}{X}(B)\subseteq \fkntar{\chi}{X}(C)$.
It follows that 
\begin{align*}
\fkntar{\vp}{X}(B) & = \evalpXB(\psi\to\chi)=C_A\evalpXB(\psi) \cup \evalpXB(\chi) \\
& = \gkntar{\psi}{X}(B) \cup \fkntar{\chi}{X}(B)  \\
& \se \gkntar{\psi}{X}(C) \cup \fkntar{\chi}{X}(C) \\
& = C_A \evalpXC(\psi) \cup \evalpXC(\chi) = \evalpXC(\psi\to\chi)\\
& =\fkntar{\vp}{X}(C). 
\end{align*}
\item[(ii)] Assume that $\vp$ is negative in $X$. We have that $\psi$ is positive in $X$ and  $\chi$ is negative in $X$.
We can apply the induction hypothesis for $\psi$ to get that 
$\fkntar{\psi}{X}(B)\subseteq \fkntar{\psi}{X}(C)$ and the induction hypothesis for  $\chi$ to get that 
$\gkntar{\chi}{X}(B)\se \gkntar{\chi}{X}(C)$.
It follows that 
\begin{align*}
\gkntar{\vp}{X}(B) & = C_A\evalpXB(\psi\to\chi)=C_A\bigg(C_A \evalpXB(\psi)\cup \evalpXB(\chi)\bigg) \\
& = \evalpXB(\psi) \cap C_A\evalpXB(\chi) = \fkntar{\psi}{X}(B) \cap \gkntar{\chi}{X}(B)  \\
& \se \fkntar{\psi}{X}(C) \cap \gkntar{\chi}{X}(C)= \evalpXC(\psi) \cap C_A\evalpXC(\chi)\\
&=C_A\bigg(C_A \evalpXC(\psi)\cup \evalpXC(\chi)\bigg) = C_A\evalpXC(\psi\to\chi)\\
& =\gkntar{\vp}{X}(C). 
\end{align*}
\ee

\item[(e)] $\vp=\exists x.\psi$. 
\be
\item[(i)] Assume that $\vp$ is positive in $X$. We have that $\psi$ is positive in $X$. 
It follows that 
\begin{align*}
\fkntar{\vp}{X}(B) & = \evalpXB(\exists x.\psi) =\bigcup_{a\in A} \evalpXx{B}{a}(\psi) = \bigcup_{a\in A} \evalpxX{a}{B}(\psi) \\
& = \bigcup_{a\in A} \fkntara{\psi}{X}(B) \se \bigcup_{a\in A}\fkntara{\psi}{X}(C)\\
&  \text{by the induction hypothesis for~}\psi \text{~applied to~} \evalxa\\
&= \bigcup_{a\in A} \evalpxX{a}{C}(\psi)=\bigcup_{a\in A} \evalpXx{C}{a}(\psi) =\evalpXC(\exists x.\psi)\\
& =\fkntar{\vp}{X}(C).
\end{align*}
\item[(ii)] Assume that $\vp$ is negative in $X$. We have that $\psi$ is negative in $X$. 
It follows that 
\begin{align*}
\fkntar{\vp}{X}(B) & = \evalpXB(\exists x.\psi) =\bigcup_{a\in A} \evalpXx{B}{a}(\psi) = \bigcup_{a\in A} \evalpxX{a}{B}(\psi)\\
& = \bigcup_{a\in A} \fkntara{\psi}{X}(B) \supseteq \bigcup_{a\in A}\fkntara{\psi}{X}(C)
\end{align*}
as, by the induction hypothesis for $\psi$ applied to $\evalxa$, we have that $\gkntara{\psi}{X}(B)\se \gkntara{\psi}{X}(C)$, hence 
$\fkntara{\psi}{X}(B)\supseteq \fkntara{\psi}{X}(C)$. Thus,
\begin{align*}
\fkntar{\vp}{X}(B) & \supseteq \bigcup_{a\in A}\fkntara{\psi}{X}(C)= \bigcup_{a\in A} \evalpxX{a}{C}(\psi)=\bigcup_{a\in A} \evalpXx{C}{a}(\psi) \\
& =\evalpXC(\exists x.\psi)=\fkntar{\vp}{X}(C).
\end{align*}
Hence, $\gkntar{\vp}{X}(B) \se \gkntar{\vp}{X}(C)$.
\ee

\item[(f)] $\vp=\mu Z.\psi$. If $Z=X$, then $\vp$ is both positive and negative in $X$. As $X\notin FV(\vp)$,
we have that $\evalpXB(\vp)=\evalpXC(\vp)=\evalp(\vp)$. It follows that 
$\fkntar{\vp}{X}(B)=\fkntar{\vp}{X}(C)=\evalp(\vp)$ and $\gkntar{\vp}{X}(B)=\gkntar{\vp}{X}(C)=C_A\evalp(\vp)$.\\

Suppose that $X\ne Z$. 
\be
\item[(i)] Assume that $\vp$ is positive in $X$.  We have that $\psi$ is positive in $X$. 
Then
\begin{align*}
\fkntar{\vp}{X}(B) & = \evalpXB(\mu Z.\psi) = \bigcap \left\{ D \subseteq A \mid \evalpXZ{B}{D}(\psi) \subseteq D \right\} \\
& = \bigcap \left\{ D \subseteq A \mid \evalpZX{D}{B}(\psi) \subseteq D \right\} \\
&= \bigcap \left\{ D \subseteq A \mid \fkntarZD{\psi}{X}(B) \subseteq D \right\}.
\end{align*}
By the induction hypothesis for $\psi$ applied to $\evalZD$, we have that $\fkntarZD{\psi}{X}(B)\se \fkntarZD{\psi}{X}(C)$, so 
$\fkntarZD{\psi}{X}(C)\se D$ implies $\fkntarZD{\psi}{X}(B)\se D$, hence 
$$\left\{ D \subseteq A \mid \fkntarZD{\psi}{X}(C) \subseteq D \right\}\se 
\left\{ D \subseteq A \mid \fkntarZD{\psi}{X}(B) \subseteq D \right\}.$$
Thus, 
$$ \bigcap  \left\{ D \subseteq A \mid \fkntarZD{\psi}{X}(B) \subseteq D \right\}
\se \bigcap \left\{ D \subseteq A \mid \fkntarZD{\psi}{X}(C) \subseteq D \right\}.$$
It follows that
\begin{align*}
\fkntar{\vp}{X}(B) & \se \bigcap \left\{ D \subseteq A \mid \fkntarZD{\psi}{X}(C) \subseteq D \right\}  
= \bigcap \left\{ D \subseteq A \mid \evalpZX{D}{C}(\psi) \subseteq D \right\} \\
& =\bigcap \left\{ D \subseteq A \mid \evalpXZ{C}{D}(\psi) \subseteq D \right\} \\
&= \evalpXC(\mu Z.\psi)= \fkntar{\vp}{X}(C).
\end{align*}
\item[(ii)] Assume that $\vp$ is negative in $X$. We have that $\psi$ is negative in $X$. 
Then 
\begin{align*}
\fkntar{\vp}{X}(B) & = \evalpXB(\mu Z.\psi) = \bigcap \left\{ D \subseteq A \mid \evalpXZ{B}{D}(\psi) \subseteq D \right\} \\
& = \bigcap \left\{ D \subseteq A \mid \evalpZX{D}{B}(\psi) \subseteq D \right\} \\
& = \bigcap \left\{ D \subseteq A \mid \fkntarZD{\psi}{X}(B) \subseteq D \right\}.
\end{align*}
By the induction hypothesis for $\psi$ applied to $\evalZD$, we have that 
$\gkntarZD{\psi}{X}(B)\se \gkntarZD{\psi}{X}(C)$, so $\fkntarZD{\psi}{X}(B)\supseteq \fkntarZD{\psi}{X}(C)$. It follows that 
$\fkntarZD{\psi}{X}(B)\se D$ implies $\fkntarZD{\psi}{X}(C)\se D$, hence 
$$\left\{ D \subseteq A \mid \fkntarZD{\psi}{X}(B) \subseteq D \right\}\se 
\left\{ D \subseteq A \mid \fkntarZD{\psi}{X}(C) \subseteq D \right\}.$$
Thus, 
$$ \bigcap  \left\{ D \subseteq A \mid \fkntarZD{\psi}{X}(C) \subseteq D \right\} \se 
\bigcap \left\{ D \subseteq A \mid \fkntarZD{\psi}{X}(B) \subseteq D \right\}.$$
It follows that
\begin{align*}
\fkntar{\vp}{X}(B) & \supseteq \bigcap \left\{ D \subseteq A \mid \fkntarZD{\psi}{X}(C) \subseteq D \right\} 
 = \bigcap \left\{ D \subseteq A \mid \evalpZX{D}{C}(\psi) \subseteq D \right\} \\
& =\bigcap \left\{ D \subseteq A \mid \evalpXZ{C}{D}(\psi) \subseteq D \right\} \\
&= \evalpXC(\mu Z.\psi)= \fkntar{\vp}{X}(C).
\end{align*}
\ee
Hence, $\gkntar{\vp}{X}(B) \se \gkntar{\vp}{X}(C)$.
\ee
}

\mbox{}

As a consequence of Knaster-Tarski Theorem~\ref{Knaster-Tarski}, it follows that if $\vp$ is positive in $X$, then $\fkntar{\vp}{X}$ has 
a least fixpoint $\mu \fkntar{\vp}{X}$ and a greatest fixpoint $\nu \fkntar{\vp}{X}$, given by:
\begin{align*}
\mu \fkntar{\vp}{X} & = \bigcap \{ B \subseteq A \mid \fkntar{\vp}{X}(B) \subseteq B\} = 
\bigcap \{ B \subseteq A \mid \evalpXB(\vp)  \subseteq B\} \\
& = \bigcap \{ B \subseteq A \mid \evalpXB(\vp)  = B\}, \\
\nu \fkntar{\vp}{X} & = \bigcup \{ B \subseteq A \mid B \subseteq \fkntar{\vp}{X}(B)\} = 
\bigcup \{ B \subseteq A \mid B \subseteq \evalpXB(\vp)\}\\
& = \bigcup \{ B \subseteq A \mid \evalpXB(\vp)  = B\}.
\end{align*}

Furthermore, by Definition~\ref{def-val-patterns}.\eqref{def-val-patterns-mu} and 
Proposition~\ref{val-patterns-nu},  we get that
\begin{align}
\mu \fkntar{\vp}{X} = \evalp(\mu X.\vp)   \quad \text{and} \quad \nu \fkntar{\vp}{X} =  \evalp(\nu X.\vp).
\end{align}

Thus, 

\bprop\label{fkntar-fp}
Let $\vp$ be a pattern and $X$ be a set variable such that $\vp$ is positive in $X$. Then, for every $(\cA, \eval)$, 
\be
\item\label{fkntar-mu-nu-fp} $\evalpX{\evalp(\mu X.\vp)}(\vp)=\evalp(\mu X.\vp)$ and $\evalpX{\evalp(\nu X.\vp)}(\vp)=\evalp(\nu X.\vp)$.
\item\label{fkntar-mu-nu-fp-least-greatest}  For every $B\se A$ such that $\evalpXB(\vp)=B$, 
$$\evalp(\mu X.\vp)\se B \se \evalp(\nu X.\vp).$$
\ee
\eprop

%% file: AML-semantics-examples.tex
\subsection{Equivalences, logical consequences, validities}

\bprop
For all patterns $\vp$, $\psi$, $\chi$,
\begin{align}
& \models \vp \dnd \vp, \label{dnd-reflexive}\\
& \models \vp \dnd \psi \quad \text{iff} \quad \models \psi \dnd \vp, \label{dnd-symmetric}\\
& \models \vp \dnd \psi \text{ and } \models \psi \dnd  \chi \quad \text{implies}
 \quad \models \vp \dnd \chi,\label{dnd-tranzitive}\\
& \models \vp \to \psi \text{ and } \models \psi \to  \chi \quad \text{implies}
\quad \models \vp \to \chi.\label{to-tranzitive}
\end{align}
\eprop
\solution{\eqref{dnd-reflexive}:  It follows by \eqref{dnd-vp-vp} and Proposition~\ref{tautology-valid}.

\eqref{dnd-symmetric}: We have that 
$\models \vp \dnd \psi$ iff $\cA \models (\vp \dnd \psi)[\eval]$ for every $(\cA, \eval)$ iff
$\evalp(\vp)=\evalp(\psi)$ for every $(\cA, \eval)$ iff $\evalp(\psi)=\evalp(\vp)$ for every $(\cA, \eval)$
iff $\cA \models (\psi \dnd \vp)[\eval]$ for every $(\cA, \eval)$  iff $\models \psi \dnd \vp$.

\eqref{dnd-tranzitive}: Let $(\cA,\eval)$. Then $\cA \models (\vp \dnd \psi)[\eval]$, hence 
$\evalp(\vp)=\evalp(\psi)$. Furthermore,  $\cA \models (\psi \dnd \chi )[\eval]$, hence 
$\evalp(\psi)=\evalp(\chi)$. It follows that $\evalp(\vp)=\evalp(\chi)$. Thus, 
$\cA \models (\vp \dnd \chi)[\eval]$.

\eqref{to-tranzitive}: Assume that $\models \vp \to \psi$ 
and  $\models \psi \to  \chi$. Let $(\cA,\eval)$. Then $\cA \models (\vp \to \psi)[\eval]$, hence 
$\evalp(\vp)\se \evalp(\psi)$. Furthermore,  $\cA \models (\psi \to \chi )[\eval]$, hence 
$\evalp(\psi) \se \evalp(\chi)$. It follows that $\evalp(\vp)\se \evalp(\chi)$. Thus, 
$\cA \models (\vp \to \chi)[\eval]$.
}

\bprop
For any pattern  $\vp$ and  any variable $x$,
\begin{align}
& \models \forall x.\vp \to \exists x.\vp, \label{forallxvp-imp-existsx-vp}\\
& \models \forall x.\vp \to \vp, \label{forallxvp-imp-x}\\
& \models \vp \to \exists x. \vp, \label{vp-imp-existsvp}\\
& \models \exists x.x, \label{existsx-x}\\
& \models \forall x.\vp \text{ iff } \models \vp \label{forallxvp-vp}
\end{align}
\eprop
\solution{We apply Proposition \ref{A-e-models}.\eqref{A-e-models-to}.

\eqref{forallxvp-imp-existsx-vp}: Let $(\cA,\eval)$. Then 
$\evalp(\forall x.\vp)=\bigcap_{a\in A} \evalpxa(\vp)\se \bigcup_{a\in A} \evalpxa(\vp)=
\evalp(\exists x.\vp)$.

\eqref{forallxvp-imp-x}: Let $(\cA,\eval)$. Then 
$\evalp(\forall x.\vp)=\bigcap_{a\in A} \evalpxa(\vp)\se \evalpx{e(x)}(\vp)=\evalp(\vp)$.

\eqref{vp-imp-existsvp}: Let $(\cA,\eval)$. Then $\evalp(\vp)=\evalpx{e(x)}(\vp)\se \bigcup_{a\in A} \evalpxa(\vp)=
\evalp(\exists x.\vp)$.

\eqref{existsx-x}: Let $(\cA, \eval)$. Then 
$\evalp(\exists x.x)=\bigcup_{a\in A} \evalpxa(x)=\bigcup_{a\in A}\{a\}=A$. 
Thus, $\cA \models (\exists x.x)[\eval]$. 

\eqref{forallxvp-vp}: By Proposition \ref{A-models}.\eqref{A-models-forall}.
}

\bprop
For any pattern  $\vp$ and  any variable $x$,
\begin{align}
& \models \exists x.\vp \dnd \vp \quad \text{ if } x\notin FV(\vp) \label{existsx-equiv-x-FV},\\
& \models  \forall x.\vp \dnd \vp \quad \text{ if } x\notin FV(\vp) \label{forallx-equiv-x-FV}.
\end{align}
\eprop
\solution{We apply Proposition \ref{A-e-models}.\eqref{A-e-models-dnd}. Let $(\cA,\eval)$. 
As $x\notin FV(\vp)$, we can apply Proposition~\ref{evaluation-patterns-free-variables} to get that
 $\evalpxa(\vp)=\evalp(\vp)$ for any $a\in A$.

\eqref{existsx-equiv-x-FV}:  We have that 
$\evalp(\exists x.\vp)=\bigcup_{a\in A} \evalpxa(\vp)=\bigcup_{a\in A}\evalp(\vp)=\evalp(\vp)$.

\eqref{forallx-equiv-x-FV}: We have that 
$\evalp(\forall x.\vp)=\bigcap_{a\in A} \evalpxa(\vp)=\bigcap_{a\in A}\evalp(\vp)=\evalp(\vp)$.
}

\bprop
For any pattern  $\vp$ and  any variable $x$,
\begin{align}
& \models \forall x.(\vp \to \psi) \to  (\forall x.\vp \to \forall x.\psi) \label{forallxvptopsi-to-forallxvptoforallxpsi}.
\end{align}
\eprop
\solution{ Let $(\cA,\eval)$. We have that 

\begin{align*}
\evalp(\forall x. \vp \to \forall x.\psi) & = C_A\evalpxa(\forall x \vp) \cup \evalpxa(\forall x \psi) 
\quad \text{by Lemma \ref{basic-val}.\eqref{basic-val-imp}}\\
& = C_A\left(\bigcap_{a\in A} \evalpxa(\vp)\right) \cup \evalpxa(\psi) \\
& = \bigcup_{a\in A} C_A\evalpxa(\vp) \cup \bigcap_{a\in A} \evalpxa(\psi)  \quad 
\text{by \eqref{de-morgan-family-CAcap}}\\
& = \bigcap_{a\in A} 
\left(\bigcup_{a\in A} C_A\evalpxa(\vp) \cup \evalpxa(\psi)\right)  \quad 
\text{by \eqref{distrib-cap-family-left}}\\
& \supseteq \bigcap_{a\in A} \big(C_A \evalpxa(\vp) \cup\evalpxa(\psi)\big)\\
& = \bigcap_{a\in A} \evalpxa(\vp \to \psi) \quad \text{by Lemma \ref{basic-val}.\eqref{basic-val-imp}}\\
& = \evalp(\forall x. (\vp \to \psi))   \quad \text{by Lemma \ref{basic-val}.\eqref{basic-val-forall}}
\end{align*}
We have got that $\evalp(\forall x. (\vp \to \psi)) \se \evalp(\forall x. \vp \to \forall x.\psi)$. Apply Proposition \ref{A-e-models}.\eqref{A-e-models-to}
to conclude that 
$\cA \models \big(\forall x.(\vp \to \psi) \to  (\forall x.\vp \to \forall x.\psi)\big)[e]$.
}

%% file: AML-semantics-examples-rules.tex
\subsection{Rules}

Let $\Gamma$ be a set of  patterns.

\subsubsection{$\ssc$}

\bprop
For all  patterns $\vp$, $\psi$, $\chi$, 
\begin{align}
\Gamma \ssc \vp \text{ and } \Gamma \ssc \vp\to \psi & \quad \Ra \quad \Gamma \ssc \psi 
\label{ssc-closure-mp}\\
\Gamma \ssc \vp \to \psi \text{ and }  \Gamma \ssc \psi \to \chi & \quad \Ra \quad \Gamma \ssc
\vp\to\chi \label{ssc-closure-syllogism}\\
\Gamma \ssc \vp \si \psi \to \chi & \quad \Ra \quad
\Gamma \ssc \vp \to (\psi \to \chi) \label{ssc-closure-exportation}\\
\Gamma \ssc \vp \to (\psi \to \chi)  & \quad \Ra \quad
\Gamma \ssc \vp \si \psi \to \chi \label{ssc-closure-importation}
\end{align}
\eprop
\solution{Let $(\cA, \eval)$.

\eqref{ssc-closure-mp}:  Assume that $\Gamma \ssc \vp$ and $\Gamma \ssc \vp\to \psi$. Then  $\bigcap_{\gamma \in \Gamma} \evalp(\gamma)
\se \evalp(\vp)$ and  $\bigcap_{\gamma \in \Gamma} \evalp(\gamma)
\se \evalp(\vp\to \psi)$, hence $\bigcap_{\gamma \in \Gamma} \evalp(\gamma) \se \evalp(\vp) \cap \evalp(\vp\to \psi)$. Remark that 
\begin{align*}
\evalp(\vp) \cap \evalp(\vp\to \psi)& =\evalp(\vp)\cap
\left(C_A \evalp(\vp)\cup\evalp(\psi)\right)=\left(\evalp(\vp)\cap C_A \evalp(\vp)\right)\cup \left(\evalp(\vp)\cap \evalp(\psi)\right) \\
& = \emptyset \cup \left(\evalp(\vp)\cap \evalp(\psi)\right) 
 = \evalp(\vp)\cap \evalp(\psi) \se
\evalp(\psi).
\end{align*} 
Hence, $\bigcap_{\gamma \in \Gamma} \evalp(\gamma)\se \evalp(\psi)$, that is $\Gamma \ssc \psi$.

\eqref{ssc-closure-syllogism}:  Assume that $\Gamma \ssc \vp \to \psi$ and $\Gamma \ssc \psi \to \chi$. Then 
$\bigcap_{\gamma \in \Gamma} \evalp(\gamma)
\se \evalp(\vp\to\psi)$ and  $\bigcap_{\gamma \in \Gamma} \evalp(\gamma)
\se \evalp(\psi\to\chi)$, hence
$\bigcap_{\gamma \in \Gamma} \evalp(\gamma) \se \evalp(\vp\to\psi) \cap \evalp(\psi\to\chi)$. 
Remark now that 
\begin{align*}
\evalp(\vp\to\psi) \cap \evalp(\psi\to\chi)
& = C_A\big(\evalp(\vp) \setminus\evalp(\psi)\big)\cap C_A\big(\evalp(\psi) \setminus\evalp(\chi)\big)\\
& \stackrel{\eqref{de-morgan-1}}{=}
C_A\bigg((\evalp(\vp) \setminus\evalp(\psi))\cup (\evalp(\psi) \setminus\evalp(\chi))\bigg)\\
&\se C_A\big(\evalp(\vp) \setminus \evalp(\chi)\big) \text{~by \eqref{A-C-se-A-B-cup-B-C} and \eqref{CA-B-se-D}}\\
& =\evalp(\vp\to\chi).
\end{align*}

\eqref{ssc-closure-exportation}, \eqref{ssc-closure-importation}: By  \eqref{exp+imp} and 
Remark \ref{ssc-sleqs-to-dnd}.\eqref{ssc-sleqs-to-dnd-sleqs}, 
we have that $\vp \to (\psi \to \chi)\sleqs \vp \si \psi \to \chi$. Apply now 
Remark \ref{ssc-sleqs-Gamma}.\eqref{ssc-sleqs-Gamma-sleqs}.
}

\subsubsection{$\lsc$}

\bprop
For all  patterns $\vp$, $\psi$, $\chi$, 
\begin{align}
\Gamma \lsc \vp \text{ and } \Gamma \lsc \vp\to \psi & \quad \Ra \quad \Gamma \lsc \psi 
\label{lsc-closure-mp}\\
\Gamma \lsc \vp \to \psi \text{ and }  \Gamma \lsc \psi \to \chi & \quad \Ra \quad \Gamma \lsc
\vp\to\chi \label{lsc-closure-syllogism}\\
\Gamma \lsc \vp \si \psi \to \chi & \quad \Ra \quad
\Gamma \lsc \vp \to (\psi \to \chi) \label{lsc-closure-exportation}\\
\Gamma \lsc \vp \to (\psi \to \chi)  & \quad \Ra \quad
\Gamma \lsc \vp \si \psi \to \chi \label{lsc-closure-importation}
\end{align}
\eprop
\solution{Let $(\cA, \eval)$ be such that $\cA\models \Gamma[\eval]$.

\eqref{lsc-closure-mp}: Assume that $\Gamma \lsc \vp$ and $\Gamma \lsc \vp\to \psi$. Then  
$\cA\models \vp [\eval]$ and $\cA\models (\vp\to\psi)[\eval]$, hence  $\evalp(\vp)\se \evalp(\psi)$. 
 It follows that 
$\evalp(\psi)=A$, hence $\cA\models \psi[\eval]$.

\eqref{lsc-closure-syllogism}: Assume that $\Gamma \lsc \vp \to \psi$ and $\Gamma \lsc \psi \to \chi$. 
Then  $\cA\models (\vp\to\psi)[\eval]$ and $\cA\models (\psi\to\chi)[\eval]$, hence 
$\evalp(\vp)\se \evalp(\psi)$ and $\evalp(\psi)\se \evalp(\chi)$.  It follows that 
$\evalp(\vp)\se \evalp(\chi)$, so $\cA\models (\vp\to\chi)[\eval]$. 

\eqref{lsc-closure-exportation}, \eqref{lsc-closure-importation}: 
By  \eqref{exp+imp} and 
Remark \ref{ssc-sleqs-to-dnd}.\eqref{ssc-sleqs-to-dnd-sleqs}, 
we have that $\vp \to (\psi \to \chi)\sleqs \vp \si \psi \to \chi$, hence, by 
Remark \ref{remark-conseq-equiv-to-dnd}.\eqref{remark-conseq-equiv-to-dnd-dnd}, 
$\vp \to (\psi \to \chi)\lleq \vp \si \psi \to \chi$. 
Apply now 
Remark \ref{lsc-lleq-Gamma}.\eqref{lsc-lleq-Gamma-lleq}
}

\subsubsection{$\gsc$}

\bprop
For all  patterns $\vp$, $\psi$, $\chi$, 
\begin{align}
\Gamma \gsc \vp \text{ and } \Gamma \gsc \vp\to \psi & \quad \Ra \quad \Gamma \gsc \psi,
\label{closure-mp-gsc}\\
\Gamma \gsc \vp \to \psi \text{ and }  \Gamma \gsc \psi \to \chi & \quad \Ra \quad \Gamma \gsc
\vp\to\chi, \label{closure-syllogism-gsc}\\
\Gamma \gsc \vp \si \psi \to \chi & \quad \Ra \quad
\Gamma \gsc \vp \to (\psi \to \chi), \label{closure-exportation-gsc}\\
\Gamma \gsc \vp \to (\psi \to \chi),  & \quad \Ra \quad
\Gamma \gsc \vp \si \psi \to \chi, \label{closure-importation-gsc}\\
\Gamma \gsc \vp \to \psi  & \quad \Ra \quad \Gamma \gsc
\chi\sau\vp \to \chi\sau\psi. \label{closure-expansion-gsc}
\end{align}
\eprop
\solution{Let $\cA$ be such that $\cA\models \Gamma$.

\eqref{closure-mp-gsc}: Assume that $\Gamma \gsc \vp$ and $\Gamma \gsc \vp\to \psi$. Let  $\eval$ be an $\cA$-evaluation.  Then   
$\cA\models \vp [\eval]$ and $\cA\models (\vp\to\psi)[\eval]$, so  $A=\evalp(\vp)\se \evalp(\psi)$. It follows that 
$\evalp(\psi)=A$, hence $\cA\models \psi[\eval]$.

\eqref{closure-syllogism-gsc}: Assume that $\Gamma \gsc \vp \to \psi$ and $\Gamma \gsc \psi \to \chi$. Let  $\eval$ be an $\cA$-evaluation.
Then $\cA\models (\vp\to\psi)[\eval]$ and $\cA\models (\psi\to\chi)[\eval]$, hence 
$\evalp(\vp)\se \evalp(\psi)$ and $\evalp(\psi)\se \evalp(\chi)$.  It follows that 
$\evalp(\vp)\se \evalp(\chi)$, so $\cA\models (\vp\to\chi)[\eval]$.

\eqref{closure-exportation-gsc}, \eqref{closure-importation-gsc}: 
By  \eqref{exp+imp} and 
Remark \ref{ssc-sleqs-to-dnd}.\eqref{ssc-sleqs-to-dnd-sleqs}, 
we have that $\vp \to (\psi \to \chi)\sleqs \vp \si \psi \to \chi$, hence, by 
Remark \ref{remark-conseq-equiv-to-dnd}.\eqref{remark-conseq-equiv-to-dnd-dnd}, 
$\vp \to (\psi \to \chi)\gleq \vp \si \psi \to \chi$. 
Apply now 
Remark \ref{gsc-gleq-Gamma}.\eqref{gsc-gleq-Gamma-gleq}.

\eqref{closure-expansion-gsc}: Assume that $\Gamma \gsc \vp \to \psi$. Let  $\eval$ be an $\cA$-evaluation. Then 
$\cA\models (\vp\to\psi)[\eval]$, hence $\evalp(\vp)\se \evalp(\psi)$. We get that 
$\evalp(\chi\sau\vp) = \evalp(\chi) \cup \evalp(\vp) \se \evalp(\chi) \cup \evalp(\psi) = \evalp(\chi\sau\psi)$. Thus,
$\cA\models (\chi\sau\vp\to\chi\sau\psi)[\eval]$.
}

\bprop
For all  patterns $\vp$, $\psi$ and variables $x$, 
\begin{align}
\Gamma \gsc \vp \to \psi &  \quad \Ra \quad \Gamma \gsc \exists x.\vp \to \psi \quad \text{ if } x\notin FV(\psi).
\label{closure-exists-quantifier-gsc}, \\
\Gamma \gsc \vp  &  \quad \Ra \quad \Gamma \gsc \forall x.\vp. \label{closure-gen}
\end{align}
\eprop
\solution{Let $\cA$ be such that $\cA\models \Gamma$.

\eqref{closure-exists-quantifier-gsc} Assume that 
$\Gamma \gsc \vp \to \psi$. Let  $\eval$ be an $\cA$-evaluation. Then for every $a\in A$, 
$\cA\models (\vp \to \psi)[\evalxa]$, so $\evalpxa(\vp)\se \evalpxa(\psi)$. Since $x\notin FV(\psi)$, we have that 
$\evalpxa(\psi)=\evalp(\psi)$. Thus, we have got that for every $a\in A$, $\evalpxa(\vp)\se \evalp(\psi)$. It follows that 
$\evalp(\exists x.\vp)=\bigcup_{a\in A} \evalpxa(\vp)\se \evalp(\psi)$, that is $\cA \models (\exists x.\vp \to \psi)[\eval]$.

\eqref{closure-gen} Assume that $\Gamma \gsc \vp$. Let  $\eval$ be an $\cA$-evaluation. Then for every $a\in A$, $\cA\models \vp[\evalxa]$, as 
$\cA\models \vp$. Thus, $\cA\models \forall x \vp$. 
}

%% file: AML-semantics-examples-application.tex
\subsection{Application}

\bprop
For all patterns $\vp$, $\psi$, $\chi$, 
\begin{align}
& \models \appln{\vp}{\bot} \dnd \bot, \label{propagation-l}\\
& \models  \appln{\bot}{\vp} \dnd \bot, \label{propagation-r}\\
& \models \appln{(\vp \sau\psi)}{\chi} \dnd \appln{\vp}{\chi}\sau \appln{\psi}{\chi}, 
\label{propagation-sau-l}\\
& \models \appln{\chi}{(\vp \sau\psi)} \dnd \appln{\chi}{\vp}\sau \appln{\chi}{\psi}, 
\label{propagation-sau-r}\\
& \models \appln{(\exists x. \vp)}{\psi} \dnd \exists x.\appln{\vp}{\psi}
\quad  \text{ if } x \notin FV(\psi), \label{propagation-exists-l}\\
& \models \appln{\psi}{(\exists x.\vp)} \dnd \appln{\exists x.\psi}{\vp}  
\quad \text{ if } x \not\in FV(\psi). \label{propagation-exists-r}
\end{align}
\eprop
\solution{Let $(\cA, \eval)$. 

\eqref{propagation-l}, \eqref{propagation-r}: Apply \eqref{B-appl-empty} to get that 
 $\evalp(\appln{\vp}{\bot})=\evalp(\vp)\appls \evalp(\bot)=
\evalp(\vp)\appls \emptyset =\emptyset=\evalp(\bot)$ and  $\evalp(\appln{\bot}{\vp})=\evalp(\bot)\appls \evalp(\vp)=
\emptyset \appls \evalp(\vp) =\emptyset=\evalp(\bot)$.

\eqref{propagation-sau-l}: We have that $\evalp(\appln{(\vp \sau\psi)}{\chi})=
\evalp(\vp \sau \psi)\appls \evalp(\chi)=(\evalp(\vp)\cup \evalp(\psi))\appls \evalp(\chi)
\stackrel{\eqref{C-cup-D-appls-B}}{=}
(\evalp(\vp)\appls \evalp(\chi)) \cup (\evalp(\psi)\appls\evalp(\chi))=
\evalp(\appln{\vp}{\chi})\cup \evalp(\appln{\psi}{\chi})
=\evalp(\appln{\vp}{\chi}\sau \appln{\psi}{\chi})$.

\eqref{propagation-sau-r}: We have that $\evalp(\appln{\chi}{(\vp \sau\psi)})=
\evalp(\chi)\appls \evalp(\vp \sau \psi) =\evalp(\chi)\appls (\evalp(\vp)\cup \evalp(\psi))
\stackrel{\eqref{B-appls-C-cup-D}}{=}
(\evalp(\chi)\appls \evalp(\vp)) \cup (\evalp(\chi)\appls\evalp(\psi))=
\evalp(\appln{\chi}{\vp})\cup \evalp(\appln{\chi}{\psi})
=\evalp(\appln{\chi}{\vp}\sau \appln{\chi}{\psi})$.

\eqref{propagation-exists-l}: We have that 
\begin{align*}
\evalp(\appln{(\exists x. \vp)}{\psi}) & =\evalp(\exists x. \vp)\appls
\evalp(\psi) = \left(\bigcup_{a\in A} \evalpxa(\vp)\right)\appls \evalp(\psi)\stackrel{\eqref{cup-family-appls-B}}{=} 
 \bigcup_{a\in A}(\evalpxa(\vp) \appls \evalp(\psi))\\
 &= \bigcup_{a\in A} \left(\evalpxa(\vp) \appls \evalpxa(\psi)\right) \quad \text{as } x \notin FV(\psi), \text{ hence } \evalpxa(\psi)=\evalp(\psi)\\
 &= \bigcup_{a\in A} \evalpxa(\appln{\vp}{\psi}) = \evalp(\exists x.\appln{\vp}{\psi}).
\end{align*}

\eqref{propagation-exists-r}: We have that 
\begin{align*}
\evalp(\appln{\psi}{(\exists x. \vp)}) & =\evalp(\psi) \appls \evalp(\exists x. \vp)
=\evalp(\psi) \appls\left(\bigcup_{a\in A} \evalpxa(\vp)\right)\stackrel{\eqref{B-appls-cup-family}}{=} 
 \bigcup_{a\in A}(\evalp(\psi)\appls \evalpxa(\vp) )\\
 &= \bigcup_{a\in A} \left(\evalpxa(\psi)\appls\evalpxa(\vp) \right) \quad \text{as } x \notin FV(\psi), \text{ hence } \evalpxa(\psi)=\evalp(\psi)\\
 &= \bigcup_{a\in A} \evalpxa(\appln{\psi}{\vp}) = \evalp(\exists x.\appln{\psi}{\vp}).
\end{align*}
}

\bprop
For all patterns $\vp$, $\psi$, $\chi$, 
\begin{align}
& \models \appln{(\vp \si\psi)}{\chi} \to \appln{\vp}{\chi}\si \appln{\psi}{\chi}, 
\label{propagation-si-l}\\
& \models \appln{\chi}{(\vp \si\psi)} \to \appln{\chi}{\vp}\si \appln{\chi}{\psi}, 
\label{propagation-si-r}\\
& \models \appln{(\forall x. \vp)}{\psi} \to \forall x.\appln{\vp}{\psi}
\quad \text{ if } x \notin FV(\psi), \label{propagation-forall-l}\\
& \models \appln{\psi}{(\forall x.\vp)} \to \appln{\forall x.\psi}{\vp}  
\quad\text{ if } x \not\in FV(\psi). \label{propagation-forall-r}
\end{align}
\eprop
\solution{Let $(\cA, \eval)$. 

\eqref{propagation-si-l}: We have that $\evalp(\appln{(\vp \si\psi)}{\chi})=
\evalp(\vp \si \psi)\appls\evalp(\chi) =(\evalp(\vp)\cap \evalp(\psi))\appls \evalp(\chi)
\stackrel{\eqref{C-cap-D-appls-B}}{\se}
(\evalp(\vp)\appls \evalp(\chi)) \cap (\evalp(\psi)\appls\evalp(\chi))=
\evalp(\appln{\vp}{\chi})\cap \evalp(\appln{\psi}{\chi})
=\evalp(\appln{\vp}{\chi}\si \appln{\psi}{\chi})$.

\eqref{propagation-si-r}: We have that $\evalp(\appln{\chi}{(\vp \si\psi)})=
\evalp(\chi)\appls \evalp(\vp \si \psi) =\evalp(\chi)\appls (\evalp(\vp)\cap \evalp(\psi))
\stackrel{\eqref{B-appls-C-cap-D}}{\se}
(\evalp(\chi)\appls \evalp(\vp)) \cap (\evalp(\chi)\appls\evalp(\psi))=
\evalp(\appln{\chi}{\vp})\cap \evalp(\appln{\chi}{\psi})
=\evalp(\appln{\chi}{\vp}\si \appln{\chi}{\psi})$.

\eqref{propagation-forall-l}: We have that 
\begin{align*}
\evalp(\appln{(\forall x. \vp)}{\psi}) & =\evalp(\forall x. \vp)\appls
\evalp(\psi) = \left(\bigcap_{a\in A} \evalpxa(\vp)\right)\appls \evalp(\psi)\stackrel{\eqref{cap-family-appls-B}}{\se} 
 \bigcap_{a\in A}(\evalpxa(\vp) \appls \evalp(\psi))\\
 &= \bigcap_{a\in A} \left(\evalpxa(\vp) \appls \evalpxa(\psi)\right) \quad \text{as } x \notin FV(\psi), \text{ hence } \evalpxa(\psi)=\evalp(\psi)\\
 &= \bigcap_{a\in A} \evalpxa(\appln{\vp}{\psi}) = \evalp(\forall x.\appln{\vp}{\psi}).
\end{align*}

\eqref{propagation-forall-r}: \begin{align*}
\evalp(\appln{\psi}{(\forall x. \vp)}) & =\evalp(\psi) \appls \evalp(\forall x. \vp)
=\evalp(\psi) \appls\left(\bigcap_{a\in A} \evalpxa(\vp)\right)\stackrel{\eqref{B-appls-cap-family}}{\se} 
 \bigcap_{a\in A}(\evalp(\psi)\appls \evalpxa(\vp) )\\
 &= \bigcap_{a\in A} \left(\evalpxa(\psi)\appls\evalpxa(\vp) \right) \quad \text{as } x \notin FV(\psi), \text{ hence } \evalpxa(\psi)=\evalp(\psi)\\
 &= \bigcap_{a\in A} \evalpxa(\appln{\psi}{\vp}) = \evalp(\forall x.\appln{\psi}{\vp}).
\end{align*}
}

\subsubsection{$\lsc$}

\bprop\label{apply-left-right-to-dnd-lsc}
For all patterns $\vp$, $\psi$, $\chi$, 
\begin{align}
 \vp\to \psi \,\,  \lsc  & \,\, \appln{\chi}{\vp}\to \appln{\chi}{\psi}, 
\label{left-apply-to-lsc}\\
 \vp\dnd \psi \,\, \lsc & \,\,  \appln{\chi}{\vp}\dnd \appln{\chi}{\psi}, 
\label{left-apply-dnd-lsc}\\
 \vp\to \psi \,\, \lsc  & \,\, \appln{\vp}{\chi}\to \appln{\psi}{\chi},
\label{right-apply-to-lsc}\\
 \vp\dnd \psi \,\, \lsc &  \,\, \appln{\vp}{\chi}\dnd \appln{\psi}{\chi}, 
\label{right-apply-dnd-lsc}\\
 \vp\to \psi \,\, \lsc  &  \,\, \appln{\chi}{\neg\psi}\to \appln{\chi}{\neg\vp}, 
\label{left-apply-neg-lsc}\\
\vp\dnd \psi\,\,  \lsc & \,\,  \appln{\chi}{\neg\psi}\dnd \appln{\chi}{\neg\vp}, 
\label{left-apply-neg-dnd-lsc}\\
 \vp\to \psi \,\, \lsc  & \,\, \appln{\neg\psi}{\chi}\to \appln{\neg\vp}{\chi},
\label{right-apply-neg-lsc}\\
 \vp\dnd \psi \,\,  \lsc & \,\,  \appln{\neg\psi}{\chi}\dnd \appln{\neg\vp}{\chi}.
\label{right-apply-neg-dnd-lsc}
\end{align}
\eprop
\solution{Let $(\cA, \eval)$.

\eqref{left-apply-to-lsc}:  Assume that $\cA \models (\vp \to \psi )[\eval]$, 
hence $\evalp(\vp)\se \evalp(\psi)$. It follows that
$\evalp(\appln{\chi}{\vp})=\evalp(\chi)\appls \evalp(\vp) \stackrel{\eqref{B-se-C-appls-D}}{\se} \evalp(\chi)\appls \evalp(\psi)=
\evalp(\appln{\chi}{\psi})$, hence $\cA \models (\appln{\chi}{\vp} \to \appln{\chi}{\psi})[\eval]$.

\eqref{left-apply-dnd-lsc}: Assume that $\cA \models (\vp \dnd \psi )[\eval]$. Then, by Proposition~\ref{A-e-models}.\eqref{A-e-models-dnd-2}, 
$\cA \models (\vp \to \psi)[\eval]$ and $\cA \models (\psi \to \vp)[\eval]$. Apply now \eqref{left-apply-to-lsc}  to get that 
$\cA \models (\appln{\chi}{\vp} \to \appln{\chi}{\psi})[\eval]$ and $\cA \models (\appln{\chi}{\psi} \to \appln{\chi}{\vp})[\eval]$, hence, by applying again
Proposition~\ref{A-e-models}.\eqref{A-e-models-dnd-2}, $\cA \models (\appln{\chi}{\vp} \dnd \appln{\chi}{\psi})[\eval]$ 

\eqref{right-apply-to-lsc}: Assume that
$\cA \models (\vp \to \psi )[\eval]$, hence  $\evalp(\vp)\se \evalp(\psi)$. It follows that
$\evalp(\appln{\vp}{\chi})=\evalp(\vp) \appls \evalp(\chi) \stackrel{\eqref{B-se-C-appls-D}}{\se}  \evalp(\psi) \appls \evalp(\chi) =
\evalp(\appln{\psi}{\chi})$, hence $\cA \models (\appln{\vp}{\chi} \to \appln{\psi}{\chi})[\eval]$.

\eqref{right-apply-dnd-lsc}: Assume that $\cA \models (\vp \dnd \psi )[\eval]$. Then, by Proposition~\ref{A-e-models}.\eqref{A-e-models-dnd-2}, 
$\cA \models (\vp \to \psi)[\eval]$ and $\cA \models (\psi \to \vp)[\eval]$. Apply now \eqref{right-apply-to-lsc}  to get that 
$\cA \models (\appln{\vp}{\chi} \to \appln{\psi}{\chi})[\eval]$ and $\cA \models (\appln{\psi}{\chi} \to \appln{\vp}{\chi})[\eval]$, hence, by applying again
Proposition~\ref{A-e-models}.\eqref{A-e-models-dnd-2}, $\cA \models (\appln{\vp}{\chi} \dnd \appln{\psi}{\chi})[\eval]$.

\eqref{left-apply-neg-lsc}: Assume that
$\cA \models (\vp \to \psi )[\eval]$, hence $\evalp(\vp)\se \evalp(\psi)$. It follows that
$\evalp(\appln{\chi}{\neg\psi})=\evalp(\chi)\appls \evalp(\neg\psi) = \evalp(\chi)\appls  C_A \evalp(\psi)
\stackrel{\eqref{B-se-C-appls-D-comp}}{\se} \evalp(\chi)\appls C_A \evalp(\vp)=
\evalp(\appln{\chi}{\neg\vp})$, hence $\cA \models (\appln{\chi}{\neg\psi}\to \appln{\chi}{\neg\vp})[\eval]$.

\eqref{left-apply-neg-dnd-lsc}: Assume that $\cA \models (\vp \dnd \psi )[\eval]$. Then, by Proposition~\ref{A-e-models}.\eqref{A-e-models-dnd-2}, 
$\cA \models (\vp \to \psi)[\eval]$ and $\cA \models (\psi \to \vp)[\eval]$. Apply now \eqref{left-apply-neg-lsc}  to get that 
$\cA \models (\appln{\chi}{\neg\psi}\to \appln{\chi}{\neg\vp})[\eval]$ and $\cA \models (\appln{\chi}{\neg\vp}\to \appln{\chi}{\neg\psi})[\eval]$, hence, by applying again
Proposition~\ref{A-e-models}.\eqref{A-e-models-dnd-2}, $\cA \models (\appln{\chi}{\neg\psi}\dnd \appln{\chi}{\neg\vp})[\eval]$.

\eqref{right-apply-neg-lsc}: Assume that
$\cA \models (\vp \to \psi )[\eval]$, hence $\evalp(\vp)\se \evalp(\psi)$. It follows that
$\evalp(\appln{\neg\psi}{\chi})=\evalp(\neg\psi) \appls\evalp(\chi) = C_A\evalp(\psi) \appls  \evalp(\chi)
\stackrel{\eqref{B-se-C-appls-D-comp}}{\se} C_A \evalp(\vp) \appls  \evalp(\chi)=\evalp(\neg\vp) \appls\evalp(\chi)=
\evalp(\appln{\neg\vp}{\chi})$, hence $\cA \models (\appln{\neg\psi}{\chi}\to \appln{\neg\vp}{\chi})[\eval]$.

\eqref{right-apply-neg-dnd-lsc}: Assume that $\cA \models (\vp \dnd \psi )[\eval]$. Then, by Proposition~\ref{A-e-models}.\eqref{A-e-models-dnd-2}, 
$\cA \models (\vp \to \psi)[\eval]$ and $\cA \models (\psi \to \vp)[\eval]$. Apply now \eqref{right-apply-neg-lsc}  to get that 
$\cA \models (\appln{\neg\psi}{\chi}\to \appln{\neg\vp}{\chi})[\eval]$ and $\cA \models (\appln{\neg\vp}{\chi}\to \appln{\neg\psi}{\chi})[\eval]$, hence, by applying again
Proposition~\ref{A-e-models}.\eqref{A-e-models-dnd-2}, $\cA \models (\appln{\neg\psi}{\chi} \dnd \appln{\neg\vp}{\chi})[\eval]$.
}

\bprop
Let $\Gamma$ be a set of patterns  and $\vp$, $\psi$, $\chi$ be patterns. Then
\begin{align}
\Gamma \lsc \vp \to \psi & \quad  \text{implies} \quad \Gamma \lsc \appln{\vp}{\chi} \to \appln{\psi}{\chi}, 
\label{framing-l-lsc-Gamma}\\
\Gamma \lsc \vp \dnd \psi & \quad  \text{implies} \quad \Gamma \lsc \appln{\vp}{\chi} \dnd \appln{\psi}{\chi}, 
\label{framing-l-dnd-lsc-Gamma}\\
\Gamma \lsc \vp \to \psi & \quad  \text{implies} \quad \Gamma \lsc \appln{\chi}{\vp}\to \appln{\chi}{\psi}, 
\label{framing-r-lsc-Gamma}\\
\Gamma \lsc \vp \dnd \psi & \quad  \text{implies} \quad \Gamma \lsc \appln{\chi}{\vp}\dnd \appln{\chi}{\psi}, 
\label{framing-r-dnd-lsc-Gamma}\\
\Gamma \lsc  \vp\to \psi &\quad  \text{implies} \quad  \Gamma \lsc \appln{\chi}{\neg\psi}\to \appln{\chi}{\neg\vp}, 
\label{left-apply-neg-lsc-Gamma}\\
\Gamma \lsc \vp\dnd \psi & \quad  \text{implies} \quad \Gamma \lsc \appln{\chi}{\neg\psi}\dnd \appln{\chi}{\neg\vp}, 
\label{left-apply-neg-dnd-lsc-Gamma}\\
 \Gamma \lsc \vp\to \psi & \quad  \text{implies} \quad \Gamma \lsc \appln{\neg\psi}{\chi}\to \appln{\neg\vp}{\chi},
\label{right-apply-neg-lsc-Gamma}\\
 \Gamma \lsc \vp\dnd \psi &  \quad  \text{implies} \quad \Gamma \lsc \appln{\neg\psi}{\chi}\dnd \appln{\neg\vp}{\chi}.
\label{right-apply-neg-sdnd-gsc-Gamma}
\end{align}
\eprop
\solution{
Apply Proposition~\ref{apply-left-right-to-dnd-lsc} and Lemma~\ref{lsc-lleq-Gamma}.\eqref{lsc-lleq-Gamma-lsc}.
}

\bprop\label{apply-left-right-to-dnd-lsc-predicate}
Let $\chi$ be a predicate pattern. Then for all patterns $\vp$, $\psi$, 
\begin{align}
\sigma \to (\vp\to \psi) \,\,  \lsc  & \,\,  \sigma \to (\appln{\chi}{\vp}\to \appln{\chi}{\psi}), 
\label{predicate-left-apply-to-lsc}\\
\sigma \to (\vp\to \psi) \,\,  \lsc  & \,\,  \sigma \to (\appln{\vp}{\chi}\to \appln{\psi}{\chi}). 
\label{predicate-right-apply-to-lsc}
\end{align}
\eprop
\solution{Apply \eqref{left-apply-to-lsc}, \eqref{right-apply-to-lsc} and 
Proposition~\ref{predicate-imp-rules-lsc-gsc-Gamma}.\eqref{predicate-imp-rules-lsc-gsc-1}.

}

\bprop
Let $\chi$ be a predicate pattern and  $\Gamma\cup \{\vp,\psi,\chi\}$ be a set of patterns
Then
\begin{align}
\Gamma \lsc \sigma \to (\vp\to \psi)  & \quad  \text{implies} \quad 
\Gamma \lsc \sigma \to (\appln{\chi}{\vp}\to \appln{\chi}{\psi}),
\label{predicate-framing-l-lsc-Gamma}\\
\Gamma \lsc \sigma \to (\vp\to \psi)  & \quad  \text{implies} \quad 
\Gamma \lsc \sigma \to (\appln{\vp}{\chi}\to \appln{\psi}{\chi}). 
\label{predicate-framing-r-lsc-Gamma}
\end{align}
\eprop
\solution{
Apply \eqref{left-apply-to-lsc}, \eqref{right-apply-to-lsc} and 
Proposition~\ref{predicate-imp-rules-lsc-gsc-Gamma}.\eqref{predicate-imp-rules-lsc-gsc-1-Gamma}.
}

\subsubsection{$\gsc$}

\bprop\label{apply-left-right-to-dnd-gsc}
For all patterns $\vp$, $\psi$, $\chi$, 
\begin{align}
 \vp\to \psi \,\,  \gsc  & \,\, \appln{\chi}{\vp}\to \appln{\chi}{\psi}, 
\label{left-apply-to-gsc}\\
 \vp\dnd \psi \,\, \gsc & \,\,  \appln{\chi}{\vp}\dnd \appln{\chi}{\psi}, 
\label{left-apply-dnd-gsc}\\
 \vp\to \psi \,\, \gsc  & \,\, \appln{\vp}{\chi}\to \appln{\psi}{\chi},
\label{right-apply-to-gsc}\\
 \vp\dnd \psi \,\, \gsc &  \,\, \appln{\vp}{\chi}\dnd \appln{\psi}{\chi}, 
\label{right-apply-dnd-gsc}\\
 \vp \to \psi \,\, \gsc  &  \,\, \appln{\chi}{\neg\psi}\to \appln{\chi}{\neg\vp}, 
\label{left-apply-neg-gsc}\\
\vp \dnd \psi\,\,  \gsc & \,\,  \appln{\chi}{\neg\psi}\dnd \appln{\chi}{\neg\vp}, 
\label{left-apply-neg-dnd-gsc}\\
 \vp \to \psi \,\, \gsc  & \,\, \appln{\neg\psi}{\chi}\to \appln{\neg\vp}{\chi},
\label{right-apply-neg-gsc}\\
 \vp \dnd \psi \,\,  \gsc & \,\,  \appln{\neg\psi}{\chi}\dnd \appln{\neg\vp}{\chi}.
\label{right-apply-neg-dnd-gsc}
\end{align}
\eprop
\solution{Apply Proposition~\ref{apply-left-right-to-dnd-lsc} and 
Proposition~\ref{remark-conseq-equiv-to-dnd}.\eqref{remark-conseq-equiv-to-imp-dnd}.
}

\bprop
Let $\Gamma$ be a set of patterns  and $\vp$, $\psi$, $\chi$ be patterns. Then
\begin{align}
\Gamma \gsc \vp \to \psi & \quad  \text{implies} \quad \Gamma \gsc \appln{\vp}{\chi} \to \appln{\psi}{\chi}, 
\label{framing-l-gsc-Gamma}\\
\Gamma \gsc \vp \dnd \psi & \quad  \text{implies} \quad \Gamma \gsc \appln{\vp}{\chi} \dnd \appln{\psi}{\chi}, 
\label{framing-l-dnd-gsc-Gamma}\\
\Gamma \gsc \vp \to \psi & \quad  \text{implies} \quad \Gamma \gsc \appln{\chi}{\vp}\to \appln{\chi}{\psi}, 
\label{framing-r-gsc-Gamma}\\
\Gamma \gsc \vp \dnd \psi & \quad  \text{implies} \quad \Gamma \gsc \appln{\chi}{\vp}\dnd \appln{\chi}{\psi}, 
\label{framing-r-dnd-gsc-Gamma}\\
\Gamma \gsc  \vp\to \psi &\quad  \text{implies} \quad  \Gamma \gsc \appln{\chi}{\neg\psi}\to \appln{\chi}{\neg\vp}, 
\label{left-apply-neg-gsc-Gamma}\\
\Gamma \gsc \vp\dnd \psi & \quad  \text{implies} \quad \Gamma \gsc \appln{\chi}{\neg\psi}\dnd \appln{\chi}{\neg\vp}, 
\label{left-apply-neg-dnd-gsc-Gamma}\\
 \Gamma \gsc \vp\to \psi & \quad  \text{implies} \quad \Gamma \gsc \appln{\neg\psi}{\chi}\to \appln{\neg\vp}{\chi},
\label{right-apply-neg-gsc-Gamma}\\
 \Gamma \gsc \vp\dnd \psi &  \quad  \text{implies} \quad \Gamma \gsc \appln{\neg\psi}{\chi}\dnd \appln{\neg\vp}{\chi}.
\label{right-apply-neg-dnd-gsc-Gamma}
\end{align}
\eprop
\solution{Apply Proposition~\ref{apply-left-right-to-dnd-gsc} and 
Lemma~\ref{gsc-gleq-Gamma}.\eqref{gsc-gleq-Gamma-gsc}.
}

\bprop\label{apply-left-right-to-dnd-gsc-predicate}
Let $\sigma$ be a predicate pattern. Then for all patterns $\vp$, $\psi$, $\chi$, 
\begin{align}
\sigma \to (\vp\to \psi) \,\,  \gsc  & \,\,  \sigma \to (\appln{\chi}{\vp}\to \appln{\chi}{\psi}), 
\label{predicate-left-apply-to-gsc}\\
\sigma \to (\vp\to \psi) \,\,  \gsc  & \,\,  \sigma \to (\appln{\vp}{\chi}\to \appln{\psi}{\chi}). 
\label{predicate-right-apply-to-gsc}
\end{align}
\eprop
\solution{Apply \eqref{left-apply-to-lsc}, \eqref{right-apply-to-lsc} and 
Proposition~\ref{predicate-imp-rules-lsc-gsc-Gamma}.\eqref{predicate-imp-rules-lsc-gsc-2}.

}

\bprop\label{framing-predicate-pattern}
Let $\sigma$ be a predicate pattern and  $\Gamma\cup \{\vp,\psi,\chi\}$ be a set of patterns
Then
\begin{align}
\Gamma \gsc \sigma \to (\vp\to \psi)  & \quad  \text{implies} \quad 
\Gamma \gsc \sigma \to (\appln{\chi}{\vp}\to \appln{\chi}{\psi}),
\label{predicate-framing-l-gsc-Gamma}\\
\Gamma \gsc \sigma \to (\vp\to \psi)  & \quad  \text{implies} \quad 
\Gamma \gsc \sigma \to (\appln{\vp}{\chi}\to \appln{\psi}{\chi}). 
\label{predicate-framing-r-gsc-Gamma}
\end{align}
\eprop
\solution{
Apply \eqref{left-apply-to-lsc}, \eqref{right-apply-to-lsc} and 
Proposition~\ref{predicate-imp-rules-lsc-gsc-Gamma}.\eqref{predicate-imp-rules-lsc-gsc-2-Gamma}.
}

%% file: AML-semantics-examples-contexts.tex
\subsection{Contexts}

\bprop\label{def-evaluation-contexts}
Let $C$ be a context, $\vp$ be a pattern, and $(\cA,\eval)$. 
\begin{align}
\evalp(C[\vp]) & = \begin{cases} \evalp(\vp) & \text{if~} C=\Box,\\
\evalp(C_1[\vp])\appls \evalp(\psi) & \text{if~} C=\ApplC C_1\psi,\\
\evalp(\psi) \appls \evalp(C_1[\vp]) & \text{if~} \ApplC \psi C_1.
\end{cases}
\end{align}
\eprop
\solution{Apply Proposition~\ref{unique-read-contexts}. We have the following cases: 
\be
\item $C=\Box$. Then $C[\vp]=\vp$, so $\evalp(C[\vp])=\evalp(\vp)$.
\item $C=\ApplC C_1\psi$. Then $C[\vp]=\appln{C_1[\vp]}{\psi}$, hence
$\evalp(C[\vp])= \evalp(C_1[\vp])\appls \evalp(\psi)$.
\item $C=\ApplC \psi C_1$. Then $C[\vp]=\appln{\psi}{C_1[\vp]}$, hence 
$\evalp(C[\vp])=\evalp(\psi) \appls \evalp(C_1[\vp])$. 
\ee
}

\bprop
Let $C$ be a context and $\vp$, $\psi$ be patterns.
\begin{align}
& \models C[\bot] \dnd \bot, \label{context-bot}\\
& \models C[\vp \vee \psi] \dnd C[\vp] \vee C[\psi], \label{context-vee}\\
& \models C[\exists x. \vp] \dnd \exists x. C[\vp] 
\quad \text{ if } x \notin FV(C), \label{context-exists}\\
& \models C[\vp \si \psi] \to C[\vp] \si C[\psi], \label{context-si}\\
& \models C[\forall x. \vp] \to \forall x. C[\vp] 
\quad \text{ if } x \notin FV(C). \label{context-forall}
\end{align}
\eprop
\solution{ 
\eqref{context-bot}: The proof is by  induction on contexts:
\be
\item $C=\Box$. Then $C[\bot]=\bot$. Apply \eqref{dnd-reflexive}. 
\item $C=\ApplC C_1\psi$ and, by the induction hypothesis for $C_1$, $\models C_1[\bot] \dnd \bot$. Since 
$C[\bot]= \appln{C_1[\bot]}{\psi}$, we can apply \eqref{right-apply-dnd-gsc} to get that 
$C_1[\bot] \dnd \bot \gsc C[\bot]\dnd \appln{\bot}{\psi}$ and Remark~\ref{gsc-gleq-models}.\eqref{gsc-gleq-models-to} to get that 
$\models C[\bot] \dnd  \appln{\bot}{\psi}$. Hence, by \eqref{propagation-r} and \eqref{dnd-tranzitive}, it follows that 
$\models C[\bot] \dnd \bot$.
\item $C=\ApplC \psi C_1$ and, by the induction hypothesis for $C_1$, $\models C_1[\bot] \dnd \bot$. Since 
$C[\bot]= \appln{\psi}{C_1[\bot]}$, we can apply \eqref{left-apply-dnd-gsc} to get that $C_1[\bot] \dnd \bot \gsc C[\bot]\dnd \appln{\psi}{\bot}$
and Remark~\ref{gsc-gleq-models}.\eqref{gsc-gleq-models-to} to get that 
$\models C[\bot] \dnd  \appln{\psi}{\bot}$. Hence, by \eqref{propagation-l} and \eqref{dnd-tranzitive}, it follows that 
$\models C[\bot] \dnd \bot$. 
\ee

\eqref{context-vee}: The proof is by  induction on contexts:
\be
\item $C=\Box$. Then $C[\vp \vee \psi]=\vp \vee \psi= C[\vp] \vee C[\psi]$. Apply \eqref{dnd-reflexive}. 
\item $C=\ApplC C_1\chi$ and, by the induction hypothesis for $C_1$, 
$\models C_1[\vp \vee \psi] \dnd C_1[\vp] \vee C_1[\psi] $. 
Since 
$C[\vp \vee \psi]= \appln{C_1[\vp \vee \psi]}{\chi}$, we can apply \eqref{right-apply-dnd-gsc} to get that 
$C_1[\vp \vee \psi] \dnd C_1[\vp] \vee C_1[\psi] \gsc  C[\vp \vee \psi]\dnd \appln{(C_1[\vp] \vee C_1[\psi])}{\chi}$
and Remark~\ref{gsc-gleq-models}.\eqref{gsc-gleq-models-to} to get that 
$\models C[\vp \vee \psi] \dnd  \appln{(C_1[\vp] \vee C_1[\psi])}{\chi}$. 
Hence, by \eqref{propagation-sau-l} and \eqref{dnd-tranzitive}, it follows that 
$\models C[\vp \vee \psi] \dnd \appln{C_1[\vp]}{\chi} \vee \appln{C_2[\vp]}{\chi}=C[\vp] \vee C[\psi]$.

\item $C=\ApplC \chi C_1$ and, by the induction hypothesis for $C_1$, 
$\models C_1[\vp \vee \psi] \dnd C_1[\vp] \vee C_1[\psi] $. 
Since 
$C[\vp \vee \psi]= \appln{\chi}{C_1[\vp \vee \psi]}$, we can apply \eqref{left-apply-dnd-gsc} to get that 
$C_1[\vp \vee \psi] \dnd C_1[\vp] \vee C_1[\psi] \gsc  C[\vp \vee \psi]\dnd \appln{\chi}{(C_1[\vp] \vee C_1[\psi])}$
and Remark~\ref{gsc-gleq-models}.\eqref{gsc-gleq-models-to} to get that 
$\models C[\vp \vee \psi] \dnd  \appln{\chi}{(C_1[\vp] \vee C_1[\psi])}$. 
Hence, by \eqref{propagation-sau-r} and \eqref{dnd-tranzitive}, it follows that 
$\models C[\vp \vee \psi] \dnd \appln{\chi}{C_1[\vp]} \vee \appln{\chi}{C_2[\vp]} = C[\vp] \vee C[\psi]$.
\ee

\eqref{context-exists}: The proof is by  induction on contexts:
\be
\item $C=\Box$. Then $C[\exists x. \vp] =\exists x. \vp= \exists x. C[\vp]$. Apply  \eqref{dnd-reflexive}.
\item $C=\ApplC C_1\psi$. As $x \notin FV(C)$ and, by Definition~\ref{def-free-var-contexts}, 
$FV(C)=FV(C_1)\cup FV(\vp)$, 
we have that $x\notin FV(C_1)$, so we can apply the induction hypothesis for $C_1$
to get that $\models C_1[\exists x. \vp] \dnd \exists x. C_1[\vp]$. 
Since  $C[\exists x. \vp]=\appln{C_1[\exists x. \vp]}{\psi}$, we can apply 
\eqref{right-apply-dnd-gsc} to get that 
$C_1[\exists x. \vp] \dnd \exists x. C_1[\vp] \gsc C[\exists x. \vp] \dnd  \appln{(\exists x. C_1[\vp])}{\psi}$
and Remark~\ref{gsc-gleq-models}.\eqref{gsc-gleq-models-to} to get that 
$\models C[\exists x. \vp] \dnd   \appln{(\exists x. C_1[\vp])}{\psi}$. 
Hence, by \eqref{propagation-exists-l} and \eqref{dnd-tranzitive}, it follows that 
$\models C[\exists x. \vp] \dnd \exists x. \appln{ C_1[\vp]}{\psi}=\exists x. C[\vp]$. 
\item $C=\ApplC \psi C_1$. As $x \notin FV(C)$ and, by Definition~\ref{def-free-var-contexts}, 
$FV(C)=FV(C_1)\cup FV(\vp)$, 
we have that $x\notin FV(C_1)$, so we can apply the induction hypothesis for $C_1$
to get that $\models C_1[\exists x. \vp] \dnd \exists x. C_1[\vp]$. 
Since  $C[\exists x. \vp]=\appln{\psi}{C_1[\exists x. \vp]}$, we can apply 
\eqref{left-apply-dnd-gsc} to get that 
$C_1[\exists x. \vp] \dnd \exists x. C_1[\vp] \gsc C[\exists x. \vp] \dnd  \appln{\psi}{(\exists x. C_1[\vp])}$
and Remark~\ref{gsc-gleq-models}.\eqref{gsc-gleq-models-to} to get that 
$\models C[\exists x. \vp] \dnd  \appln{\psi}{(\exists x. C_1[\vp])}$. 
Hence, by \eqref{propagation-exists-r} and \eqref{dnd-tranzitive}, it follows that 
$\models C[\exists x. \vp] \dnd  \exists x.\appln{\psi}{C_1[\vp]}=\exists x. C[\vp]$. 
\ee

\eqref{context-si}: The proof is by  induction on contexts:
\be
\item $C=\Box$. Then $C[\vp \si \psi]=\vp \si \psi= C[\vp] \si C[\psi]$. Apply \eqref{dnd-reflexive} and 
Remark~\ref{valid-basic}.\eqref{valid-basic-dnd}. 

\item $C=\ApplC C_1\chi$ and, by the induction hypothesis for $C_1$, 
$\models C_1[\vp \si \psi] \to C_1[\vp] \si C_1[\psi] $. 
Since 
$C[\vp \si \psi]= \appln{C_1[\vp \si \psi]}{\chi}$, we can apply \eqref{right-apply-to-gsc} to get that
$C_1[\vp \si \psi] \to C_1[\vp] \si C_1[\psi] \gsc C[\vp \si \psi] \to \appln{(C_1[\vp] \si C_1[\psi])}{\chi}$
and Remark~\ref{gsc-gleq-models}.\eqref{gsc-gleq-models-to}
 to get that 
$\models C[\vp \si \psi] \to  \appln{(C_1[\vp] \si C_1[\psi])}{\chi}$. 
Hence, by \eqref{propagation-si-l} and \eqref{to-tranzitive}, it follows that 
$\models C[\vp \si \psi] \to \appln{C_1[\vp]}{\chi} \si \appln{C_2[\psi]}{\chi} = C[\vp] \si C[\psi]$.
\item $C=\ApplC \chi C_1$ and, by the induction hypothesis for $C_1$, 
$\models C_1[\vp \si \psi] \to C_1[\vp] \si C_1[\psi] $. 
Since 
$C[\vp \si \psi]= \appln{\chi}{C_1[\vp \si \psi]}$, we can apply \eqref{left-apply-to-gsc} 
to get that
$C_1[\vp \si \psi] \to C_1[\vp] \si C_1[\psi] \gsc C[\vp \si \psi] \to \appln{\chi}{(C_1[\vp] \si C_1[\psi])}$
and Remark~\ref{gsc-gleq-models}.\eqref{gsc-gleq-models-to}
to get that 
$\models C[\vp \si \psi] \to  \appln{\chi}{(C_1[\vp] \si C_1[\psi])}$. 
Hence, by \eqref{propagation-si-r} and \eqref{to-tranzitive}, it follows that 
$\models C[\vp \si \psi] \to \appln{\chi}{C_1[\vp]} \si \appln{\chi}{C_2[\psi]} = C[\vp] \si C[\psi]$.
\ee

\eqref{context-forall}: The proof is by  induction on contexts:
\be
\item $C=\Box$. Then $C[\forall x. \vp] =\forall x. \vp= \forall x. C[\vp]$. Apply  \eqref{dnd-reflexive} and 
Remark~\ref{valid-basic}.\eqref{valid-basic-dnd}. 
\item $C=\ApplC C_1\psi$. As $x \notin FV(C)$ and, by Definition~\ref{def-free-var-contexts}, 
$FV(C)=FV(C_1)\cup FV(\vp)$, 
we have that $x\notin FV(C_1)$, so we can apply the induction hypothesis for $C_1$
to get that $\models C_1[\forall x. \vp] \to \forall x. C_1[\vp]$. 
Since  $C[\forall x. \vp]=\appln{C_1[\forall x. \vp]}{\psi}$, we can apply 
\eqref{right-apply-to-gsc} to get that 
$C_1[\forall x. \vp] \to \forall x. C_1[\vp] \gsc C[\forall x. \vp] \to \appln{(\forall x. C_1[\vp])}{\psi}$ 
and Remark~\ref{gsc-gleq-models}.\eqref{gsc-gleq-models-to} to get that 
$\models C[\forall x. \vp] \to   \appln{(\forall x. C_1[\vp])}{\psi}$. 
Hence, by \eqref{propagation-forall-l} and \eqref{to-tranzitive}, it follows that 
$\models C[\forall x. \vp] \to \forall x. \appln{ C_1[\vp]}{\psi}=\forall x. C[\vp]$. 
\item $C=\ApplC \psi C_1$. As $x \notin FV(C)$ and, by Definition~\ref{def-free-var-contexts}, 
$FV(C)=FV(C_1)\cup FV(\vp)$, 
we have that $x\notin FV(C_1)$, so we can apply the induction hypothesis for $C_1$
to get that $\models C_1[\forall x. \vp] \to \forall x. C_1[\vp]$. 
Since  $C[\forall x. \vp]=\appln{\psi}{C_1[\forall x. \vp]}$, we can apply 
\eqref{left-apply-to-gsc} to get that 
$C_1[\forall x. \vp] \to \forall x. C_1[\vp] \gsc C[\forall x. \vp] \to \appln{\psi}{(\forall x. C_1[\vp])}$ 
and 
Remark~\ref{gsc-gleq-models}.\eqref{gsc-gleq-models-to} to get that 
$\models C[\forall x. \vp] \to  \appln{\psi}{(\forall x. C_1[\vp])}$. 
Hence, by \eqref{propagation-forall-r} and \eqref{to-tranzitive}, it follows that 
$\models C[\forall x. \vp] \to  \forall x.\appln{\psi}{C_1[\vp]}=\forall x. C[\vp]$. 
\ee
}

\subsubsection{$\lsc$}

\bprop\label{framing-context-left-right-to-dnd-lsc}
Let $C$ be a context. Then for every patterns, $\vp$, $\psi$, 
\begin{align}
\vp \to \psi \,\,  \lsc  & \,\, C[\vp] \to C[\psi], \label{framing-context-lsc}\\
\vp \dnd \psi\,\,  \lsc  & \,\, C[\vp] \dnd C[\psi]. \label{framing-context-dnd-lsc}
\end{align}
\eprop
\solution{
\eqref{framing-context-lsc}: Let $(\cA,\eval)$ be such that $\cA \models (\vp \to \psi)[\eval]$. We prove that 
 $\cA \models (C[\vp] \to C[\psi])[\eval]$ by induction on $C$.

\be
\item $C=\Box$. Then $C[\vp] \to C[\psi] =\vp \to \psi$ and the conclusion is obvious.
\item $C=\ApplC C_1\chi$. By the induction hypothesis, 
we have that $\cA \models (C_1[\vp] \to C_1[\psi])[\eval]$. Apply now \eqref{right-apply-to-lsc} to get that 
$\cA \models (\appln{C_1[\vp]}{\chi} \to \appln{C_1[\psi]}{\chi})[\eval]$, that is $\cA \models (C[\vp] \to C[\psi])[\eval]$.
\item $C=\ApplC \chi C_1$. By the induction hypothesis, 
we have that $\cA \models (C_1[\vp] \to C_1[\psi])[\eval]$.   Apply now \eqref{left-apply-to-lsc} to get that 
$\cA \models (\appln{\chi} {C_1[\vp]}\to \appln{\chi}{C_1[\psi]})[\eval]$, that is $\cA \models (C[\vp] \to C[\psi])[\eval]$.
\ee

\eqref{framing-context-dnd-lsc}: Apply \eqref{framing-context-lsc} and 
Proposition~\ref{A-e-models}.\eqref{A-e-models-dnd-2}.
}

\bprop
Let $C$ be a context, $\Gamma$ be a set of patterns and $\vp$, $\psi$ be patterns. Then
\begin{align}
\Gamma \lsc \vp \to \psi & \quad \text{implies} \quad \Gamma \lsc C[\vp] \to C[\psi], 
\label{framing-context-Gamma-lsc}\\
\Gamma \lsc \vp\dnd \psi & \quad \text{implies} \quad \Gamma \lsc C[\vp] \dnd C[\psi]. 
\label{framing-context-dnd-Gamma-lsc}
\end{align}
\eprop
\solution{Apply Proposition~\ref{framing-context-left-right-to-dnd-lsc} and 
Lemma~\ref{lsc-lleq-Gamma}.\eqref{lsc-lleq-Gamma-lsc}.
}

\subsubsection{$\gsc$}

\bprop\label{framing-context-left-right-to-dnd-gsc}
Let $C$ be a context. Then for every patterns, $\vp$, $\psi$, 
\begin{align}
\vp \to \psi \,\,  \gsc  & \,\, C[\vp] \to C[\psi], \label{framing-context-gsc}\\
\vp \dnd \psi\,\,  \gsc  & \,\, C[\vp] \dnd C[\psi]. \label{framing-context-dnd-gsc}
\end{align}
\eprop
\solution{Apply Proposition~\ref{framing-context-left-right-to-dnd-lsc} and 
Proposition~\ref{remark-conseq-equiv-to-dnd}.\eqref{remark-conseq-equiv-to-imp-dnd}.
}

\bprop
Let $C$ be a context, $\Gamma$ be a set of patterns and $\vp$, $\psi$ be patterns. Then
\begin{align}
\Gamma \gsc \vp \to \psi & \quad \text{implies} \quad \Gamma \gsc C[\vp] \to C[\psi], 
\label{framing-context-Gamma-gsc}\\
\Gamma \gsc \vp\dnd \psi & \quad \text{implies} \quad \Gamma \gsc C[\vp] \dnd C[\psi]. 
\label{framing-context-dnd-Gamma-gsc}
\end{align}
\eprop
\solution{Apply Proposition~\ref{framing-context-left-right-to-dnd-gsc} and 
Lemma~\ref{gsc-gleq-Gamma}.\eqref{gsc-gleq-Gamma-gsc}.
}

\subsubsection{\prule{Singleton Variable}}

\blem\label{vp-empty-Cvp-empty}
Let  $C$ be a context and $\vp$ be a pattern. Then for every $(\cA, \eval)$,
$$
\evalp(\vp)=\emptyset \quad  \text{implies} \quad \evalp(C[\vp])=\emptyset.
$$
\elem
\solution{Since $\evalp(\vp)=\emptyset=\evalp(\bot)$, we have that $\cA \models (\vp\dnd \bot)[\eval]$. Apply 
\eqref{framing-context-dnd-lsc} to get that $\cA \models (C[\vp]\dnd C[\bot])[\eval]$, so $\evalp(C[\vp])=\evalp(C[\bot])$.
As, by  \eqref{context-bot}, $\models C[\bot]\dnd \bot$, it follows that $\evalp(C[\bot])=\evalp(\bot)=\emptyset$. Thus, 
we have got that $\evalp(C[\vp])=\emptyset$.
}

\bprop
For every contexts $C_1$, $C_2$, pattern $\vp$ and element variable $x$,
\begin{align}
\models \neg (C_1[x \si \vp] \si C_2[x \si \neg \vp]). \label{Singleton-Variable}
\end{align}
\eprop
\solution{Denote $\theta=C_1[x \si \vp] \si C_2[x \si \neg \vp]$. 
Let  $(\cA, \eval)$. We have to prove that $\cA\models (\neg\theta)[\eval]$, that is
$\evalp(\neg\theta)=A$, hence that $\evalp(\theta)=\emptyset$.
We have two cases:
\be
\item $\eval(x)\in \evalp(\vp)$. Then 
$\evalp(x \si \neg \vp)=\evalp(x)\cap \evalp(\neg\vp)= 
\{\eval(x)\} \cap C_A\evalp(\vp)=\emptyset$. 
Apply Lemma~\ref{vp-empty-Cvp-empty} to get that $\evalp(C_2[x \si \neg \vp])=\emptyset$. It follows that 
$\evalp(\theta)=\evalp(C_1[x \si \vp])\cap \evalp(C_2[x \si \neg \vp]) = \evalp(C_1[x \si \vp])\cap \emptyset =\emptyset$.
\item $\eval(x)\notin \evalp(\vp)$. Then 
$\evalp(x \si \vp)=\{\eval(x)\} \cap \evalp(\vp)=\evalp(x)\cap \evalp(\vp)=\emptyset$. 
Apply Lemma~\ref{vp-empty-Cvp-empty} to get that $\evalp(C_1[x \si \vp])=\emptyset$. It follows that 
$\evalp(\theta)=\evalp(C_1[x \si \vp])\cap \evalp(C_2[x \si \neg \vp]) =  \emptyset \cap \evalp(C_2[x \si \neg \vp])
 =\emptyset$.
\ee
}

%% file: AML-semantics-examples-set-variables.tex
\subsection{Set variables}

Let $\Gamma$ be a set of patterns.

\subsubsection{$\gsc$}

\bprop\label{set-variable-substitution-prop-1}
Let $\vp, \psi$ be patterns and $X$  be a set variable 
such that $X$ is free for $\psi$ in $\vp$. Then 
\begin{align*}
\vp & \gsc \subfX{\psi}{\vp}.
\end{align*}
\eprop
\solution{Let $\cA$ be  a model of $\vp$ and $\eval$ be an arbitrary $\cA$-evaluation. We have to prove that 
 $\cA\models \subfX{\psi}{\vp}[\eval]$. By Proposition~\ref{eval-subfvpXdelta}, we 
 have that $\evalp(\subfX{\psi}{\vp})=\evalpX{\evalp(\psi)}(\vp)$. 
Since  $\cA\models \vp$, we have in particular that
$\cA \models \vp[\evalX{\evalp(\psi)}]$. Thus, $\cA\models \subfX{\psi}{\vp}[\eval]$.
}

\bprop\label{set-variable-substitution-prop-2}
Let $\Gamma\cup\{\vp, \psi\}$ be a set of patterns and $X$  be a set variable 
such that $X$ is free for $\psi$ in $\vp$. Then 

\begin{align*}
\Gamma \gsc \vp & \quad  \text{implies} \quad  \Gamma \gsc \subfX{\psi}{\vp}.
\end{align*}
\eprop
\solution{Apply Proposition~\ref{set-variable-substitution-prop-1} and 
Lemma~\ref{gsc-gleq-Gamma}.\eqref{gsc-gleq-Gamma-gsc}.
}

\bprop\label{set-variable-substitution-free-1}
For all patterns $\vp$, $\psi$ and any set variable $X$,
\begin{align*}
 \vp &  \gsc \fsubvpX{\psi}. 
\end{align*}
\eprop
\solution{
We have the following two cases:
\be
\item $X$ is free for $\psi$ in $\vp$. Then $\fsubvpX{\psi}=\subfX{\psi}{\vp}$, so we apply 
Proposition~\ref{set-variable-substitution-prop-1} to get the conclusion.
\item $X$ is not free for $\psi$ in $\vp$. Then $\fsubvpX{\psi}=\subfX{\psi}{\theta}$ with
$$\theta=\subb{U_1}{Z_1}\subb{U_2}{Z_2}\ldots \subb{U_p}{Z_p}\subb{u_1}{z_1}
\subb{u_2}{z_2}\ldots \subb{u_k}{z_k}\vp,$$
where
\be
\item $u_1,\ldots, u_k$ are the element variables and $U_1,\ldots, U_p$ are the set variables that occur bound in $\vp$ and also 
occur in $\psi$;
\item  $z_1, \ldots, z_k$ are new element variables and $Z_1, \ldots, Z_p$  are 
new set variables, that do not occur in $\vp$ or $\psi$. 
\ee
Applying repeatedly Propositions~\ref{bounded-subst-equiv-x}, \ref{bounded-subst-equiv-X}, we get that 
\begin{equation}\label{fsX-1}
\models \theta \dnd\vp.
\end{equation}  

Let $\cA$ be  a model of $\vp$. By  \eqref{fsX-1}, we get that $\cA\models \theta$. 
Apply Proposition~\ref{set-variable-substitution-prop-1} (with $\vp=\theta$) to get that 
$\cA\models \subfX{\psi}{\theta}=\fsubvpX{\psi}$.
\ee
}

\bprop\label{set-variable-substitution-free-2}
For all patterns $\vp$, $\psi$,
\begin{align*}
\Gamma \gsc \vp &  \quad \text{implies} \quad  \Gamma \gsc \fsubvpX{\psi}. 
\end{align*}
\eprop
\solution{Apply Proposition~\ref{set-variable-substitution-free-1} and 
Lemma~\ref{gsc-gleq-Gamma}.\eqref{gsc-gleq-Gamma-gsc}.
}

%% file: AML-semantics-examples-mu-nu.tex
\subsection{$\mu$}

\bprop
Let  $\vp$  be a pattern and $X$ be  a set variable  such that  $\vp$ is positive in $X$ and $X$ is free for $\mu X.\vp$ in $\vp$. Then
\begin{align}
\models \subfX{\mu X. \vp}{\vp} \dnd  \mu X.\vp. 
\label{pre-fixpoint-dnd}
\end{align}
\eprop
\solution{Let $(\cA,\eval)$. We have that 
\begin{align*}
\evalp\left(\subfX{\mu X. \vp}{\vp}\right) &= \evalpX{\evalp(\mu X. \vp)}(\vp) \quad \text{by Proposition~\ref{eval-subfvpXdelta}}\\
& = \evalp(\mu X. \vp)  \quad \text{by Proposition~\ref{fkntar-fp}.\eqref{fkntar-mu-nu-fp}}
\end{align*}
}	

\subsubsection{$\lsc$}

\bprop\label{Knaster-Tarski-rule-lsc-prop-1}
Let $\vp, \psi$ be patterns and $X$  be a set variable 
such that $X$ is free for $\psi$ in $\vp$. Then 
\begin{align*}
 \subfX{\psi}{\vp} \to \psi & \lsc  \mu X.\vp \to \psi.
\end{align*}
\eprop
\solution{Let $(\cA,\eval)$ be such that $\cA \models (\subfX{\psi}{\vp} \to \psi)[\eval]$, so 
$\evalp\left(\subfX{\psi}{\vp}\right)\se \evalp(\psi)$. 
By Propositions~\ref{eval-subfvpXdelta}, we have that 
$\evalp\left(\subfX{\psi}{\vp}\right)=\evalpX{\evalp(\psi)}(\vp)$. We get that 
$\evalpX{\evalp(\psi)}(\vp)\se \evalp(\psi)$, hence 
$\evalp(\psi)\in \{ B \subseteq A \mid \evalpX{B}(\vp)\subseteq B\}$. It follows
that 
\begin{align*}
\evalp(\mu X.\vp)=\bigcap \{ B \subseteq A \mid \evalpX{B}(\vp) \subseteq B\}\se \evalp(\psi), 
\end{align*}
so $\cA \models (\mu X.\vp \to \psi)[\eval]$. 
}

\bprop\label{Knaster-Tarski-rule-lsc-prop-2}
Let $\Gamma\cup\{\vp, \psi\}$ be a set of patterns and $X$  be a set variable 
such that $X$ is free for $\psi$ in $\vp$. Then 
\begin{align*}
\Gamma \lsc \subfX{\psi}{\vp} \to \psi & \quad \text{implies} \quad \Gamma \lsc \mu X.\vp \to \psi.
\end{align*}
\eprop
\solution{Apply Proposition~\ref{Knaster-Tarski-rule-lsc-prop-1} and 
Lemma~\ref{lsc-lleq-Gamma}.\eqref{lsc-lleq-Gamma-lsc}.
}

\subsubsection{$\gsc$}

\bprop\label{Knaster-Tarski-rule-gsc-prop-1}
Let $\vp, \psi$ be patterns and $X$  be a set variable 
such that $X$ is free for $\psi$ in $\vp$. Then 
\begin{align*}
 \subfX{\psi}{\vp} \to \psi & \gsc  \mu X.\vp \to \psi.
\end{align*}
\eprop
\solution{Apply Proposition~\ref{Knaster-Tarski-rule-lsc-prop-1} and Proposition~\ref{remark-conseq-equiv-to-dnd}.\eqref{remark-conseq-equiv-to-imp-dnd}.
}

\bprop\label{Knaster-Tarski-rule-gsc-prop-2}
Let $\Gamma\cup\{\vp, \psi\}$ be a set of patterns and $X$  be a set variable 
such that $X$ is free for $\psi$ in $\vp$. Then 
\begin{align*}
\Gamma \gsc \subfX{\psi}{\vp} \to \psi & \quad \text{implies} \quad \Gamma \gsc \mu X.\vp \to \psi.
\end{align*}
\eprop
\solution{Apply Proposition~\ref{Knaster-Tarski-rule-gsc-prop-1} and 
Lemma~\ref{gsc-gleq-Gamma}.\eqref{gsc-gleq-Gamma-gsc}.
}

%% file: AML-syntax.tex
\section{Proof system}

Let $\AMLsig=(EVar, SVar, \Sigma)$ be a signature.

We present in the sequel a Hilbert type proof system for AML, that we denote $\PSAML$, obtained 
by combining the ones from \cite{CheRos19} and \cite{BerCheHorMizPenTus22}. $\PSAML$ is also given in Section~\ref{AML-proof-system}.

\bdfn
The \defnterm{axioms} of $\PSAML$ are the following patterns \\[2mm]
\begin{tabular}{ll}
\prule{Tautology} & $\vp$ \quad if $\vp$ is a tautology\\[2mm]
\prule{$\exists$-Quantifier} & $\subfx{y}{\vp} \to \exists x.\varphi$ \quad  if $x$ is free 
for $y$ in $\vp$
\\[2mm]
\prule{Propagation$_\bot$} &
$\appln{\vp}{\bot} \to \bot$ \qquad $ \appln{\bot}{\vp} \to \bot$
\\[2mm]
\prule{Propagation$_\vee$} & 
$\appln{(\vp \sau\psi)}{\chi} \to \appln{\vp}{\chi}\sau \appln{\psi}{\chi}$ \qquad 
$\appln{\chi}{(\vp \sau\psi)} \to \appln{\chi}{\vp}\sau \appln{\chi}{\psi}$
\\[2mm]
\prule{Propagation$_\exists$} &
$\appln{(\exists x. \vp)}{\psi} \to \exists x. \appln{\vp}{\psi}$, \qquad 
$\appln{\psi}{(\exists x.\vp)} \to \appln{\exists x. \psi}{\vp}$\\[1mm]
& if $x \not\in FV(\psi)$
\\[2mm]
\prule{Pre-Fixpoint} &
$\subfX{\mu X. \vp}{\vp} \to  \mu X.\vp$ \quad  if $\vp$ is  positive in $X$ \\[1mm]
& \qquad \qquad \qquad \qquad\qquad and  $X$ is free for $\mu X. \vp$ in $\vp$
\\[2mm]
\prule{existence} &
$\exists x.x$,
\\[2mm]
\prule{Singleton Variable} &
$\neg (C_1[x \wedge \varphi] \wedge C_2[x \wedge \neg \varphi])$.
\et
\edfn

\bdfn
The \defnterm{deduction rules} (or \defnterm{inference rules}) of $\PSAML$ are the following: \\[2mm]
\begin{tabular}{ll}
\prule{modus ponens} &
$ \ds \frac{\vp \quad \vp \to \psi}{\psi}$
\\[5mm]
\prule{$\exists$-quantifier rule} &
$ \ds \frac{\vp \to \psi}{\exists x.\vp \to \psi}$  \quad if $x \not\in FV(\psi)$
\\[5mm]
\prule{framing} &
$ \ds \frac{\vp \to \psi}
{\appln{\vp}{\chi} \to \appln{\psi}{\chi}} \qquad \frac{\vp \to \psi}{\appln{\chi}{\vp}\to \appln{\chi}{\psi}}$
\\[5mm]
\prule{Set Variable Substitution} & 
$ \ds \frac{\vp}{\subfX{\psi}{\vp}}$ \quad   if $X$ is free for $\psi$ in $\vp$
\\[5mm]
\prule{Knaster-Tarski} &
$ \ds \frac{\subfX{\psi}{\vp} \to \psi }{\mu X.\vp \to \psi}$ \quad if $X$ is free for $\psi$ in $\vp$
\\[5mm]
\et
\edfn

Thus, we have two types of inference rules:
\be 
\item $\ds \frac{\vp}{\chi}$. 
 This inference rule is read as follows: from $\vp$ infer $\chi$. $\vp$ is said to be the \defnterm{premise} of the rule and $\chi$ is the 
\defnterm{conclusion} of the rule.
\item $\ds \frac{\vp \quad \psi}{\chi}$. 
This inference rule is read as follows: from $\vp$ and $\psi$ infer $\chi$. $\vp$ and $\psi$ are said to be the \defnterm{premise} of the rule and $\chi$ is the 
\defnterm{conclusion} of the rule.
\ee

\mbox{}

Let $\Gamma$ be a set of patterns.

\bdfn\label{def-Gamma-thms}
The  \defnterm{$\Gamma$-theorems}  are the patterns inductively defined as follows:
\be
\item Every  axiom is a $\Gamma$-theorem.
\item Every pattern of $\Gamma$ is a $\Gamma$-theorem.
\item For every inference rule, the following holds: If its premises  are  $\Gamma$-theorems, then its conclusion 
is a $\Gamma$-theorem.
\item Only the patterns  obtained by applying the above rules are $\Gamma$-theorems.
\ee
\edfn

\bdfn\label{def-Gamma-thms-equiv}[Alternative definition for $\Gamma$-theorems\,]\\
The set of \defnterm{$\Gamma$-theorems}  is the intersection of all 
sets $\Delta$ of patterns that have the following properties:
\be
\item $\Delta$ contains all the axioms.
\item $\Delta$ contains all the patterns from $\Gamma$, that is $\Gamma \se \Delta$.
\item For every inference rule, the following holds: If $\Delta$ contains its premise(s), then its conclusion is also in $\Delta$. 
\ee
\edfn

The set of $\Gamma$-theorems is denoted by $Thm(\Gamma)$. If $\varphi$ is a $\Gamma$-theorem, then we also say that $\vp$  is \defnterm{deduced from the hypotheses} $\Gamma$.

As an immediate consequence of Definition~\ref{def-Gamma-thms-equiv}, we get the induction principle for $\Gamma$-theorems.

\begin{proposition}\label{induction-pspat}[Induction principle on $\Gamma$-theorems\,]\\
Let $\Delta$  be a set of patterns  satisfying the following properties:
\be
\item $\Delta$ contains all the axioms.
\item $\Gamma \se \Delta$. 
\item For every inference rule, the following holds: If $\Delta$ contains its premise(s), then its conclusion is also in $\Delta$. 
\ee
Then $Thm(\Gamma)\se \Delta $.
\end{proposition}

\bntn
Let $\Gamma, \Delta$ be sets of patterns and $\vp$ be a pattern. We use the following notations\\[1mm]
\bt{lll}
$\Gamma\vdash \varphi$ &=& $\vp$ is a $\Gamma$-theorem.\\[2mm]
$\vdash \vp$ &= & $\emptyset \vdash \vp$,\\[1mm]
$\Gamma\vdash \Delta$ & $\Lra$ & $\Gamma\vdash \varphi$ for any $\vp\in\Delta$.
\et
\entn

\bdfn
A pattern $\vp$ is called a \defnterm{theorem} if $\vdash \vp$.
\edfn

\begin{proposition}\label{Gamma-thm-prop-basic}\vspace*{-2mm}
Let $\Gamma, \Delta$ be sets of patterns.
\be
\item\label{basic-Gamma-se-Delta}  Assume that $\Delta \subseteq \Gamma$. Then for every pattern $\vp$, $Thm(\Delta) \se Thm(\Gamma)$, that is
\bce $\Delta \vdash\vp$ implies  $\Gamma \vdash\vp$. 
\ece
\item\label{basic-Thm-se-Gamma-Thm}  For every pattern $\vp$,  $Thm(\emptyset) \se Thm(\Gamma)$, that is
\bce
$\vdash\vp$ implies $\Gamma\vdash\vp$.
\ece
\item\label{gamma-delta-thm} Assume that $\Gamma\vdash\Delta$. Then  for every pattern $\vp$, $Thm(\Delta) \se Thm(\Gamma)$, that is
\bce 
$\Delta\vdash \vp$ implies
$\Gamma\vdash\vp$.
\ece
\item For every pattern $\vp$, $Thm(Thm(\Gamma))=Thm(\Gamma)$, that is
\bce  
$Thm(\Gamma)\vdash \vp$ \ddaca $\Gamma \vdash \vp$.
\ece
\ee
\end{proposition}
\solution{
\be
\item  As $\Delta \subseteq \Gamma$,  one  proves immediately by induction on $\Delta$-theorems that $Thm(\Delta) \se Thm(\Gamma)$. 
\item Apply \eqref{basic-Gamma-se-Delta} with $\Delta=\emptyset$.
\item As, by hypothesis, $\Delta \subseteq Thm(\Gamma)$, one proves immediately by induction on $\Delta$-theorems that 
$Thm(\Delta) \se Thm(\Gamma)$.
\item $\La$ As, by definition, $\Gamma\se Thm(\Gamma)$, we can apply \eqref{basic-Gamma-se-Delta} to get that 
$Thm(\Gamma)\se Thm(Thm(\Gamma))$. 

$\Ra$ We have that $\Gamma\vdash Thm(\Gamma)$, so we can apply  \eqref{gamma-delta-thm}  with $\Delta=Thm(\Gamma)$ 
to get that $Thm(Thm(\Gamma))\se Thm(\Gamma)$. 
\ee 
}

\subsection{$\Gamma$-proofs}

\bdfn
A \defnterm{$\Gamma$-proof} (or \defnterm{proof from the hypotheses $\Gamma$})  is
a sequence of patterns $\theta_1, \ldots, \theta_n$ such that  for all
 $i\in\{1,\ldots,n\}$, one of the following holds:
\be
\item $\theta_i$ is an axiom.
\item $\theta_i\in\Gamma$.
\item $\theta_i$ is the conclusion of an inference rule whose premise(s) are previous pattern(s).
\ee
An $\emptyset$-proof  is called simply a  \defnterm{proof}.
\edfn

\bdfn
Let $\vp$ be  a pattern. A \defnterm{$\Gamma$-proof of $\vp$} or a
\defnterm{proof  of $\varphi$ from the hypotheses $\Gamma$ } is a $\Gamma$-proof  
 $~\theta_1$, $\ldots$, $\theta_n$ such that  $\theta_n=\varphi$. 
\edfn

\bprop\label{Gamma-dem-demonstratie}
For any pattern $\vp$, 
\bce 
$\Gamma \vdash \varphi $ \quad iff \quad  there exists a $\Gamma$-proof of $\vp$.
\ece
\eprop
\solution{Let us denote $\Theta=\{\vp \in \setpat \mid \text{there exists a~} \Gamma\text{-proof of~}\vp\}$. 

$\Ra$ We prove by induction on $\Gamma$-theorems that $Thm(\Gamma)\se \Theta$:
\be
\item $\vp$ is an axiom or a member of $\Gamma$. Then $\vp$ is a $\Gamma$-proof of $\vp$. Hence, $\vp\in \Theta$.
\item Let $\ds\frac{\psi}{\vp}$ be an inference rule such that $\psi\in \Theta$. Then there exists a $\Gamma$-proof
$\theta_1, \ldots, \theta_n=\psi$ of $\psi$. It follows that $\theta_1, \ldots, \theta_n=\psi, \theta_{n+1}=\vp$ is a
$\Gamma$-proof of $\vp$. Thus, $\vp\in \Theta$.
\item Let $\ds \frac{\psi \quad \chi}{\vp}$ be an inference rule such that $\psi, \chi\in \Theta$. 
Then there exist a $\Gamma$-proof $\theta_1, \ldots, \theta_n=\psi$ of $\psi$ and a $\Gamma$-proof 
$\delta_1, \ldots, \delta_p=\chi$ of $\chi$. It follows that 
$\theta_1, \ldots, \theta_n=\psi, \theta_{n+1}=\delta_1, \ldots, \theta_{n+p}=\delta_p=\chi, 
\theta_{n+p+1}=\vp$ is a $\Gamma$-proof of $\vp$. Thus, $\vp\in \Theta$.
\ee

$\La$ Assume that $\vp$ has a $\Gamma$-proof $\theta_1, \ldots, \theta_n=\vp$. We prove by induction on $i$ that for all 
$i=1,\ldots,n$,  $\Gamma \vdash\theta_i$. As a consequence, $\Gamma \vdash \theta_n=\vp$.
\be
\item $i=1$. Then $\theta_1$ must be an axiom or a member of $\Gamma$. Then obviously $\Gamma \vdash\theta_1$.
\item Assume that the induction hypothesis is true for all $j=1, \ldots, i$. If 
$\theta_{i+1}$ is an axiom or a member of $\Gamma$, then obviously $\Gamma \vdash \theta_{i+1}$. Assume that $\theta_{i+1}$ is the conclusion of
an inference rule whose premise(s) are previous pattern(s). If the inference rule is
of type $\ds \frac{\theta_j}{\theta_{i+1}}$ with $j\le i$, then by the induction hypothesis we have that $\Gamma\vdash \theta_j$. 
By the definition of $\Gamma$-theorems, it follows that $\Gamma\vdash \theta_{i+1}$.
If the inference rule is
of type $\ds \frac{\theta_j \quad \theta_k}{\theta_{i+1}}$ with $j, k\le i$, then by the induction hypothesis we have that $\Gamma\vdash \theta_j$ and 
$\Gamma\vdash \theta_k$. 
By the definition of $\Gamma$-theorems, it follows that $\Gamma\vdash \theta_{i+1}$.
\ee
}

%% file: AML-syntax-derived-rules.tex
\subsection{Derived rules}

In the sequel we present different rules that will help us in proving $\Gamma$-theorems. 
These rules are of the  form
\bce
If $A_1$, $A_2$, \ldots, $A_n$ ($n\geq 1$) hold, then $B$ holds. 
\ece
and will be written as $\ds \frac{A_1 \quad A_2 \quad \ldots \quad A_n}{B}$.

\mbox{}

Let $\Gamma$ be a set of patterns.

\bprop
For every patterns $\vp$, $\psi$, 
\begin{align}\label{Gamma-vp-vp-dnd-psi-taut-Gamma-psi}
\begin{prftree}
{\Gamma \vdash \vp\quad}{\tautsign\!\! \vp \dnd \psi }
{\Gamma \vdash \psi}
\end{prftree}
\end{align}
\eprop
\solution{$\,$\\
\bt{llll}
(1) & $\Gamma$ & $\vdash \vp$ & hypothesis \\[1mm]
(2) & & $\vdash \vp \dnd \psi$ & \prule{tautology}: hypothesis \\[1mm]
(3) & & $\vdash (\vp \dnd \psi)\to (\vp \to\psi)$ &  \prule{tautology}: \eqref{weakening-si} \\[1mm]
(4) & & $\vdash \vp \to\psi$ & \prule{modus ponens}: (2), (3) \\[1mm]
(5) & $\Gamma$ & $\vdash \vp \to\psi$ & Proposition~\ref{Gamma-thm-prop-basic}.\eqref{basic-Thm-se-Gamma-Thm}  \\[1mm]
(6) & $\Gamma$ & $\vdash \psi$ & \prule{modus ponens}: (1), (5).
\et 
}

%% file: AML-soundness.tex
\section{Soundness Theorem}

\subsection{$\ssc$}

We consider the proof system 
\begin{align*}
\PSAML_{\ssc} & = \PSAML-\{\prule{$\exists$-quantifier rule}, \prule{framing}, 
\prule{set variable substitution}, 
\prule{knaster-tarski}\}.
\end{align*}

\bthm\label{soundness-ssc}
Let $\Gamma\cup\{\vp\}$ be a set of patterns. Then
\bce 
$\Gamma\vdash \vp$ implies $\Gamma\ssc\vp$.
\ece
\ethm
\solution{Let us denote $\Theta=\{\vp\in Pattern \mid \Gamma\ssc\vp\}$. We prove that $Thm(\Gamma)\se \Theta$
by induction on $\Gamma$-theorems.\\

\textbf{Axioms:}

\prule{Tautology}: By Proposition ~\ref{tautology-valid} and Remark~\ref{basic-ssc-gamma}.\eqref{basic-ssc-gamma-valid}.

\prule{$\exists$-Quantifier}: By \eqref{exists-axiom} and Remark~\ref{basic-ssc-gamma}.\eqref{basic-ssc-gamma-valid}.

\prule{Propagation$_\bot$}: By \eqref{propagation-l}, \eqref{propagation-r} and Remark~\ref{basic-ssc-gamma}.\eqref{basic-ssc-gamma-valid}.

\prule{Propagation$_\vee$}: By \eqref{propagation-sau-l}, \eqref{propagation-sau-r} and Remark~\ref{basic-ssc-gamma}.\eqref{basic-ssc-gamma-valid}.

\prule{Propagation$_\exists$}: By \eqref{propagation-exists-l}, \eqref{propagation-exists-r} and Remark~\ref{basic-ssc-gamma}.\eqref{basic-ssc-gamma-valid}.

\prule{Pre-Fixpoint}: By \eqref{pre-fixpoint-dnd} and Remark~\ref{basic-ssc-gamma}.\eqref{basic-ssc-gamma-valid}.

\prule{Existence}: By \eqref{existsx-x} and Remark~\ref{basic-ssc-gamma}.\eqref{basic-ssc-gamma-valid}.

\prule{Singleton Variable}: By \eqref{Singleton-Variable} and Remark~\ref{basic-ssc-gamma}.\eqref{basic-ssc-gamma-valid}.\\

\textbf{Rules:}

\prule{modus ponens}: Assume that $\vp, \vp\to\psi\in \Theta$, hence $\Gamma\ssc\vp$ and $\Gamma\ssc\vp\to\psi$. Apply \eqref{ssc-closure-mp} to get that
$\Gamma\ssc\psi$, hence $\psi\in \Theta$. 
}

\subsection{$\lsc$}

We consider the proof system 
\begin{align*}
\PSAML_{\lsc} & = \PSAML-\{\prule{$\exists$-quantifier rule}, \prule{set variable substitution}\}.
\end{align*}

\bthm\label{soundness-lsc}
Let $\Gamma\cup\{\vp\}$ be a set of patterns. Then
\bce 
$\Gamma\vdash \vp$ implies $\Gamma\lsc\vp$.
\ece
\ethm
\solution{Let us denote $\Theta=\{\vp\in Pattern \mid \Gamma\lsc\vp\}$. We prove that $Thm(\Gamma)\se \Theta$
by induction on $\Gamma$-theorems. \\

\textbf{Axioms:}

If $\vp$ is an axiom, we have by Theorem~\ref{soundness-ssc} that $\Gamma\ssc\vp$.  
Apply now Proposition~\ref{Gamma-ssc-imp-lsc-imp-gsc} to get that $\Gamma\lsc\vp$, hence $\vp\in\Theta$. \\

\textbf{Rules:}

\prule{modus ponens}: Assume that $\vp, \psi\in \Theta$, hence $\Gamma\lsc\vp$ and $\Gamma\lsc\vp\to\psi$. Apply \eqref{lsc-closure-mp} to get that
$\Gamma\lsc\psi$, hence $\psi\in \Theta$. 

\prule{framing}: Assume that $\vp \to \psi \in \Theta$, hence $\Gamma\lsc \vp \to \psi$. Apply  \eqref{framing-l-lsc-Gamma} to get that
$\Gamma \lsc \appln{\vp}{\chi} \to \appln{\psi}{\chi}$, hence $\appln{\vp}{\chi} \to \appln{\psi}{\chi}\in \Theta$, and \eqref{framing-r-lsc-Gamma} to get that
$\Gamma \lsc \appln{\chi}{\vp}\to \appln{\chi}{\psi}$, hence $\appln{\chi}{\vp}\to \appln{\chi}{\psi}\in \Theta$. 

\prule{Knaster-Tarski}: Assume that $X$ is free for $\psi$ in $\vp$ and that $\subfX{\psi}{\vp} \to \psi\in \Theta$, 
hence $\Gamma\lsc\subfX{\psi}{\vp} \to \psi$. Apply Proposition~\ref{Knaster-Tarski-rule-lsc-prop-2} 
to get that $\Gamma\lsc \mu X.\vp \to \psi$,
hence $\mu X.\vp \to \psi\in\Theta$.
}

\subsection{$\gsc$}

\bthm\label{soundness-gsc}
Let $\Gamma\cup\{\vp\}$ be a set of patterns. Then
\bce 
$\Gamma\vdash \vp$ implies $\Gamma\gsc\vp$.
\ece
\ethm
\solution{
The proof is by induction on $\Gamma$-theorems.\\

\textbf{Axioms:}

If $\vp$ is an axiom, we have by Theorem~\ref{soundness-ssc} that $\Gamma\ssc\vp$.  
Apply now Proposition~\ref{Gamma-ssc-imp-lsc-imp-gsc} to get that $\Gamma\gsc\vp$, hence $\vp\in\Theta$. \\

\textbf{Rules:}

\prule{modus ponens}: Assume that $\vp, \psi\in \Theta$, hence $\Gamma\gsc\vp$ and $\Gamma\gsc\vp\to\psi$. Apply \eqref{closure-mp-gsc} to get that
$\Gamma\gsc\psi$, hence $\psi\in \Theta$. 

\prule{$\exists$-quantifier rule}: Assume that $x \not\in FV(\psi)$ and  $\vp \to \psi \in \Theta$, hence $\Gamma\gsc \vp \to \psi$. 
Apply \eqref{closure-exists-quantifier-gsc} to get that $\Gamma\gsc \exists x.\vp \to \psi$, hence $\exists x.\vp \to \psi\in \Theta$. 

\prule{framing}: Assume that $\vp \to \psi \in \Theta$, hence $\Gamma\gsc \vp \to \psi$. Apply  \eqref{framing-l-gsc-Gamma} to get that
$\Gamma \gsc \appln{\vp}{\chi} \to \appln{\psi}{\chi}$, hence $\appln{\vp}{\chi} \to \appln{\psi}{\chi}\in \Theta$, and \eqref{framing-r-gsc-Gamma} to get that
$\Gamma \gsc \appln{\chi}{\vp}\to \appln{\chi}{\psi}$, hence $\appln{\chi}{\vp}\to \appln{\chi}{\psi}\in \Theta$. 

\prule{Set Variable Substitution}: Assume that $X$ is free for $\psi$ in $\vp$  and  $\vp \in \Theta$, hence $\Gamma\gsc \vp$. 
Apply Proposition~\ref{set-variable-substitution-free-2} to get that $\Gamma\gsc \subfX{\psi}{\vp}$, hence $\subfX{\psi}{\vp}\in \Theta$.

\prule{Knaster-Tarski}: Assume that $X$ is free for $\psi$ in $\vp$ and that $\subfX{\psi}{\vp} \to \psi\in \Theta$, 
hence $\Gamma\gsc\subfX{\psi}{\vp} \to \psi$. Apply Proposition~\ref{Knaster-Tarski-rule-gsc-prop-2} 
to get that $\Gamma\gsc \mu X.\vp \to \psi$,
hence $\mu X.\vp \to \psi\in\Theta$.

}

%% file: AML-semantics-def-equal-tot.tex
\section{Adding definedness, equality, totality, membership}

In the sequel, we consider the signature $\AMLsigdef=(EVar, SVar, \Sigma)$ satisfying the following:
\bce 
$\defsymb\in \Sigma$; $\defsymb$ is called the \defnterm{definedness symbol}.
\ece 

We introduce the derived  connective $\defmapn$, $\totmapn$, $\eqdef$ by the following abbreviations:
\begin{align}
\ceil{\vp} & := \appln{\defsymb}{\vp},\\
\floor{\vp} & := \neg\ceil{\neg\vp},\\
\vp \eqdef \psi & := \floor{\vp\dnd\psi},\\
x \in \vp & := \ceil{x\si\vp}.
\end{align}

We assume that $\eqdef$ has higher precedence than $\si, \sau$, $\ra$, $\dnd$.

\bdfn
A $\AMLsigdef$-structure $\cA$ is obtained by defining $\defsymb^\cA \se A$ such that 
$$
\defsymb^\cA \appls \{a\} = A \quad \text{for all }a \in A. 
$$
\edfn

\blem\label{basic-prop-AMLsigdef}
Let $\cA$ be a $\AMLsigdef$-structure. Then
\be
\item\label{basic-prop-AMLsigdef-1} $A \appls \{a\} = A$.
\item\label{basic-prop-AMLsigdef-2} $\defsymb^\cA \appls B = A$ for all $\emptyset \ne B \se A$.
\ee
\elem
\solution{
\be
\item We have that, for all  $a \in A$,  
$$ A = \defsymb^\cA \appls \{a\} \stackrel{\eqref{B-se-C-appls-D}}{\se} A \appls \{a\} \se A.$$ 
Thus, $A \appls \{a\} = A$.
\item Let $\emptyset \ne B \se A$. There exists $a\in B$. Then 
$$A = \defsymb^\cA \appls \{a\} \stackrel{\eqref{B-se-C-appls-D}}{\se} \defsymb^\cA \appls B \se A.$$ 
Thus, $\defsymb^\cA \appls B = A$.
\ee
}

In the sequel, we write  ``Let $(\cA, \eval)$'' instead of ``Let $\cA$ be a $\AMLsigdef$-structure 
and $\eval$ be an $\cA$-valuation''.

\bprop\label{evalp-ceil-floor-vp}
Let $(\cA, \eval)$. For any pattern $\vp$, 
\begin{align}
\evalp(\ceil{\vp}) & = \begin{cases}
\emptyset & \text{if }\evalp(\vp)=\emptyset, \\
A & \text{if }\evalp(\vp) \ne \emptyset,
\end{cases}  \label{evalp-ceil-vp}\\
\evalp(\floor{\vp}) & = \begin{cases}
A & \text{if } \evalp(\vp) = A, \\
\emptyset & \text{if } \evalp(\vp) \ne A.
\end{cases} \label{evalp-floor-vp}
\end{align}
\eprop
\solution{

We get that 
\begin{align*}
\evalp(\ceil{\vp}) & = \evalp(\appln{\defsymb}{\vp})= \evalp(\defsymb)\appls \evalp(\vp) = 
\defsymb^\cA  \appls \evalp(\vp)\\
& = \begin{cases}
\emptyset & \text{if }\evalp(\vp)=\emptyset, \quad \text{by \eqref{B-appl-empty}}\\
A & \text{if }\evalp(\vp) \ne \emptyset, \quad \text{by Lemma \ref{basic-prop-AMLsigdef}.\eqref{basic-prop-AMLsigdef-2}}
\end{cases} \\[2mm]
\evalp(\floor{\vp}) & = \evalp(\neg \ceil{\neg\vp}) = A \setminus \evalp(\ceil{\neg\vp}) 
\stackrel{\eqref{evalp-ceil-vp}}{=} \begin{cases}
A \setminus \emptyset & \text{if }\evalp(\neg\vp)=\emptyset\\
A \setminus  A & \text{if }\evalp(\neg\vp) \ne \emptyset
\end{cases} \\
& = \begin{cases}
A  & \text{if }\evalp(\neg\vp)=\emptyset\\
\emptyset & \text{if }\evalp(\neg\vp) \ne \emptyset
\end{cases} 
 = \begin{cases}
A  & \text{if } A \setminus \evalp(\vp)=\emptyset\\
\emptyset & \text{if } A \setminus \evalp(\vp) \ne \emptyset
\end{cases} 
= \begin{cases}
A  & \text{if } \evalp(\vp) = A\\
\emptyset & \text{if } \evalp(\vp) \ne A
\end{cases}
\end{align*}
}

As immediate consequences of Proposition \ref{evalp-ceil-floor-vp}, we get that

\bprop\label{def-tot-vp-predicate}
For any pattern $\vp$,  $\ceil{\vp}$ and $\floor{\vp}$ are predicate patterns.
\eprop

\bprop\label{eqdef-predicate}
For any patterns $\vp$, $\psi$, $\vp \eqdef \psi$ is a predicate pattern.
\eprop
\solution{Apply Proposition \ref{def-tot-vp-predicate} and the definition of $\eqdef$.
}

\bprop
For any element variable $x$, 
\beq
\models \ceil{x}.\label{ceil-x-valid}
\eeq
\eprop
\solution{Let $(\cA, \eval)$. As $\evalp(x)\ne \emptyset$, we get that $\evalp(\ceil{x})=A$.
}

\bprop
For any pattern $\vp$,  
\bea
\models & \vp \to \ceil{\vp}, \label{ax-vp-to-defvp-new-system}\\
\models & \floor{\vp} \to \vp. \label{totvp-to-vp}
\eea
\eprop
\solution{Let $(\cA, \eval)$. We have that 
\bea
\cA \models (\vp \to\ceil{\vp})[\eval] & \text{iff} &  \evalp(\vp) \se \evalp(\ceil{\vp}), 
\text{ which is true, by \eqref{evalp-ceil-vp}}, \\
\cA \models (\floor{\vp} \to \vp)[\eval] & \text{iff} &  \evalp(\floor{\vp}) \se \evalp(\vp), 
\text{ which is true, by \eqref{evalp-floor-vp}}.
\eea
}

\bprop
For any distinct variables $x$, $y$, 
\begin{align}
& \models \exists x.(x \eqdef y) \label{exists-x-x=y-new-system}
\end{align}
\eprop
\solution{Let $(\cA,\eval)$ and $b:= \eval(y)$. Then 
$$\evalp(\exists x. (x \eqdef y))=\bigcup_{a\in A} \evalpxa(x \eqdef y) \supseteq \evalpxb(x \eqdef y).$$
As $$\evalpxb(x\dnd y) = A \setminus (\evalpxb(x) \Delta \evalpxb(y))= A \setminus (\{b\}\Delta \{b\})=A,$$
we can apply \eqref{evalp-floor-vp} to get that 
$\evalpxb(x \eqdef y)= \evalpxb(\floor{x\dnd y})=A$. 

It follows that $\evalp(\exists x (x \eqdef y))=A$, that is $\cA\models (\exists x (x \eqdef y))[\eval]$.
}

\bprop
For any patterns $\vp$, $\psi$,
\begin{align}
\appln{\ceil{\vp}}\psi \to \ceil{\vp}, \label{ax-propagation-def-1-new-system} \\
\appln{\psi}{\ceil{\vp}} \to \ceil{\vp}. \label{ax-propagation-def-2-new-system} 
\end{align}
\eprop
\solution{Let $(\cA,\eval)$. We have the following cases:
\be 
\item $\evalp(\vp) = \emptyset$, hence $\evalp(\ceil{\vp})=\emptyset$.
Then $\evalp(\appln{\ceil{\vp}}\psi)= \evalp(\ceil{\vp})\appls \evalp(\psi)=\emptyset \appls \evalp(\psi) = \emptyset$ and 
 $\evalp(\appln{\psi}{\ceil{\vp}})= \evalp(\psi) \appls \evalp(\ceil{\vp})=\evalp(\psi) \appls \emptyset= 
 \emptyset$. Thus, 
\eqref{ax-propagation-def-1-new-system}, \eqref{ax-propagation-def-2-new-system} hold.
\item $\evalp(\vp) \ne  \emptyset$, hence $\evalp(\ceil{\vp})=A$. Then obviously 
\eqref{ax-propagation-def-1-new-system}, \eqref{ax-propagation-def-2-new-system} hold.
\ee
}

\bprop
\begin{align}
& \models  \ceil{\bot} \dnd \bot. \label{ax-bot-dnd-ceilbot-new-system}
\end{align}
\eprop
\solution{Let $(\cA,\eval)$. We have that $\evalp(\ceil{\bot})= \emptyset$, as $\evalp(\bot)= \emptyset$. 
}

\bprop
For any pattern $\vp$ and variable $x$, 
\begin{align}
& \models  \neg(x\in \vp)\sau \neg(x\in \neg\vp). \label{singleton-simple-valid}
\end{align}
\eprop
\solution{
Let $(\cA,\eval)$. 
Denote 
$$\theta=\neg(x\in \vp)\sau \neg(x\in \neg\vp)=\neg\ceil{x\si\vp} \sau \neg\ceil{x\si\neg\vp}.$$
We have to prove that $\cA\models (\theta)[\eval]$, that is, $\evalp(\theta)=A$. 

Remark that
\bua
\evalp(\theta)  & = & \evalp(\neg\ceil{x\si\vp} \sau \neg\ceil{x\si\neg\vp}) =
 \evalp(\neg\ceil{x\si\vp}) \cup \evalp(\neg\ceil{x\si\neg\vp}) \\
 & = & C_A\evalp(\ceil{x\si\vp}) \cup C_A\evalp(\ceil{x\si\neg\vp}) \stackrel{\eqref{de-morgan-2}}{=} 
 C_A\big(\evalp(\ceil{x\si\vp}) \cap \evalp(\ceil{x\si\neg\vp})\big)
 \eua

Thus, $\evalp(\theta)=A$ iff $\evalp(\ceil{x\si\vp}) \cap \evalp(\ceil{x\si\neg\vp})=\emptyset$.
 
 We have two cases:
\be
\item $\eval(x)\in \evalp(\vp)$. Then 
$\evalp(x \si \neg \vp)=\evalp(x)\cap \evalp(\neg\vp)= 
\{\eval(x)\} \cap C_A\evalp(\vp)=\emptyset$. Hence, by  \eqref{evalp-ceil-vp}, 
$\evalp(\ceil{x\si\neg \vp})=\emptyset$. In particular, $\evalp(\theta)=A$.
\item $\eval(x)\notin \evalp(\vp)$. Then 
$\evalp(x \si \vp)=\evalp(x)\cap \evalp(\vp)= 
\{\eval(x)\} \cap \evalp(\vp)=\emptyset$. Hence, by  \eqref{evalp-ceil-vp}, 
$\evalp(\ceil{x\si\vp})=\emptyset$. In particular, $\evalp(\theta)=A$.
\ee
}

%% file: appendix-proof-system-AML.tex
\section{AML proof system $\PSAML$ }\label{AML-proof-system}

\begin{figure*}[h]
\centering

\begin{tabular}{ll}
\hline
\hline\\
\prule{Tautology} & $\vp$ \qquad \qquad \qquad \quad \, if $\vp$ is a tautology \\[4mm]
\prule{$\exists$-Quantifier} & $\subfx{y}{\vp} \to \exists x.\varphi$ \quad if $x$ is free 
for $y$ in $\vp$\\[4mm]
\hline\\
\prule{modus ponens} &
$\ds \frac{\vp \quad \vp \to \psi}{\psi}$\\[4mm]
\prule{$\exists$-quantifier rule} &
$
\ds \frac{\vp \to \psi}{\exists x.\vp \to \psi}$  \qquad \qquad  if $x \not\in FV(\psi)$
\\[4mm]
\hline\\
\prule{Propagation$_\bot$} &
$\appln{\vp}{\bot} \to \bot$ \quad $ \appln{\bot}{\vp} \to \bot$
\\[4mm]
\prule{Propagation$_\vee$} & 
$\appln{(\vp \sau\psi)}{\chi} \to \appln{\vp}{\chi}\sau \appln{\psi}{\chi}$ \quad 
$\appln{\chi}{(\vp \sau\psi)} \to \appln{\chi}{\vp}\sau \appln{\chi}{\psi}$
\\[4mm]
\prule{Propagation$_\exists$} &
$\appln{(\exists x. \vp)}{\psi} \to \exists x. \appln{\vp}{\psi}$ \qquad 
$\appln{\psi}{(\exists x.\vp)} \to \appln{\exists x. \psi}{\vp}$ \quad  if $x \not\in FV(\psi)$
\\[4mm]
\prule{framing} &
$ \ds \frac{\vp \to \psi}
{\appln{\vp}{\chi} \to \appln{\psi}{\chi}} \qquad \frac{\vp \to \psi}{\appln{\chi}{\vp}\to \appln{\chi}{\psi}}$
\\[4mm]
\hline\\
\prule{Set Variable Substitution} & 
$ \ds \frac{\vp}{\subfX{\psi}{\vp}}$ \qquad  \qquad \quad\,\,  if $X$ is free for $\psi$ in $\vp$
\\[4mm]
\hline\\
\prule{Pre-Fixpoint} &
$\subfX{\mu X. \vp}{\vp} \to  \mu X.\vp$ \quad  if $\vp$ is  positive in $X$ and  $X$ is free for $\mu X. \vp$ in $\vp$
\\[4mm]
\prule{Knaster-Tarski} &
$ \ds \frac{\subfX{\psi}{\vp} \to \psi }{\mu X.\vp \to \psi}$ \quad \qquad \,\, if $X$ is free for $\psi$ in $\vp$
\\[4mm]
\hline\\
\existence &
$\exists x.x$
\\[4mm]
\prule{Singleton Variable} &
$\neg (C_1[x \wedge \varphi] \wedge C_2[x \wedge \neg \varphi])$\\[4mm]
\hline
\hline
\end{tabular}
\caption{Proof system $\PSAML$ for AML}
\label{fig:PSAML}
\vspace*{-2ex}
\end{figure*}

\newpage

%% file: appendix-set-theory.tex
\section{Set theory}\label{app-set-theory}

Let $A$, $B$ be sets. We use the following notations:
\be
\item $A\cup B$ for the union of $A$ and $B$.
\item $A\cap B$ for the intersection of $A$ and $B$.
\item $A\setminus B$ for the difference between $A$ and $B$.
\item $A\diffsym B$ for the symmetric difference of $A$ and $B$.
\item $2^A$ for the powerset of $A$. 
\item $C_AB$ for the complementary of $B$, when $B\se A$.
\item $Fun(A,B)$ for the set of functions from $A$ to $B$.  
\item $A^n$ for the set $A\times A\times \ldots \times A$, where $n\geq 2$. 
\ee

Let $f:A\to B$ be a mapping. For every $C\se A$, we denote by $f|_C$ the restriction of $f$ to $C$. Thus,
$f|_C:C\to B, \, (f|_C)(x)=f(x)$ for every $x\in C$.

\subsection{Expressions over a set}

An \defnterm{expression over} $A$ is a finite sequence of elements from $A$. We denote an expression over $A$ of
length $n\in \N^*$ by $a_1a_2\ldots a_n$, where $a_i\in A$ for every $i=1,\ldots,n$. The empty expression (of length $0$)
is denoted by $\lambda$. The concatenation of exoressions over $A$ is defined as follows: 
if $a=a_1\ldots a_n$ and $b=b_1\ldots b_k$, then $ab = a_1\ldots a_nb_1\ldots b_k$.

Let $a=a_0a_1\ldots a_n $ be an expression over $A$, where $a_i\in A$ 
for all $i=0,\ldots, k-1$. 
\be
\item If $0\le i\le j\le k-1$, then the expression $a_i\ldots a_j$ is called 
the $(i,j)$-\defnterm{subexpression} of $a$.
\item We say that an expression $b$ \defnterm{occurs} in $a$ if there exists 
$0\le i\le j\le k-1$ such that $b$ is the $(i,j)$-subexpression of $a$.
\item A \defnterm{proper initial segment} of $a$ is an expression $a_0a_1\ldots a_i$, where
$0\le i < n-1$. 
\ee

\subsection{Set-theoretic properties used in the lecture notes}

\bprop
Let $A, B, C,D$ be sets. Then
\begin{align}
B \se C \text{~and~} B \se D & \text{~implies~}  B\se C\cap D \label{B-se-CcapD}\\
A\setminus B = \emptyset & \text{~iff~} A\se B \label{AsetminusB-AseB}\\
A\cup (B\cup C) = (A\cup B)\cup C,\\
A\cup (B\cap C) & = (A\cup B) \cap (A\cup C),\\
A\cap (B\cup C) & = (A\cap B) \cup (A\cap C),\\
(A\cap D)\cup (B\cap C) & = (A\cup B) \cap (A\cup C) \cap (D\cup B) \cap (D\cup C),\\
(A\cup D)\cap (B\cup C) & = (A\cap B) \cup (B\cap C)\cup (D\cap B) \cup (D\cap C), \\
A\diffsym B & =\emptyset  \quad \text{iff} \quad A=B \label{app-sets-diff-dym-empty},\\
A\setminus C & \se (A\setminus B)\cup (B\setminus C) \label{A-C-se-A-B-cup-B-C}
\end{align}
\eprop

\bprop
Let $B, D \se A$. Then
\begin{align}
B\cup C_AB & = A, \label{B-cup-CAB=B}\\
C_A(C_AB) &= B \label{CA-idemp}\\
C_A(B\cup D) &= C_AB\cap C_AD \label{de-morgan-1}\\
C_A(B\cap D) &= C_AB\cup C_AD \label{de-morgan-2}\\
C_A(C_AB\cup C_AD) &= B\cap D \label{de-morgan-si-sau}\\
B\setminus D &= B\cap C_AD, \label{B-D=B-capCAC}\\
C_A(B\setminus D) &= C_A B\cup D \label{CAB-D=B-cap-CAC}\\
B \se D & \text{~iff~} C_A D\se C_A B \label{CA-B-se-D}\\
C_AB =\emptyset & \text{~iff~} B=A \label{CAB-empty}\\
C_AB =A & \text{~iff~} B=\emptyset \label{CAB-A}, \\
C_A(B\setminus D)\cap C_A(D\setminus B) &= C_A(B\diffsym D) \label{app-sets-diff-symm-1}
\end{align} 
\eprop

\subsection{Families of sets}

Let $(B_i)_{i\in I}$ be a family of subsets of $A$.

\bfact
If $I=\emptyset$, then $\bigcap_{i\in I} B_i=A$.
\efact

\bprop
\begin{align} 
C_A\left(\bigcup_{i\in I} B_i\right) & = \bigcap_{i\in I} C_AB_i
\label{de-morgan-family-1}\\
C_A\left(\bigcap_{i\in I} B_i\right) & = \bigcup_{i\in I} C_AB_i
\label{de-morgan-family-CAcap}\\
A \cup \bigcap_{i\in I} B_i  & = \bigcap_{i\in I} (A \cup B_i) \label{distrib-cap-family-left}
\end{align} 
\eprop

%% file: appendix-set-theory-app.tex
\subsection{Application}

In the sequel, $A$ is a nonempty set and $\_\appls\_:A\times A\to 2^A$ is a binary \defnterm{application} function.

We extend the application function $\_\appls\_$  as follows:
$$\_\appls\_: 2^A \times 2^A\to 2^A, \quad 
B \appls C =\bigcup_{b\in B, c\in C} b \appls c.
$$

\bprop
Let $B, C, D\se A$ and $(A_i)_{i\in I}$ be a family of subsets of $A$.
\begin{align}
B\appls \emptyset & = \emptyset \appls B=\emptyset, \label{B-appl-empty}\\
(C\cup D)  \appls B & = (C\appls B) \cup (D\appls B), \label{C-cup-D-appls-B}\\
B\appls (C\cup D) & = (B\appls C) \cup (B\appls D), \label{B-appls-C-cup-D}\\
(C\cap D)  \appls B & \se  (C\appls B) \cap (D\appls B), \label{C-cap-D-appls-B}\\
B\appls (C\cap D) & \se (B\appls C) \cap (B\appls D), \label{B-appls-C-cap-D}\\
\left(\bigcup_{i\in I}A_i\right) \appls B & = \bigcup_{i\in I}(A_i\appls B),  \label{cup-family-appls-B}\\
B  \appls \left(\bigcup_{i\in I}A_i\right) & = \bigcup_{i\in I}(B\appls A_i),  \label{B-appls-cup-family}\\
\left(\bigcap_{i\in I}A_i\right) \appls B & \se \bigcap_{i\in I}(A_i\appls B),  \label{cap-family-appls-B}\\
B  \appls \left(\bigcap_{i\in I}A_i\right) & \se \bigcap_{i\in I}(B\appls A_i),  \label{B-appls-cap-family}\\
B \se C & \text{~implies~} B\appls D \se C\appls D \text{~and~} D\appls B \se D\appls C 
\label{B-se-C-appls-D}\\
B \se C  \text{~and~} D\se E & \text{~implies~} B\appls D \se C\appls E  \label{B-se-C-appls-D-se E}\\
B \se C & \text{~implies~} C_AB\appls D \supseteq C_AC\appls D \text{~and~} D\appls C_AB \supseteq D\appls C_AC \label{B-se-C-appls-D-comp}
\end{align}
\eprop
\solution{
\eqref{B-appl-empty}: Obviously.

\eqref{C-cup-D-appls-B}:  $(C\cup D)\appls B =\bigcup_{c\in C\cup D,b\in B}  c \appls b =
\left(\bigcup_{c\in C, b\in B} c \appls b \right)\cup \left(\bigcup_{c\in D, b\in B} c \appls b \right) = (C\appls B) \cup (D\appls B) $.

\eqref{B-appls-C-cup-D}:  $B\appls (C\cup D)=\bigcup_{b\in B, c\in C\cup D} b \appls c =
\left(\bigcup_{b\in B, c\in C} b \appls c \right)\cup \left(\bigcup_{b\in B, c\in D} b \appls c \right) = (B\appls C) \cup (B\appls D)$.

\eqref{C-cap-D-appls-B}:  Let $a\in (C\cap D)\appls B$. Then $a=e\appls b$ with $e\in C\cap D$ and $b\in B$. 
It follows that $a\in (C\appls B) \cap (D\appls B) $.

\eqref{B-appls-C-cap-D}: Let $a\in B\appls (C\cap D)$. Then $a=b\appls e$ with $e\in C\cap D$ and $b\in B$. 
It follows that 
$a\in (B\appls C) \cap (B\appls D)$.

\eqref{cup-family-appls-B}: $\left(\bigcup_{i\in I}A_i\right) \appls B = \bigcup_{a\in \bigcup_{i\in I}A_i, b\in B} a\appls b = \bigcup_{i\in I} \bigcup_{a\in A_i, b\in B}
a\appls b = \bigcup_{i\in I}(A_i\appls B)$. 

\eqref{B-appls-cup-family}: $B\appls \left(\bigcup_{i\in I}A_i\right) = \bigcup_{b\in B, a\in \bigcup_{i\in I}A_i} b\appls a = 
\bigcup_{i\in I} \bigcup_{a\in A_i, b\in B}
b\appls a = \bigcup_{i\in I}(B\appls A_i)$. 

\eqref{cap-family-appls-B}: Let $a\in \left(\bigcap_{i\in I}A_i\right) \appls B$. Then $a=e\appls b$ with $b\in B$ and  $e\in \bigcap_{i\in I}A_i$,
hence $a\in A_i$ for all $i\in I$. It follows that $a\in \bigcap_{i\in I}(A_i\appls B)$. 

\eqref{B-appls-cap-family}: Let $a\in B\appls\left(\bigcap_{i\in I}A_i\right)$. Then $a=b\appls e$ with $b\in B$ and  $e\in \bigcap_{i\in I}A_i$,
hence $a\in A_i$ for all $i\in I$. It follows that $a\in \bigcap_{i\in I}(B\appls A_i)$. 

\eqref{B-se-C-appls-D}: $B\appls D =\bigcup_{b\in B, d\in D} b \appls d \se 
\bigcup_{b\in C, d\in D} b \appls d = C\appls D$ and, similarly, $D\appls B \se D\appls C$. 

\eqref{B-se-C-appls-D-se E}: Apply \eqref{B-se-C-appls-D} twice to get that $B\appls D\se C\appls D\se C\appls E$. 

\eqref{B-se-C-appls-D-comp} As $B\se C$, we have that $C_AB\supseteq C_AC$. Apply \eqref{B-se-C-appls-D}.
}

%% file: appendix-KnasterTarski.tex
\subsection{Knaster-Tarski Theorem}

In the sequel $A$ is a nonempty set and $F:2^A \to 2^A$ is a mapping. 

\bdfn
A subset $D$ of $A$ is a \defnterm{fixpoint} of $F$ if $F(D)=D$.
\edfn

\bdfn
A subset $D$ of $A$ is 
\be
\item \defnterm{the least fixpoint} of $F$ if  $D$ is a fixpoint of $F$ and for every fixpoint 
$D'$ of $F$, we have that $D\se D'$.
\item  \defnterm{the greatest fixpoint} of $F$ if  $D$ is a fixpoint of $F$ and for every fixpoint 
$D'$ of $F$, we have that $D\supseteq D'$.
\ee
\edfn

\bdfn
$F$ is said to be \defnterm{monotone} if for every $B, C\se A$, 
$$ B\se C \quad  \text{implies} \quad F(B)\se F(C).$$
\edfn

\begin{theorem}[Knaster-Tarski~{\cite{Tar55}}]\label{Knaster-Tarski}\,\\
Let $F:2^A \to 2^A$ be a monotone function. Define 
\begin{align}
\mu F = \bigcap \{B \subseteq A \mid F(B) \subseteq B\} \quad \text{and} \quad \nu F = \bigcup \{C \subseteq A \mid C \subseteq F(C)\}.
\end{align}

Then $\mu F$ is the least fixpoint of $F$ and $\nu F$ is the greatest fixpoint of $F$. Furthermore, 
\begin{align}
\mu F =\bigcap \{D \subseteq A \mid F(D) = D\} \quad \text{and} \quad \nu F = \bigcup \{D \subseteq A \mid F(D) = D\}. \label{mu-nu-fixpoint-cup-cap}
\end{align}
\end{theorem}
\solution{Let us denote $\cB=\{B \subseteq A \mid F(B) \subseteq B\}$ and $\cC=\{ C \subseteq A \mid C \subseteq F(C)\}$. \\[1mm]
\textbf{Claim 1:}  $F(\mu F)=\mu F$.\\[1mm]
\textbf{Proof of claim:} ``$\subseteq$" For every $B\in\cB$, we have 
that $\mu F\se B$, hence $F(\mu F)\se F(B)\se B$. 
It follows that $F(\mu F)\se \bigcap \cB=\mu F$. 

''$\supseteq$" As $F(\mu F)\se \mu F$, we have that $F(F(\mu F))\se F(\mu F)$, so $F(\mu F)\in \cB$.
It follows that $\mu F\se F(\mu F)$. \hfill  $\blacksquare$\\

\textbf{Claim 2:}  $\mu F$ is the least fixpoint of $F$.\\[1mm]
\textbf{Proof of claim:} If $D$ is another fixpoint of $F$, we have, in particular, that $F(D)\se D$, 
so  $D\in \cB$. It follows that  $\mu F\se D$. \hfill $\blacksquare$\\

\textbf{Claim 3:}  $F(\nu F)=\nu F$.\\[1mm]
\textbf{Proof of claim:} ``$\supseteq$"  For every $C\in \cC$, we have that $C\se \nu F$.
It follows that  $F(C)\se F(\nu F)$, so 
$C\se F(C)\se F(\nu F)$.
It follows that $\nu F = \bigcup \cC\se F(\nu F)$. 

``$\subseteq$" As $\nu F \se F(\nu F)$, we have that $F(\nu F) \se F(F(\nu F))$, so $F(\nu F)\in \cC$.
It follows that $F(\nu F) \se \nu F$. \hfill  $\blacksquare$\\

\textbf{Claim 4:}  $\nu F$ is the greatest fixpoint of $F$. \\[1mm]
\textbf{Proof of claim:} If $D$ is another fixpoint of $F$, we have, in particular, that $D\se F(D)$, 
so  $D\in \cC$. It follows that  $D\se \nu F$. \hfill  $\blacksquare$\\

Let us denote $\calD=\{D \subseteq A \mid F(D) = D\}$. \\

\textbf{Claim 5:}  $\mu F=\bigcap \calD$.\\[1mm]
\textbf{Proof of claim:} ``$\supseteq$" We have, by Claim 1,  that $\mu F\in \calD$, hence $\mu F\supseteq \bigcap \calD$. 

''$\subseteq$" For every $D\in\calD$, we have, by Claim 2, that $\mu F \se D$.
Hence, $\mu F\subseteq \bigcap \calD$. \hfill  $\blacksquare$\\

\textbf{Claim 6:}  $\nu F=\bigcup \calD$.\\[1mm]
\textbf{Proof of claim:} ``$\subseteq$" We have, by Claim 3,  that $\nu F\in \calD$, hence $\nu F\subseteq \bigcup \calD$. 

''$\supseteq$" For every $D\in\calD$, we have, by Claim 4, that $D \se \nu F$. Hence, $\bigcup \calD \se \nu F$. 
}

Thus, if $F:2^A \to 2^A$ is monotone, then $F(\mu F)=\mu F$, $F(\nu F)=\nu F$ and for every 
$B\se A$ such that $F(B)=B$, 
$$
\mu F\se B \se \nu F. 
$$

\newpage